\documentstyle[12pt]{article}
\newlength{\TZ}
\newcommand{\beq}{\begin{equation}}
\newcommand{\eeq}{\end{equation}}

\newcommand{\spup}[1]{\mbox{$\sigma^{+}_{#1}$}}
\newcommand{\spdo}[1]{\mbox{$\sigma^{-}_{#1}$}}

\newcommand{\crea}[2]{\mbox{$\hat{#1}^{\dagger}_{#2}$}}
\newcommand{\anni}[2]{\mbox{$\hat{#1}_{#2}$}}
\newcommand{\nac}[1]{\mbox{$\tilde{n}_{#1}(t)$}}

\newcommand{\spx}[1]{\mbox{$\sigma^{x}_{#1}$}}
\newcommand{\spz}[1]{\mbox{$\sigma^{z}_{#1}$}}
\newcommand{\spy}[1]{\mbox{$\sigma^{y}_{#1}$}}
\newcommand{\ave}[1]{\mbox{$< #1>$}}
\newcommand{\comm}[2]{\left[\, #1 \, , \, #2 \, \right]}
\newcommand{\anti}[2]{\left\{\, #1 \, , \, #2 \, \right\}}
\newcommand{\be}{\begin{equation}}
\newcommand{\bel}[1]{\begin{equation}\label{#1}}
\newcommand{\ee}{\end{equation}}
\newcommand{\bea}{\begin{eqnarray}}
\newcommand{\ba}{\begin{array}}
\newcommand{\eea}{\end{eqnarray}}
\newcommand{\ea}{\end{array}}

\newcommand{\BEQ}{\begin{equation}}    
\newcommand{\BEA}{\begin{eqnarray}}
\newcommand{\EEQ}{\end{equation}}      
\newcommand{\EEA}{\end{eqnarray}}
\newcommand{\eps}{\epsilon}                      
\newcommand{\lmb}{\lambda}                       
\newcommand{\sig}{\sigma}                        
\newcommand{\vph}{\varphi}                       
\newcommand{\rar}{\rightarrow}                   
\newcommand{\lrar}{\leftrightarrow}              
\newcommand{\es}{\emptyset}                      
\newcommand{\wit}[1]{\widetilde{#1}}             
\newcommand{\ket}[1]{\left|#1\right\rangle}      
\newcommand{\bra}[1]{\left\langle #1\right|}     
\newcommand{\lvk}[1]{{\stackrel{\leftarrow}{#1}}}
\newcommand{\zeile}[1]{\vskip #1 \baselineskip}  
\newcommand{\vekz}[2]
     {\mbox{${\begin{array}{c} #1  \\ #2 \end{array}}$}}
\newcommand{\matz}[4]
     {\mbox{${\begin{array}{cc} #1 & #2  \\ #3 & #4 \end{array}}$}}
 
\newcommand{\build}[3]{\mathrel{\mathop{\kern 0pt#1}\limits_{#2}^{#3} }}
 
\newcommand{\appsection}[2]{\setcounter{equation}{0}\setcounter{subsection}{0}
                            \section*{Appendix #1. #2}
\renewcommand{\theequation}{#1.\arabic{equation}}
              \renewcommand{\thesection}{#1} }
\catcode`\@=11
\def\numberbysection{\@addtoreset{equation}{section}
        \def\theequation{\thesection.\arabic{equation}}}
\numberbysection

\begin{document}
\baselineskip 0.3in
%
%
\begin{titlepage}
\null
\vskip 1cm
\begin{center}
\vskip 0.5in
{\Large \bf Reaction-Diffusion Processes from Equivalent 
Integrable Quantum Chains}
\vskip 0.5in
Malte Henkel$^{a}$, Enzo Orlandini$^{b,c}$ and Jaime Santos$^b$ \\[.3in]
$^a${\em Laboratoire de
Physique des Mat\'eriaux\footnote{Unit\'e de Recherche associ\'ee 
au CRNS no. 155},
Universit\'e Henri Poincar\'e Nancy I, BP 239, \\
F - 54506 Vand{\oe}uvre-l\`es-Nancy Cedex, France}, \\
$^b${\em Theoretical Physics, Department of Physics, \\
University of Oxford, 1 Keble Road, Oxford OX1 3NP, UK}
\\ $^c${\em Service de Physique Th\'eorique, CEN Saclay, F - 91191 
Gif-sur-Yvette Cedex, France\footnote{address after 1st october 1996}}
\end{center}
\zeile{2}
%
\begin{abstract}
One-dimensional reaction-diffusion systems are mapped through 
a similarity transformation onto integrable (and {\em a priori} 
non-stochastic) quantum chains. Time-dependent properties of these 
chemical models can then be found exactly. The reaction-diffusion processes
related to free fermion systems with site-independent interactions are
classified. The time-dependence of the mean particle density is calculated. 
Furthermore new integrable stochastic 
processes related to the Heisenberg XXZ chain are identified
and the relaxation times for the particle density and density
correlation for these systems are found. 
\end{abstract}
\zeile{3}
\end{titlepage}
 
\newpage
%
%
\section{Introduction}

The physics of interacting particles out of thermodynamic equilibrium
is an intriguing and challenging problem. For important 
reaction-diffusion processes which display a wide variety of interesting
and unexpected phenomena, see \cite{Marr96,Priv96,McKa95,Schm95,Droz95} 
and references therein. 
They are not only of experimental importance but
also pose a theoretical problem of considerable difficulty as there is no
general framework for their treatment, such as 
detailed balance for equilibrium systems.
While the classic approach, involving phenomenological rate equations, is 
sufficient to the describe the time-dependent behavior of non-equilibrium 
systems in sufficiently high spatial dimensions, 
in low dimensions the diffusive 
mixing is ineffective and new phenomena appear. 

As an example, consider a system
consisting of a single species of particles $A$ undergoing diffusion and
subject to the two chemical reactions $A+A\rar A$ with rate
$r_c$ and $A+A\rar \mbox{\rm inert}$ with rate $r_a$. For this system
one expects, starting from an initial particle density $\rho_0$,
at late times an algebraic decay of the mean particle 
concentration \cite{Tous83}
\BEQ \label{eq:ADens}
\rho_A (t) \sim {\cal A} t^{-y} \;\;\;\; ; \;\; t\rar\infty
\EEQ
In one-dimensional systems, one has $y=\frac{1}{2}$ 
independently of the branching ratio $r_c/r_a$ as opposed
to $y=1$ obtained from a (mean field) rate equation. The importance of
fluctuations which invalidate the rate equation approach is underlined
by recent experimental results for the exponent $y$ in effectively
one-dimensional systems. In these experiments, the particles are
excitons carried by long polymers and the effective one-dimensionality is
achieved by either constraining the polymer to move in small pores of a
convenient medium when the excitons move between the polymers 
\cite{Pras89,Kope90} or by the structure of the polymer
itself when the excitons move along the polymer \cite{Kroo93}. For mixed 
annihilation-coagulation (both $r_a,r_c \neq 0$) one finds $y=0.52-0.59$
\cite{Pras89} and $y=0.47(3)$ \cite{Kope90}, while for the pure
coagulation case ($r_a=0$), $y\simeq 0.48$ \cite{Kroo93}. In the latter
case, the numerical value of the exponent $y$ is also shown to be
independent of the initial density $\rho_0$ \cite{Kroo93}, which confirms
exponent universality. Furthermore, in that particular $1D$ system, 
the amplitude $\cal A$ is independent of $\rho_0$ as 
well and it was realized \cite{Henk95} that this theoretical prediction is 
found experimentally \cite{Kroo93} as well. This means that for 
low-dimensional systems with short-ranged interactions, 
simple kinetic equations as first investigated by Smoluchowski \cite{Smol17}
are no longer adequate to describe the strong fluctuations
occurring in these systems and leading to so-called {\em anomalous kinetics}
\cite{Koto88}. More powerful techniques to include these
effects must be sought. 
 
Theoretical results of this nature can be obtained by a variety of means,
e.g. the renormalization group (see e.g. \cite{Card96}), 
improved mean-field approaches or exact 
calculations, see \cite{Priv96} for a collection of reviews. 
It is the aim of this paper to explore some of the possibilities
of obtaining exact results. 
A convenient starting point for an exact description
of a stochastic process is the formulation of the problem in terms of a 
master equation. The underlying assumption of this approach, viz. the absence
of memory effects, is at least approximately true for many real systems.
A master equation (see section 2) is a first order linear differential
equation in time for the probability distribution of the stochastic variables
describing the system. Using standard procedures, 
see \cite{Glau63,Kada68,Doi76,Gras80} it may be conveniently
written as a matrix equation and thus its 
solution is turned into an eigenvalue
problem\footnote{For models with discrete-time dynamics one obtains an 
eigenvalue problem relating the probability distribution of the system at time
$t=n+1$ to the distribution at time $t=n$}. The crucial point is that it has 
been realized in recent years that for many models of interest the matrix
appearing in this eigenvalue problem is the quantum Hamiltonian 
(or the transfer matrix respectively, in the case of discrete time dynamics) 
of some integrable quantum chain 
\cite{Feld70,Alex78,Kand90,Gwa92,Schu93,Alca93,Alca94,Schu95,Dahm95,Kim95,Sand94,Hinr95a,Bark96,Toka96,Albe96,Albe96a,Hinr96}. 
This insight has made available the tool 
box of integrable models (e.g. Bethe ansatz \cite{Matt,Gaud83,Baxt82}) 
for these interacting particle 
systems far from equilibrium and has led to many new exact results for their 
dynamical and stationary properties. 
Clearly, the models tractable by such means are usually rather simple, but
the observed universality of many quantities of interest make these models
both theoretically and experimentally interesting.

Most integrable stochastic models studied so 
far are directly given in terms of 
an integrable quantum chain or transfer matrix. In these cases, 
``just by looking'' at the
problem one recognizes some known integrable system. However, it has been 
observed recently that one may obtain from a given stochastic quantum 
Hamiltonian by a similarity transformation the Hamiltonian of some other 
stochastic process, see \cite{Kreb95,Simo95,Henk95,Priv95}. 
This relates a system which is not obviously
integrable to another system which {\it is} and in this way exact
results may be obtained. At the same time it was also realized that indeed one
may obtain by a similarity transformation a stochastic Hamiltonian from some
integrable Hamiltonian which does {\em not} describe a stochastic process
\cite{Schu95,Henk95}. This is implicit in many earlier treatments of
reaction-diffusion processes, but has not yet been exploited systematically.
It shows that many more stochastic systems than previously thought are
actually integrable. 

In this paper,
our strategy is as follows. We start from some ({\it a priori} 
non-stochastic) integrable quantum chain $H$. The examples we shall 
consider can be analyzed using Bethe ansatz
and/or free fermion techniques. Next we perform a similarity transformation
and ask under which conditions on the parameters of the original
integrable quantum chain $H$ 
and on the transformation matrix one obtains a stochastic
Hamiltonian. In this way we identify several classes of new {\it 
integrable} stochastic models.
Finally, we use the knowledge provided by the
integrability to obtain some exact results for these processes, in particular,
the average density at time $t$, and the (longest) relaxation time in the
system.

Pursuing this program in full generality appears to be a formidable task. 
Therefore we shall restrict ourselves to systems which are (a) translationally
invariant, (b) involve particles of only one kind with hard core repulsion
preventing double occupancy of a single site and (c) which interact only
via effective nearest neighbor interaction. We shall make at various points
comments on generalizations, e.g. to other boundary conditions, but we shall
not work out any details in these cases. Finally, we shall consider only 
similarity 
transformations which are translationally invariant and which preserve the
locality of all observables. This excludes e.g. sublattice transformations
\cite{Barm93} or duality transformations \cite{Racz85,Sant96} which also
give rise to transformations between different stochastic Hamiltonians
or between non-stochastic and stochastic Hamiltonians. We would like to stress
that we shall consider only continous-time dynamics. However, all our results
apply also to models with suitably chosen discrete time dynamics.

The paper is organized as follows: in section 2 we review the Hamiltonian
formulation for non-equilibrium stochastic systems and its connection with
integrable quantum Hamiltonians of magnetic systems. We also specify the
similarity transformations which will be used to link stochastic systems
with integrable ones and comment briefly on the relationship with
continuum field theory. In section 3, we give the full classification of the
stochastic systems which can be reduced to a free fermion Hamiltonian with
site-independent interactions. In 
section 4, we find stochastic systems which have the same spectrum
as a XXZ chain with real matrix elements. In section 5, we use the link
with the free fermion case to calculate explicitly the time-dependent particle
density. Section 6 gives our conclusions. More technical material is relegated
to the appendices. In appendix A, we extended the standard technique of 
Lieb-Schultz-Mattis for the diagonalization of a hermitian free fermion
Hamiltonian to the non-hermitian case. Appendix B gives further details for
the classification of those stochastic systems similar to a free fermion
Hamiltonian. In appendix C we recall the relationship of one of the stochastic
systems related to free fermions to the biased voter model. Appendix D
gives some further details for the construction of stochastic processes
from the XXZ chain. Appendix E describes the details of the caculation of the
long-time behaviour of particle densities and two-point correlators from both
the Bethe ansatz and the truncated equations of motion. Appendix F recalls the
relationship between the Hamiltonian spectra for periodic and free boundary 
conditions.

\section{Quantum chain formulation of stochastic processes} 

In this section, we review the reformulation of a non-equilibrium stochastic 
system defined by some master equation in terms 
of the spectral properties of an associated 
quantum Hamiltonian $H$. To be specific, we only consider systems defined
on a chain with $L$ sites and two allowed states per site. We represent
the states of the system in terms of spin configurations $\{ \sig \} =
\{ \sig_1, \sig_2, \ldots , \sig_L \}$ where $\sig=+1$ corresponds to an
empty site and $\sig=-1$ corresponds to a site occupied by a single particle. 
We are interested in the probability distribution function $P(\sig;t)$
of the configurations $\{\sig\}$. Our starting point is the master 
equation for the $P(\sig;t)$
\BEQ \label{Master}
\partial_t P(\sig;t) = \sum_{\tau\neq\sig} \left[
w(\tau\rar\sig) P(\tau;t) - w(\sig\rar\tau) P(\sig;t) 
\vekz{ }{\,} \!\!\!\!\!\right]
\EEQ
where $w(\tau\rar\sig)$ are the transition rates between the configurations
$\{\tau\}$ and $\{\sig\}$ and are assumed to be given from the phenomenology. 
In order to rewrite this as a matrix problem, we introduce a state vector
\BEQ
|P\rangle = \sum_{\sig} P(\sig;t) |\sig\rangle
\EEQ
and eq.~(\ref{Master}) becomes
\BEQ
\partial_t \ket{P} = - H \ket{P}
\EEQ
where the matrix elements of $H$ are given by
\BEQ
\bra{\sig} H \ket{\tau} = -w(\tau\rar\sig) \;\; \mbox{\rm if }
\tau \neq \sig \;\; , \;\; 
\bra{\sig} H \ket{\sig } = \sum_{\tau\neq\sig} w(\sig\rar\tau)
\EEQ
The operator $H$ describes a stochastic process since all the
elements of the columns add up to zero. This conservation of probability,
viz. $\sum_{\sig} P(\sig;t) =1$, is equivalent to the relation
\BEQ
\bra{s} H = 0
\EEQ
where $\bra{s} = \sum_{\sig} \bra{\sig}$ is a {\em left} eigenvector of $H$
with eigenvalue 0. Correspondingly, 
$H$ has at least one right eigenvector with 
eigenvalue 0. Such a vector does not evolve in time and therefore corresponds 
to a steady state distribution of the system. Since in general $H$ is not 
symmetric, this steady state vector may be highly non-trivial. Note that all
this is completely general and applies to any stochastic process defined by
a master equation. With a view on the processes that we shall study we call
$H$ a {\em quantum Hamiltonian} and this formulation of the master equation
the {\em quantum Hamiltonian formalism}. 
The reason for this choice of language
is the fact that for the processes studied below (and, in fact, many other
processes as well) $H$ is the Hamiltonian of some quantum system such as
the Heisenberg $XXZ$ Hamiltonian. The steady state of a stochastic
system corresponds in this mapping to the ground state of the quantum system. 
 
Now, it is straightforward to translate the well-known theorems \cite{Hyver}
about the solution of the master equation (\ref{Master}) to the Hamiltonian 
formulation at hand. In particular, the real parts of the eigenvalues $E_i$
of $H$ are non-negative, $\Re E_i \geq E_0 =0$, and there are no purely
imaginary eigenvalues, which excludes the possibility of undamped oscillations.
Furthermore, if the initial distribution
$P_0(\{\sig\}) = P(\sig;0) = \bra{\sig}P_0\rangle$ satisfies 
$0\leq P_0(\{\sig\})\leq 1$ and $\bra{s}P_0\rangle=1$, it follows that since
\BEQ
\ket{P} = \exp ( - H t ) \ket{P_0}
\EEQ
the relations $0\leq \bra{\sig}P\rangle\leq 1$ and $\bra{s}P\rangle =1$ hold 
for all times $t$. This guarantees the probabilistic interpretation.
Then time-dependent averages of an observable $F$ are given by 
the matrix element
\BEQ
<F>(t) = \sum_{\sig} F(\sig ) P(\sig;t)  
= \bra{s} F \ket{P} = \bra{s} F \exp(-H t)\ket{P_0}
\label{AVE}
\EEQ 
 
Conversely, if a matrix $\wit{H}$ satisfies the conditions
\BEQ \label{StoCon}
\bra{s} \wit{H} = 0 \;\; , \;\; \bra{\sig} \wit{H} \ket{\tau} \leq 0 \;\; 
\mbox{\rm for all $\sig\neq\tau$}
\EEQ
it follows that $\wit{H}$ is the generator of a stochastic 
process \cite{Hyver}. These conditions will play a major role later on. We
shall refer to the first as {\em probability conservation condition} and to
the second as {\em positivity conditions}.  

In what follows, we shall be mainly 
interested in averages of particle numbers
$n_{j}$ at site $j$ and their correlators. These can be expressed in the
quantum spin formulation in terms of the projector
\BEQ \label{NOp}
\tilde{n}_{j} = \frac{1}{2} \left( 1 - \sig_j^z \right) = 
\left( \matz{0}{0}{0}{1} \right)_j
\EEQ

and one-point and two-point functions of the $n_j$ are then expressed as
\footnote{We stress that the structure of these matrix elements is quite
distinct from expectation values in 
ordinary quantum mechanics where one always
considers matrix elements between right and left states that are hermitean 
conjugates.}
\BEQ \label{OneTwo}
<\tilde{n}_j>(t) = \bra{s} \tilde{n}_j \ket{P} \;\; , \;\; 
<\tilde{n}_j \tilde{n}_{\ell}>(t) = \bra{s} \tilde{n}_j 
\tilde{n}_{\ell} \ket{P}
\EEQ 

Two basic situations are readily distinguished from the spectrum of $H$. If
in the limit of infinite lattice size $L\rar\infty$ the lowest excited states
have a finite gap to the ground state energy $E_0=0$, then the averages
(\ref{OneTwo}) will approach their steady state values exponentially fast with
time. On the other hand, if there is in the $L\rar\infty$ limit a continuous 
spectrum down to $E_0=0$, one expects an 
algebraic approach of the correlators to the steady state. 
  
The recent interest in this setup comes from the integrability of quantum
Hamiltonians in one dimension 
\cite{Alex78,Gwa92,Alca93,Alca94,Schu95,Dahm95,Kim95}\footnote{Or equally, 
from the integrability of
transfer matrices for discrete time dynamics \protect\cite{Kand90,Schu93}}. 
As the paradigmatic example, consider the asymmetric simple exclusion process
where particles are diffusing to the right, 
$A+\emptyset\rar\emptyset+A$ with rate 
$1/q$ and diffusing to the left $\emptyset+A\rar A+\emptyset$ with rate $q$. 
The Hamiltonian generating the time 
evolution of the system can, for free boundary conditions, 
be written in the form \cite{Alca94}
\BEQ
H = -\frac{1}{2} \sum_{j=1}^{L-1} \left[ \sig_j^x \sig_{j+1}^x 
+\sig_j^y \sig_{j+1}^y + 
\frac{1}{2}\left( q+q^{-1} \right) \left( \sig_j^z\sig_{j+1}^z -1 \right)
- \frac{1}{2} \left( q-q^{-1} \right) \left( \sig_j^z - \sig_{j+1}^z 
\right) \right]
\EEQ
where $\sig^{x,y,z}$ are Pauli matrices. 
Through various mappings this model arises in a variety of contexts,
it is related e.g. to the $1D$ Kardar-Parisi-Zhang equation 
or to the noisy Burgers equation, see e.g. \cite{Schm95} for a review. 
$H$ can be analysed exactly using the Bethe ansatz \cite{Gwa92}. 
It is identical to the integrable quantum group $U_q SU(2)$ Hamiltonian
\cite{Kiri88,Pasq90} and gives a clear physical 
interpretation to the quantum deformation parameter $q$ as hopping
asymmetry \cite{Alca94}. 
The quantum group symmetry may be used for the calculation of 
correlation functions \cite{Sand94}.
It is well-known that the spectrum of $H$
has a finite gap for $q\neq 1$. For $q=1$ however, corresponding to symmetric
diffusion, the energy levels scale with the system size $E_i \sim L^{-2}$. 
In contrast to conventional equilibrium systems, 
the spectral properties of $H$ depend
on the boundary conditions. For periodic boundary conditions, it can be shown
that for {\em all} values of $q$, there is no gap for the infinite system
and $E_i \sim L^{-\theta}$, where $\theta$ varies between $\frac{3}{2}$ and
2 depending on the particle density \cite{Gwa92,Henk94,Kim95}. The basic
mechanism behind this is explained in appendix~F. 

In fact, integrable quantum Hamiltonians can be found for much more general
systems. We restrict our attention here to models with 
nearest-neighbor interactions
and only consider a single species of particles. Unfortunately, there
is no standard notation for the various rates. In table~\ref{tab1}
we give the possible reaction-diffusion processes together with their rates, 
providing at the same time a dictionary between the notations 
used by different authors. In this paper, we shall use the notation
employed in \cite{Henk95}. 
\begin{table}
\begin{center}
\begin{tabular}{|l|c|c|c|c|c|} \hline
diffusion to the left & $\emptyset + A \rar A +\emptyset$ & $D_L$    &$a_{32}$
& $w_{1,1}(1,0)$ & $\Gamma_{10}^{01}$ \\
diffusion to the right& $A+\emptyset \rar \emptyset + A$  & $D_R$    &$a_{23}$
& $w_{1,1}(0,1)$ & $\Gamma_{01}^{10}$ \\
pair annihilation     & $A+A\rar \emptyset + \emptyset$   & $2\alpha$&$a_{14}$
& $w_{1,1}(0,0)$ & $\Gamma_{00}^{11}$ \\
coagulation to the right& $A+A\rar\emptyset + A$          &$\gamma_R$&$a_{24}$
& $w_{1,0}(0,1)$ & $\Gamma_{01}^{11}$ \\
coagulation to the left & $A+A\rar A+\emptyset $          &$\gamma_L$&$a_{34}$
& $w_{0,1}(1,0)$ & $\Gamma_{10}^{11}$ \\
death at the left  & $A+\emptyset\rar\emptyset +\emptyset$&$\delta_L$&$a_{13}$
& $w_{1,0}(0,0)$ & $\Gamma_{00}^{10}$ \\
death at the right & $\emptyset+A\rar\emptyset +\emptyset$&$\delta_R$&$a_{12}$
& $w_{0,1}(0,0)$ & $\Gamma_{00}^{01}$ \\
decoagulation to the left & $\emptyset + A\rar A+A $      & $\beta_L$&$a_{42}$
& $w_{1,0}(1,1)$ & $\Gamma_{11}^{01}$ \\
decoagulation to the right & $A+\emptyset\rar A+A  $      & $\beta_R$&$a_{43}$
& $w_{0,1}(1,1)$ & $\Gamma_{11}^{10}$ \\
birth                   & $\emptyset+\emptyset\rar A+A$   & $2\nu$   &$a_{41}$
& $w_{1,1}(1,1)$ & $\Gamma_{11}^{00}$ \\
creation at the right&$\emptyset+\emptyset\rar\emptyset+A$&$\sigma_R$&$a_{21}$
& $w_{0,1}(0,1)$ & $\Gamma_{01}^{00}$ \\
creation at the left&$\emptyset+\emptyset\rar A+\emptyset$&$\sigma_L$&$a_{31}$
& $w_{1,0}(1,0)$ & $\Gamma_{10}^{00}$ \\ \hline
\multicolumn{2}{|l|}{Rates defined according to reference} & \cite{Henk95} &
\cite{Schu95} & \cite{Alca94} & \cite{Pesc95} \\ \hline
\end{tabular}
\caption{Two-sites reaction-diffusion processes and their rates as defined
by various authors. \label{tab1}}
\end{center} \end{table}

For the time being and for purposes of illustration let us consider besides 
diffusion only those reactions which irreversibly reduce
the number of particles (that is, $\beta_{L,R}=\sigma_{L,R}=\nu=0$). 
For temporary convenience we rescale $\alpha \rar D \alpha$ and define also
\BEQ 
D = \sqrt{ D_L D_R } \;\; , \;\; \gamma = \frac{ \sqrt{\gamma_L\gamma_R} }{D}
\;\; , \;\; \delta = \frac{ \sqrt{\delta_L\delta_R} }{D} \;\; , \;\;
q = \sqrt{ \frac{D_L}{D_R} } = \sqrt{ \frac{\gamma_L}{\gamma_R} } =
\sqrt{ \frac{\delta_L}{\delta_R} } 
\EEQ
\BEQ \label{hDelta}
\Delta = \frac{1}{2} \left( q+q^{-1}\right)(1+\delta-\gamma) - \alpha 
\;\; , \;\;
h = \frac{1}{2}\left( 2\alpha + \gamma\left(q+q^{-1}\right) \right) .
\EEQ
Note that the ratio of the left and right rates is taken to be the same for
diffusion, coagulation and death processes. 
Then the quantum Hamiltonian for the system may be written in the form
\BEQ \label{HDacd}
H =  D \left( \, H_{XXZ}(h,\Delta,\delta) + 
H_{\alpha} + H_{\gamma} + H_{\delta}\vekz{ }{ }\!\!\! \right)
\EEQ
where $H_{XXZ}(h,\Delta,\delta)$ is the standard $XXZ$ quantum chain, 
written down again for free boundary conditions 
\BEA
H_{XXZ}(h,\Delta,\delta) &=& -\frac{1}{2} \sum_{j=1}^{L-1} 
\left[ \sig_j^x \sig_{j+1}^x 
+\sig_j^y \sig_{j+1}^y + \Delta \left( \sig_j^z\sig_{j+1}^z -1 \right)
\right. \nonumber \\
&+& h \left. \left( \sig_j^z + \sig_{j+1}^z -2 \right)
- \frac{1}{2} (1-\delta) \left( q-q^{-1} \right) 
\left( \sig_j^z - \sig_{j+1}^z \right) \right]
\EEA
which contains the diagonal and diffusive matrix elements while the particle
annihilation terms are contained in 
\BEA
H_{\alpha} &=& -2\alpha \sum_{j=1}^{L-1} q^{-2j-1} \sig_{j}^{+} \sig_{j+1}^{+}
\nonumber \\
H_{\gamma} &=& -\gamma \sum_{j=1}^{L-1} q^{-j}
\left( \tilde{n}_j\sig_{j+1}^+ + q^{-1}\sig_{j}^+ \tilde{n}_{j+1}\right) 
\label{ParReac} \\
H_{\delta} &=& -\delta \sum_{j=1}^{L-1} q^{-j} 
\left( q^{-2}(1-\tilde{n}_j)\sig_{j+1}^+ + 
q\sig_{j}^+ (1-\tilde{n}_{j+1})\right)
\nonumber 
\EEA
and $\sig^{\pm}=\frac{1}{2}(\sig^x\pm i\sig^y)$ are the one-particle 
annihilation/creation operators. 

Remarkably, 
the spectrum of $H$ is independent of the terms contained in (\ref{ParReac}),
that is $\mbox{\rm spec}(H) = 
\mbox{\rm spec}(D H_{XXZ}(h,\Delta,\delta))$ \cite{Alca94}. To see this, 
recall that the XXZ Hamiltonian conserves the number of particles while the
reaction terms irreversibly decrease the total particle number. 
Thus, $H$ can be
written in a block diagonal form
\BEQ
H = \left( \begin{array}{ccccc} 
{\cal N}_0 & X_{\delta} & X_{\alpha} & & \\
 & {\cal N}_1 & X_{\gamma,\delta} & X_{\alpha} &\\
 & & {\cal N}_2 & X_{\gamma,\delta}& \ddots \\
 & & & \ddots & \ddots\end{array} \right)
\EEQ
where ${\cal N}_n$ refers to the $n$-particle states and 
$X$ are the reaction matrix
elements. Because of the identity
\BEQ
\det \left( \begin{array}{cc} {\cal A} & X \\ 0 & {\cal B} \end{array}\right) 
= \det {\cal A} \det {\cal B}
\EEQ
it is clear that the elements of (\ref{ParReac}) do not enter into the 
characteristic polynomial\footnote{Alternatively, this can also be seen
through a similarity transformation of the form (\ref{BTrans}), where $B$
is diagonal. While this does not change $H_{XXZ}$, the rates occurring in
(\ref{ParReac}) become arbitrary, since they contain the elements of $B$.
Thus $\mbox{\rm spec}(H)$ must be independent of them \cite{Alca94}.} of $H$.

The phase diagram for $H_{XXZ}(h,\Delta,\delta)$ is 
known \cite{John72,Taka73}. 
For our purposes, we need the following. From (\ref{hDelta}), only the
portion of the phase diagram where $h+\Delta\geq 1$ is important for us.
The spectrum always has a finite gap when $h+\Delta >1$, 
which is realized whenever $\delta\neq 0$ or $q\neq 1$. Then the ground
state is a trivial ferromagnetic frozen state. 
The spectrum is gapless for $\Delta+h=1$,
where the system undergoes a Pokrovsky-Talapov transition. This situation
occurs for $\delta=0$ and $q=1$. We have thus identified the
cases where the model approaches the steady state exponentially
(non-vanishing gap) or algebraically (gapless). A special point is
$\Delta=0$ where $H_{XXZ}$ becomes a free fermion system and thus in certain
cases not only the spectrum may be obtained, but also correlation functions
can be calculated explicitly (see below).

While large classes of reaction-diffusion problems are now recognised as
being integrable, see \cite{Alca93,Schu95,Dahm95} for lists including systems 
with 
more than one kind of particles\footnote{We remark that up to degeneracy, 
the spectrum of $H$ for the processes $A+\es\lrar\es+A$, $B+\es\lrar\es+B$
with rate 1, $A+A\rar B$, $B+B\rar A$ with rate $\alpha$ and
$A+B\rar\es+\es$ with rate $2\alpha$, is the one 
of $H_{XXZ}(\alpha,1-\alpha,0)$
\cite{Alca94}. For $\alpha=1$, one recovers a free fermion system and closed
equations of motion can be found and solved \cite{Priv92}, leading for
example to $\rho_A(t) \sim \rho_{B}(t) \sim 1/\sqrt{t}$.}, 
it is still far from obvious
how to exploit the algebraic structure hidden beneath it in order to get
explicit expressions for the desired particle number correlators for
$\Delta \neq 0$. 
A very interesting alternative was proposed
in \cite{Pesc95}, where subsets of observables were identified for which
closed equation of motion can be obtained, thus leading to partially 
integrable systems. 
In any case, it is important to realize that the knowledge about the spectrum
(and, in certain cases such as the free fermion case $\Delta=0$, about
correlators) does not come
from the assumption that the Hamiltonian is stochastic, but from its 
integrability. It is therefore natural to ask whether one can find new 
integrable non-equilibrium systems from transformations of known integrable 
quantum chains. 

Specifically, we shall investigate the transformation 
\cite{Kreb95,Simo95,Henk95}
\BEQ \label{BTrans}
\wit{H} = {\cal B} H {\cal B}^{-1} \;\; , 
\;\; {\cal B} = \bigotimes_{j=1}^{L} B_j
\EEQ
where $H$ is a quantum Hamiltonian with known properties, $\wit{H}$ a
stochastic Hamiltonian and $B_j$ is the transformation matrix $B$ acting on 
the site $j$. From now on, we shall focus on translationally invariant
systems, i.e. we shall consider periodic boundary conditions.
To study the effect of the transformation $\cal B$, it is sufficient
to consider the effect on a two-particle Hamiltonian $H_{j,j+1}$,
that is we write
\BEQ \label{HZwei}
H = \sum_{j=1}^{L} {\bf 1}_1 \otimes \cdots \otimes {\bf 1}_{j-1} \otimes
H_{j,j+1} \otimes {\bf 1}_{j+2} \cdots \otimes {\bf 1}_L
\EEQ
where ${\bf 1}_{\ell}$ is the unit matrix acting on the site $\ell$. For the
model (\ref{HDacd}) with left-right symmetry, that is $q=1$, we have for
example
\BEQ \label{HMat}
H_{j,j+1} = D \left( \begin{array}{cccc}
0 &  -\delta &  -\delta & -2\alpha \\
0 &  1+\delta & -1 & -\gamma \\
0 & -1 &  1+\delta & -\gamma \\
0 &  0 &  0 & 2(\alpha+\gamma) \end{array} \right)
\EEQ
Transformations of the type (\ref{BTrans}) have been mainly studied when
both $H$ and $\wit{H}$ represent stochastic systems. Besides, an equivalent
way of relating stochastic systems was developed \cite{Balb95} 
in the context of the Lagrangian formulation of the model (see below), which is 
more suitable for field theory techniques. In particular, these
transformations
allow to treat the general model (\ref{HMat}), since the observables
in the two systems are simply related 
\cite{Henk95,Kreb95,Simo95,Priv95,Balb95}. The results for the 
long-time behaviour of the one- and two-point functions for translationally
invariant initial conditions are collected in table~\ref{tab2}. For the
derivation, see appendix~E. 
\begin{table} \begin{center}
\begin{tabular}{|c|cc|} \hline 
$\delta$ & $C_1$ & $C_2$  \\ \hline
0 & $t^{-1/2}$ & $t^{-3/2}$  \\
$<\alpha+\gamma$ & $\exp(-2\delta t)$ & $t^{-1/2} \exp(-4\delta t)$  \\
$>\alpha+\gamma$ & $\exp(-2\delta t)$ & $\exp{\left[ -4\delta t +
2(\Delta+1/\Delta -2)t\right]}$ \\ 
\hline
\end{tabular}
\caption[Generics]{Generic long time behaviour of the one-point function 
$C_1(t) = \sum_{j}
<n_j>(t)$ and the two-point function $C_{2}(t) = \sum_{j}<n_{j}n_{j+r}>(t)$
in the system eq.~(\ref{HMat}) for $r$ finite, translation invariant 
initial conditions and  finite initial particle density. 
The abbreviation $\Delta=1+\delta-\alpha-\gamma$ is used \label{tab2}.
 }
\end{center} \end{table}

{}From the spectrum of $H$, we would naively expect exponential factors 
$e^{-2 k\delta t}$ to be present in the $k$-point correlator $C_k$, 
but we also see that for example algebraic prefactors are not
readily predicted from the spectrum alone. This expectation, however,
is only valid provided there are no bound states in the system. 
This is so as long the model stays sufficiently close to the case
$\delta=\alpha+\gamma-1$, where the exact spectrum of $H$ can be found
from free fermion techniques. In general, however, 
one can show from the Bethe
ansatz \cite{Yang66,Matt,Gaud83,Baxt82} that there does exist a 
bound state in the
two-particle sector, which has the energy 
$4\delta+4-2\Delta-2/\Delta$.\footnote{We 
thank G.M. Sch\"{u}tz for a useful discussion on this point.}
While the results given in table~\ref{tab2} are valid for homogeneous
initial conditions, further power law prefactors may appear for other
types of initial conditions (an example of this will be discussed in
section 5).\footnote{The above analysis assumes that the associated
amplitudes are non-vanishing. This is not always so, as can be seen in
the calculation of the correlator $<n_j(t) n_{\ell}(0)>$ for a model
in which the {\em only} non-vanishing rates are $\alpha$ and $\nu$. Although
a bound state is known to be present, the associated amplitude can be
shown to vanish. We thank R. B. Stinchcombe for pointing this out to us.} 

A few comments are in order. 
When $\alpha,\gamma$ are arbitrary
and $\delta=0$, the model can be reduced \cite{Simo95,Henk95} 
to the cases $\gamma=0$ or $\alpha=0$ which had so far been the only ones
amenable to exact treatment, see \cite{uralt,Lush86}. Thus the particle number
correlators can be found exactly for all times. An equivalent transformation
was subsequently given in the continuum field theory formulation of the
theory \cite{Balb95}. For $\delta\neq 0$, the same
technique allows the transformation to a system which can be interpreted
as describing radioactively decaying particles\footnote{The radioactive
decay of particles is described in this setting by the conditions
$\alpha=0$ and $\gamma=\delta$.} which in addition undergo
diffusion and either annihilation or death processes. Then
it can be shown that in the long-time limit the equations of motion for
the $n$-point functions decouple and the long-time
behaviour of the correlators can be extracted \cite{Henk95}. Furthermore,
for left-right nonsymmetric rates ($q\neq 1$) by the same 
reduction technique it can be
shown that for $\delta_{L,R}=0$, the coagulation rates $\gamma_{L,R}$ can be
eliminated if and only if the free-fermion condition is satisfied
\cite{Henk95} and the correlators are then again known exactly for all times. 

Here, we concentrate on transformations between an integrable, but not 
necessarily stochastic, quantum chain $H$ and a stochastic Hamiltonian
$\wit{H}$. Some examples of this have 
already been given \cite{Kreb95,Henk95}. 
We attempt to treat the transformation of type (\ref{BTrans}) systematically
starting from two types of an integrable quantum chain\footnote{We
limit ourselves to these two types of integrable systems for pragmatic
reasons. In fact, for the second type of integrable system we consider here
we merely use the integrability of the spectrum and do not even ask whether
or not the entire system might be integrable. A sufficient criterion to
establish integrability of the full Hamiltonian $H$ is provided by showing
that the $H_{j,j+1}$ satisfy a Hecke algebra. It can be shown that for 
$\delta=0$ in the model (\ref{HDacd}) the {\it entire} 
non-symmetric Hamiltonian matrix and not just the spectrum-determining part
is integrable \cite{Alca94}. See \cite{Alca93} for a list of integrable
stochastic Hamiltonians through the Hecke algebra technique.}
\begin{enumerate}
\item The most general translationally invariant Hamiltonian which be may
written in terms of free fermions. We shall give in the next section the 
complete list of stochastic systems 
with spatially constant reaction rates which can be
transformed into this form.
\item The most general Hamiltonian with 
the spectrum of the $XXZ$ chain and with
real matrix elements. Here we shall obtain a new 3-parameter manifold 
(after normalization) of integrable stochastic processes.
\end{enumerate}

Before starting this, however, it might be useful to recall briefly how 
results obtained from the quantum Hamiltonian formulation of 
reaction-diffusion systems
presented here compare with the Lagrangian techniques, as developed in
\cite{Doi76,Gras80,Peli85}. See \cite{Card96} for a recent review. 
As for the quantum Hamiltonian formulation, the starting point is the
master equation. The creation and destruction of particles is conveniently
described in a Fock space formalism, where the state vector becomes
\BEQ
\ket{P(t)} = \sum_{\sigma} p(n_1,n_2,\ldots;t) 
\left( {a^{\dag}}_1\right)^{n_1}
\left( {a^{\dag}}_2\right)^{n_2} \cdots \ket{0}
\EEQ
where ${a^{\dag}}_{i}$ creates a particle at site $i$ and $p(n_1,n_2,\ldots;t)$
is the probability of having $n_1$ particles at site 1, $n_2$ particles at site
2 and so on at time $t$. Consider for example the case when only
diffusion and annihilation occur. Then the Hamiltonian is
\BEQ
H = D \sum_{(i,j)} ( {a^{\dag}}_i - {a^{\dag}}_j ) 
\left( a_i - a_j \right) - 
2\alpha \sum_i \left( a_i^2 - {a^{\dag}}_i^2 a_i^2 \right) 
\rar \int \left[ D ( \nabla a^{\dag} ) ( \nabla a )
-2\alpha \left( a^2 - {a^{\dag}}^2 a^2 \right) \right] d^dx
\EEQ
where a formal continuum limit is taken\footnote{In distinction to the
quantum Hamiltonians written down above, one usually considers `bosonic'
operators, where the number of particles per site is unrestricted. While
that does not appear to be an essential feature when restricting attention
to purely destructive reactions, this may be different when considering
processes which create new particles, see \cite{Dick89,Henk97}.}. 
One then goes over to a path
integral representation involving fields $a(x,t)$ and $a^*(x,t)$ and an
action $S[a,a^*]$. In order to reduce the calculation of avarages to the
usual QFT expectation values, it is customary to perform a shift
$a^* = 1+\bar{a}$. Then the action becomes \cite{Doi76,Gras80,Peli85}
\BEQ
S[a,\bar{a}] = \int \left[ \bar{a}\partial_t a + D \nabla \bar{a} \nabla a
+4\alpha \bar{a} a^2 +2\alpha \bar{a}^2 a^2 \right] dt d^dx
\EEQ
Remarkably, since the particle number can only decrease, the diagrammatics
of the associated field theory is greatly simplified. In particular, there
are no loop corrections to the propagator and thus no wave function
renormalization \cite{Doi76,Gras80,Peli85} and only the reaction rate
$\alpha$ needs to be renormalized. This can be done by summing up the
corresponding diagrams to all orders, with the result that the one-loop
beta function is exact. From the renormalization group equations
it then follows that the upper critical dimension is at $d^* =2$ and that
the exponent $y$ describing the decay of the mean particle number 
(\ref{eq:ADens}) is given by $y=\min(1,d/2)$, in agreement with previous
heuristic arguments \cite{Tous83,Kang84}. In particular, it transpires that
the whole probability distribution for fluctuations becomes universal
in the late time regime \cite{Card96}. 

What is the analogue of the similarity transformation (\ref{BTrans}) in the
Lagrangian formulation? To see this, consider the process with only
diffusion, annihilation and coagulation present \cite{Balb95}. The action 
takes the form, where $V$ describes the interactions of the particles 
undergoing a reaction
\BEQ
S_{\lambda} = \int \! dt dx\, \bar{a}\left( \dot{a} - D \nabla^2 a\right)
+\int \! dt dx dx'\, V(|x-x'|) \left[ \bar{a}(x)\bar{a}(x') + \lambda 
\bar{a}(x)\right] a(x) a(x') 
\EEQ
such that for $\lambda=1$ only coagulation and for $\lambda=2$ only 
annihilation reactions occur. Following \cite{Balb95}, consider the
potentials $V_{\alpha,\gamma}$ for pure annihilation and coagulation, 
respectively, to be of the form $V_{\alpha}(x) = \alpha V(x)$ and
$V_{\gamma}(x) = \gamma V(x)$. Then 
\BEQ
\lambda = \frac{2\alpha+\gamma}{\alpha+\gamma}
\EEQ
Thus, through the transformation $a \rar \lambda a, \bar{a}\rar\lambda^{-1}
\bar{a}$ the 
coagulation-annihilation model can be reduced to pure coagulation. For the
quantum Hamiltonian, the corresponding transformation matrix is
$B = \left(\matz{1}{1-\lambda}{0}{\lambda}\right)$ \cite{Henk95,Simo95}. 
In both formalisms, one arrives at the same relationship between 
the $n$-point correlators of the two models \cite{Henk95,Simo95,Balb95}
\BEA
& ~ & \langle n(x_1, t_1) \ldots n(x_n,t_n)\rangle_{\lambda} = 
\bra{s} a(x_1,t_1) \ldots a(x_n,t_n) \ket{P(t)}_{\lambda} \nonumber \\
& = & \lambda^{-n} \bra{s} a(x_1,t_1) \ldots a(x_n,t_n) \ket{P(t)'}_{1} = 
\lambda^{-n}\langle n(x_1, t_1) \ldots n(x_n,t_n)\rangle_{1}'
\EEA
where the primes refer to the rescaling of the initial densities and
correlators (since the ave
rage of $a$ corresponds to a particle density). 

This transformation can be extended to more general systems \cite{Rey96}.

\section{Stochastic systems from free fermions}

In this section we give the classification of those stochastic systems
which can be obtained from a free fermion Hamiltonian through the 
transformation (\ref{BTrans}).

\subsection{Calculation of the matrix elements}

Our starting point is the quantum chain
\BEQ \label{Hsigma}
H = - \sum_{j=1}^L \left[ 
D_1 \sig_j^+ \sig_{j+1}^- + D_2 \sig_j^- \sig_{j+1}^+ 
+ \eta_1 \sig_j^+ \sig_{j+1}^+ + \eta_2 \sig_j^- \sig_{j+1}^-
+ h_1 \sig_j^z + h_2 \sig_{j+1}^z - C\right]
\EEQ
which is diagonalizable through a Jordan-Wigner transformation followed by 
a Bogoliubov transformation or
by a canonical transformation as described in appendix~A.
Here, $D_{1,2}, \eta_{1,2}, h_{1,2}$ are free parameters.
In the following, we write $h=h_1 + h_2$, $g=h_1 - h_2$ and $\eta^2 = 
\eta_1\eta_2$. We shall consider in this section $\eta \neq 0$. The case
$\eta = 0$ is a special case of a Hamiltonian with $XXZ$ spectrum and will
be considered in the next section. Using (\ref{HZwei}), we can write down 
the matrix describing the nearest-neighbor interactions
\BEQ
H_{j,j+1} = - \left( \begin{array}{cccc} 
h-C& 0 & 0 & \eta_1 \\
0 & g-C & D_1 & 0 \\
0 & D_2 & -g-C & 0 \\
\eta_2 & 0 & 0 & -h-C \end{array} \right)
\EEQ
and we look for a matrix $\wit{H}$ through the transformation
\BEQ \label{HTilde}
\wit{H} = {\cal B} H {\cal B}^{-1} \;\; , 
\;\; {\cal B} = \bigotimes_{j=1}^{L} B_j
\EEQ
and describing a stochastic system. Following the logic described in the
preceding section we want to calculate correlation functions of the stochastic
system in terms of correlators of the non-stochastic model which we can solve.
We therefore rewrite a correlator 
\BEQ
\langle s| \tilde{n}_{x_1} (t_1) \dots \tilde{n}_{x_k}(t_k) | P_0 \rangle = 
\langle s | \tilde{n}_{x_1} 
\exp{\left(-\wit{H} (t_1-t_2)\right)} \tilde{n}_{x_2} 
\dots \tilde{n}_{x_k} 
\exp{\left(-\wit{H}t_k\right)} |P_0\rangle
\EEQ
in terms of operators transformed using $B$, 
thus reducing this correlator to a matrix element computable through
free fermion techniques. Inverting this strategy, one may first calculate
matrix elements in the free fermion system, and then obtain the corresponding
quantity for all stochastic processes related to this system by choosing an
appropriate $B$. This is the strength of this approach, 
illustrated in section 5
for the density  $\tilde{\rho}(t)$ as a function of time for 
an uncorrelated initial 
state with initial density $\tilde{\rho}_0$. 
A {\it single} calculation gives $\tilde{\rho}(t)$
for all stochastic processes related to the 
free fermion system (\ref{Hsigma}).

In order to find a convenient parametrization for $B$ we first fix the 
determinant of $B$ to be 1 which is no loss of generality since 
$B$ must be non-singular and $\wit{H}$ is independent of $\det B$. 
We can then write $B$ in a parametrization analogous to the Euler angles 
(alternatives are discussed in appendix~B)
\BEQ \label{Euler}
B = \left( \matz{\sqrt{a}}{0}{0}{\sqrt{a^{-1}}} \right) 
\left( \matz{\cosh \phi}{\sinh \phi}{\sinh \phi}{\cosh \phi} \right)
\left( \matz{b}{0}{0}{b^{-1}} \right)
\EEQ
with transformation parameters $a,b,\phi$. Of these, we can always arrange
for $b=1$, since otherwise, its effect could be absorbed by redefining
$\eta_1 \rar b^4 \eta_1$ and $\eta_2 \rar b^{-4} \eta_2$. So we are left with
just two transformation parameters $a$ and $\phi$. Below, we shall work with
$Y$ defined by
\BEQ
Y = e^{2\phi}
\EEQ

In order to ensure that $\wit{H}$ is stochastic we shall have to
satisfy the conditions (\ref{StoCon}). 
Let $\wit{C}_m =\sum_{k=1}^{4} (\wit{H}_{j,j+1})_{k,m}$ 
denote the sum of the elements of the $m$-th column of $\wit{H}$. 
Then, up to boundary terms in $\sig_{1,L}^z$ (which are absent for 
periodic boundary conditions) and up to an additive constant in $\wit{H}$ 
(which will be fixed later by specifying the constant $C$),
probability conservation (\ref{StoCon}) is implemented by demanding that
\BEQ \label{ProCon}
\wit{C}_1 = \wit{C}_4 = \frac{1}{2} \left( \wit{C}_2 + \wit{C}_3 \right)
\EEQ
These are solved by fixing $\eta_{1,2}$
\BEQ
\eta_1 = \frac{(D_1+D_2+2h)(1-a+Y+aY)^2}{2 (-1+a+Y+aY)^2} \;\; , \;\;
\eta_2 = \frac{(D_1+D_2-2h)(-1+a+Y+aY)^2}{2 (1-a+Y+aY)^2}
\EEQ
The vanishing of one of the denominators is not dangerous here, because
then either $\eta_1$ or $\eta_2$ would have to vanish. Systems of this kind
are discussed in section 4. 

For the discussion of the positivity conditions on the rates it is convenient
to introduce the parity symmetric and parity
antisymmetric combinations of reaction rates $\vph_{L,R}$ by
\BEQ \label{SymmCon}
\vph = \frac{1}{2} \left(\vph_{L} + \vph_{R}\right) \;\; , \;\; 
\vph' = \frac{1}{2} \left( \vph_{L} - \vph_{R}\right)
\EEQ
respectively, where $\vph$ is taken to be a collective symbol for any
member of the set 
$\{\beta,\gamma,\delta,\sigma,D\}$, see table~\ref{tab1}. 
In order to satisfy the positivity conditions for the stochastic rates
appearing in $\wit{H}$ one might think that all off-diagonal elements in
$\wit{H}_{j,j+1}$ have to be negative or zero. The actual conditions on these
matrix elements are, however, weaker. To see this, it is useful to write
the nearest-neighbor interactions in 
$\wit{H}$ in form of the matrix
\BEQ \label{HxyVoll}
\wit{H}_{j,j+1} = \left( \begin{array}{cccc}
\wit{\sig}_L+\wit{\sig}_R+2\wit{\nu} & -\wit{\delta}_R+y & 
-\wit{\delta}_L - y & -2\wit{\alpha} \\
-\wit{\sig}_R+x & \left(\wit{\delta}_R+\wit{\beta}_L+\wit{D}_L\right) +x-y &
-\wit{D}_R & -\wit{\gamma}_R-y \\
-\wit{\sig}_L-x & -\wit{D}_L &
\left( \wit{\delta}_L+\wit{\beta}_R+\wit{D}_R\right)-x+y & 
-\wit{\gamma}_L+y \\
-2\wit{\nu} & -\wit{\beta}_L -x & -\wit{\beta}_R+x & 
2\wit{\alpha}+\wit{\gamma}_R+\wit{\gamma}_L 
\end{array} \right)
\EEQ
where $x,y$ are free parameters. This matrix form corresponds to the
decomposition 
\BEQ \label{Hxy}
\wit{H}_{j,j+1} = \wit{H}_{j,j+1}^{\mbox{\rm phys}} 
+ x \left( \sigma_{j+1}^- -\sigma_{j}^- \right) 
+ y \left( \sigma_{j+1}^+ -\sigma_{j}^+ \right)
\EEQ
where $\wit{H}_{j,j+1}$ is the transformed matrix which in general is {\it
non-stochastic}, $\wit{H}_{j,j+1}^{\mbox{\rm phys}}$ contains the physical 
reaction rates and is stochastic and the last two terms parametrized by $x$ 
and $y$ are divergence terms. For periodic boundary conditions, these extra 
terms cancel, while for other boundaries, they correspond to surface fields. 

In this paper, we study exclusively periodic boundary conditions. 
Thus it is sufficient to impose the 
positivity conditons (\ref{StoCon}) on the
parity symmetric combinations of the matrix elements of $\wit{H}$ and on
two other combinations of transformed matrix elements in which the terms
containing $x,y$ cancel. This reduces the number of inequalities one has
to satisfy by two. It is important to keep this distinction between the
transformed matrix elements (which occur in $\wit{H}_{j,j+1}$) 
and the actual rates of the stochastic system (which occur in
$\wit{H}_{j,j+1}^{\mbox{\rm phys}}$) which have all to be positive.

Next, we have to write down the elements of the 
transformed matrix $\wit{H}_{j,j+1}$ in terms 
of the matrix elements of $H_{j,j+1}$ and the 
transformation parameters. To simplify 
expressions, we make use of the freedom to 
take the normalization of time scale
\BEQ \label{3.12}
D_1 + D_2  = 2
\EEQ
and we shall consider the other possible case, when $D_1+D_2=0$, in 
appendix~B. Using the convention of (\ref{SymmCon}), we get for
the parity symmetric matrix elements
\BEA
-2\wit{\alpha} &=& a^2 (-1+a-Y^2 -a Y^2) \cdot F_1 \cdot N \nonumber \\
-\wit{\gamma}  &=& \frac{1}{2} a (a^2-1) (Y^2 -1) \cdot F_1 \cdot N 
\nonumber \\
-2\wit{\nu}    &=& a^{-1} (-1+a+Y^2+a Y^2) \cdot F_2 \cdot N \nonumber \\
-\wit{\sig}    &=& \frac{1}{2a} (a^2 -1) (Y^2 -1) \cdot F_2 \cdot N 
\nonumber \\
-\wit{\delta}  &=& a(Y^2-1)(1-a+Y^2+a Y^2) \cdot F_3 \cdot N \nonumber \\
-\wit{\beta}   &=& - (Y^2-1)(-1+a+Y^2+aY^2) \cdot F_3 \cdot N \nonumber \\
-\wit{D}       &=& \frac{1}{2} F_4 \cdot N \label{SymmH}
\EEA
where
\BEA
F_1 &=& h-2ah+a^2h+4Y-4aY+6hY^2-2a^2hY^2+4Y^3+4aY^3+hY^4+2ahY^4+a^2hY^4
\nonumber \\
F_2 &=& h-2ah+a^2h+4aY-4a^2Y-2hY^2+6a^2hY^2-4aY^3-4a^2Y^3+hY^4+2ahY^4+a^2hY^4
\nonumber \\
F_3 &=& h-2ah+a^2h+2Y-2a^2Y+hY^2+2ahY^2+a^2hY^2 \nonumber \\
F_4 &=& h-2ah+2a^3h-a^4h+8aY-16a^2Y+8a^3Y-hY^2+6ahY-6a^3hY^2+a^4hY^2
\nonumber \\
    &-& hY^4-6ahY^4+6a^3hY^4+a^4hY^4-8aY^5-16a^2Y^5-8a^3Y^5
    \nonumber \\
    &+& hY^6+2ahY^6-2a^3hY^6-a^4hY^6 
\nonumber \\
N^{-1} &=& Y (1-a+Y+aY)^2 (-1+a+Y+aY)^2 
\EEA
It is remarkable that the matrix elements can be factorised in terms of only
very few building blocks. This observation is going to be essential in the
classification of the possible stochastic systems. Whether this factorization
property has a deeper algebraic meaning is still open. 
Note that if we use the scaling $\frac{1}{2}(D_1+D_2)=-1$, we recover the
same expressions with $Y\rar -Y$, which gives nothing new. The
parity-antisymmetric matrix elements are 
\BEA
\wit{\gamma}' - y &=& (D_1 -D_2 + 2g) \frac{1}{8Y} a (Y^2-1) \nonumber \\
\wit{\sig}' + x   &=& (D_1 -D_2 + 2g) \frac{1}{8aY} (Y^2 -1) \nonumber \\
\wit{\delta}' + y &=& (D_1 -D_2 - 2g) \frac{1}{8Y} a (Y^2 -1)\nonumber \\
-\wit{\beta}' -x  &=& (D_1 -D_2 - 2g) \frac{1}{8aY} (Y^2 -1) \nonumber \\
\wit{D}'          &=& -(D_1 -D_2) \frac{1}{4Y} (Y^2 +1) \label{AntiH}
\EEA
In order to get a stochastic system from the matrix elements
(\ref{SymmH}, \ref{AntiH}), we must implement the positivity conditions
(\ref{StoCon}) on the parity symmetric rates 
$\wit{\alpha}, \wit{\gamma}, \wit{\nu}, \wit{\sigma}, \wit{\delta}, 
\wit{\beta}, \wit{D}$. These conditions completely determine
the transformation matrix $B$. The physical systems thus found will
have left-right symmetric reaction-diffusion rates. Then, in a second
step and using the extra freedom coming from adding the divergence
terms to $\wit{H}_{j,j+1}^{\mbox{\rm phys}}$ as done in (\ref{Hxy}),
we shall identify free fermion systems with left-right biased diffusion
and reaction rates. 

We now turn to the analysis of the positivity conditions (\ref{StoCon}). 

\subsection{Reality of the transformation parameters $a,Y$}

Before embarking on checking the positivity of the symmetric rates
(\ref{SymmH}), it is useful to break the problem into smaller pieces.
It is a necessary condition to positivity that all matrix elements
of $\wit{H}$ are real. This {\em reality condition} allows us to distinguish
a few subcases. First, for $Y^2 =1$, the transformation from $H$ to
$\wit{H}$ merely amounts to the diagonal transformation encountered before
\cite{Gryn94}. So from now on we always assume $Y\neq 1,-1$. Second, 
since the ratios $\wit{\gamma}'/\wit{\sigma}'$ and 
$\wit{\delta}'/\wit{\beta}'$ must be real, it follows 
from (\ref{AntiH}) that
\BEQ
a^2 ~~\mbox{\rm real}
\EEQ
and we distinguish the following.
\begin{enumerate}
\item {\bf $a$ real.} Then, since $2\wit{\alpha}/\wit{\gamma}$ is real, it
follows that
\BEQ
\frac{a (-1+a-Y^2-a Y^2)}{(a^2-1)(Y^2-1)} ~~\mbox{\rm real}
\EEQ
and it follows that $Y^2$ must be real. In the first case, $Y$ itself
is real, and reality of $\wit{H}$ is guaranteed if $h,g$ and $D_1-D_2$ 
are real. In the
second case, $Y$ is purely imaginary and reality of the elements of $\wit{H}$
is assured if $D_1-D_2$, $h$ and $g$ are all imaginary. 
\item {\bf $a$ imaginary.} We write $a=i A$ with $A$ real. Since
$2 \wit{\nu}/\wit{\sig}$ must be real, it follows from (\ref{SymmH}) that
\BEQ
\frac{-1+a+Y^2+a Y^2}{(a^2-1) (Y^2 -1)} ~~\mbox{\rm real}
\EEQ
which can be rewritten as
\BEQ
\Re \frac{Y^2 +1}{Y^2 -1} = 0
\EEQ
or, letting $Y^2 = P + i Q$, we find that $P^2 + Q^2 =1$ and thus have
\BEQ
\left| Y^2 \right|^2 =1  ~~\mbox{\rm if $a$ is imaginary}
\EEQ
\end{enumerate}
In summary, we have to consider the following cases separately
\BEQ \label{aYFaelle}
{\rm A}) \;\; \mbox{\rm  $a$ real and $Y$ real } , \;\; 
{\rm B}) \;\; \mbox{\rm $a$ real and $Y$ imaginary } , \;\;
{\rm C}) \;\; \mbox{\rm $a$ imaginary and $|Y|=1$}
\EEQ
and we now ask when the conditions 
$\wit{\alpha}, \wit{\beta}, \wit{\gamma}, \wit{\delta}, \wit{\nu}, 
\wit{\sigma}, \wit{D} \geq 0$ can be simultaneously met. 

\subsection{A no-go theorem}

In this section, we prove the lemma stated below. The reader not interested
in the details of the proof can go on to the next section, where the 
implications will be discussed. 

\noindent {\bf Lemma:} {\it For the parity-symmetric reaction rates
$\wit{\alpha}, \wit{\gamma}, \wit{\nu}$ and $\wit{\sig}$ from 
eq.~(\ref{SymmH}) with the scaling $\frac{1}{2}(D_1+D_2)=1$, the inequalities 
\BEQ \label{Ineq}
\wit{\alpha} > 0 \;\; , \;\; 
\wit{\gamma} > 0 \;\; , \;\; 
\wit{\nu} > 0    \;\; , \;\;
\wit{\sig} > 0
\EEQ
cannot be simultaneously satisfied.} 

\noindent {\bf Proof:} We show this by examining each of the cases in 
(\ref{aYFaelle}) separately. For the sake of clarity and brevity, we 
are not going to go through
the inequalities explicitly, but rather prefer to illustrate graphically
what is happening. Once having understood from the graphics the result to
be proved, it is a straightforward matter to check the relevant inequalities. 
The matrix elements depend on the three free
parameters $a,Y$ and $h$. We are interested in localising the regions
where the rates $\wit{\vph}$ (where as in (\ref{SymmCon}), $\wit{\vph}$ 
can stand
for any of the rates $\wit{\alpha},\wit{\gamma},\wit{\nu},\wit{\sig}$)
have a constant sign. These regions are bounded
by those lines along which the $\wit{\vph}$ have a simple zero or a simple
pole. We now fix $h$ and look
for the curves $a=a(Y)$ such that the rates $\wit{\vph}$ vanish. These curves
are readily identified from the factorized form of the matrix elements
in eq.~(\ref{SymmH}). 

\noindent {\bf Case A: $a,Y$ both real.} \\
For $h$ fixed, the changes of sign of 
$\wit{\alpha}$ are along the curves
\BEA
a &=& \frac{1+Y^2}{1-Y^2} \nonumber \\
a_{\pm} &=& \left[ h\left(1-Y^2\right) \right]^{-1} \left[ h(1+Y^2) +
2 Y\left(1\pm\sqrt{1-h^2}\right) \right] \nonumber \\
Y &=& 0 
\EEA
$\wit{\gamma}$ changes sign along the curves
\BEA
a_{\pm} &=& \left[ h\left(1-Y^2\right) \right]^{-1} \left[ h (1+Y^2) +
2 Y\left(1\pm\sqrt{1-h^2}\right) \right] \nonumber \\
a &=& \pm 1 \;\; , \;\; a = 0 \nonumber \\
Y &=& \pm 1 \;\; , \;\; Y=0
\EEA
$\wit{\nu}$ changes sign along the curves
\BEA
a &=& \frac{1-Y^2}{1+Y^2} \nonumber \\
a_{\pm} &=& \left[ h-4Y+6hY^2-4Y^3+hY^4 \right]^{-1} 
\left[ \left(1-Y^2\right) \left( h\left(1+Y^2\right) -2Y\left(
1\pm\sqrt{1-h^2}\right) \right) \right] \nonumber \\
a &=& 0 \nonumber \\
Y &=& 0
\EEA
$\wit{\sig}$ changes sign along the curves
\BEA
a_{\pm} &=& \left[ h-4Y+6hY^2-4Y^3+hY^4 \right]^{-1} 
\left[ \left(1-Y^2\right) \left( h\left(1+Y^2\right) -2Y\left(
1\pm\sqrt{1-h^2}\right) \right) \right] \nonumber \\
a &=& \pm 1 \;\; , \;\; a = 0 \nonumber \\
Y &=& \pm 1 \;\; , \;\; Y=0 
\EEA 
Next, we have to distinguish between the cases $h^2 > 1$ and $h^2 \leq 1$.
In the first case, the $a_{\pm}$ given above are complex and will not play
a role in dividing the $(a,Y)$ plane into regions where the $\wit{\vph}$
are positive or negative, respectively. In fact, in this case the boundaries
are completely independent of $h$. In figure~1, we show the excluded
(shaded) regions for the four rates 
$\wit{\alpha},\wit{\gamma},\wit{\nu},\wit{\sig}$.
Comparing them, it follows that no region in the $(a,Y)$ plane remains where
the four inequalities (\ref{Ineq}) are all satisfied.  

In the second case, the $a_{\pm}$ are all real and must be taken into account.
In order to illustrate the mutual exclusion of the allowed regions, we show
in figure~2 the excluded (shaded) regions for the four rates for the special
value $h=\frac{1}{2}\sqrt{3}$. In comparing the excluded regions for the
different rates, note that because of the factor structure of the matrix 
elements, the boundary curves in the different figures coincide.  
Again, it is apparent that (\ref{Ineq}) cannot be
satisfied. 

\noindent {\bf Case B: $a$ real and $Y$ imaginary.} \\
We write $Y=i y$ and $h = i k$ where $y,k$ are both real. We fix $k$ and look
for the curves in the $(a,y)$ plane where the $\wit{\vph}$ have simple zeroes
or poles. For $\wit{\alpha}$, these occur at
\BEA
a &=& \frac{1-y^2}{1+y^2} \nonumber \\
a &=& \left[ k(1+y^2)\right]^{-1} 
\left[ k(1-y^2) +2y (1\pm \sqrt{1+k^2})\right] \nonumber \\
y &=& 0
\EEA
$\wit{\gamma}$ changes sign at
\BEA
a &=& \left[ k(1+y^2)\right]^{-1} 
\left[ k(1-y^2) +2y (1\pm \sqrt{1+k^2})\right] \nonumber \\
a &=& \pm 1 \;\; , \;\; a=0 \nonumber \\
y &=& 0
\EEA
$\wit{\nu}$ changes sign at
\BEA
a &=& \frac{1-y^2}{1+y^2} \nonumber \\
a &=& \left[ k-4y-6ky^2+4y^3+ky^4 \right]^{-1} 
\left[ (1+y^2)\left( k(1-y^2) -2y(1\pm\sqrt{1+k^2}) \right) \right] 
\nonumber \\
a &=& 0 \nonumber \\
y &=& 0
\EEA
$\wit{\sig}$ changes sign at
\BEA
a &=& \left[ k-4y-6ky^2+4y^3+ky^4 \right]^{-1} 
\left[ (1+y^2)\left( k(1-y^2) -2y(1\pm\sqrt{1+k^2}) \right) \right] 
\nonumber \\
a &=& \pm 1 \;\; , \;\; a=0 \nonumber \\
y &=& 0
\EEA
In analogy what was done in case A, one can show that each region in the
$(a,y)$ plane is excluded by at least one of the four rates and thus 
(\ref{Ineq}) cannot be satisfied. 

\noindent {\bf Case C: $a$ imaginary and $|Y|=1$.} \\
We write $a=i A$ and $Y=\cos q + i \sin q$ with $A,q$ real. We fix $h$ 
and look for the simple zeroes or poles of the rates $\wit{\vph}$. 
For $\wit{\alpha}$, these occur at
\BEA
A_{\pm} &=& h^{-1} \csc q \cdot \left( 1\pm\sqrt{1-h^2} +h \cos q\right)
\nonumber \\
A &=& \cot q
\EEA
For $\wit{\gamma}$, they occur at
\BEA
A_{\pm} &=& h^{-1} \csc q \cdot \left( 1\pm\sqrt{1-h^2} +h \cos q\right)
\nonumber \\
A &=& 0 \nonumber \\
q &=& 0, \pm\pi
\EEA
For $\wit{\nu}$, they occur at
\BEA
A_{\pm} &=& \left[ -3h +4\cos q - h\cos 2q \right]^{-1} 
\left[ 2 \sin q \cdot \left( -1 \pm\sqrt{1-h^2}+h \cos q\right) \right]
\nonumber \\
A &=& -\tan q \nonumber \\
A &=& 0
\EEA
and for $\wit{\sig}$, they occur at
\BEA
A_{\pm} &=& \left[ -3h +4\cos q - h\cos 2q \right]^{-1} 
\left[ 2 \sin q \cdot \left( -1 \pm\sqrt{1-h^2}+h \cos q\right) \right]
\nonumber \\
A &=& 0 \nonumber \\
q &=& 0, \pm\pi
\EEA
For $h^2 > 1$, the $A_{\pm}$ are complex and can be ignored. 
For $h^2 \leq 1$, the $A_{\pm}$ are real. To illustrate the mutual
exclusion of the positivity conditions, we display in figure~3
the graphs for the case $h^2 >1$. Again, it is apparent that
(\ref{Ineq}) cannot be all satisfied. The case $h^2 < 1$ is treated similarly. 

This proves the assertion. \hfill q.e.d. 

\subsection{The parity-symmetric stochastic systems}

The lemma just proven severely restricts the possibilities for the choice of
the transformation parameters $a,Y$ and still find a stochastic quantum
Hamiltonian $\wit{H}$. In fact, it follows that at least one of the rates
$\wit{\alpha},\wit{\gamma},\wit{\nu},\wit{\sig}$ has to vanish. But because
of the explicit factorization (\ref{SymmH}) of the matrix elements, this means
that two of the four rates considered will be zero. Comparing with 
(\ref{SymmH}), we see that one of the following conditions must be implemented
\BEQ
F_1 = 0 \;\; , \;\; F_2 = 0 \;\; , \;\; -1+a-Y^2-aY^2 =0 \;\; , \;\;
-1+a+Y^2 +aY^2 =0 \;\; , \;\; a^2 =1
\EEQ
We turn to them one by one. 

\noindent {\bf 1. $F_1=0$.} This means that $\wit{\alpha}=\wit{\gamma}=0$. 
We solve this for $h$ and get
\BEA
h &=& -4Y (1-a+Y^2+aY^2) N_1 \\
N_1^{-1} &=& 1-2a+a^2+6Y^2-2a^2Y^2+Y^4+2aY^4+a^2Y^4 \nonumber
\EEA
and the non-vanishing rates then become
\BEA
-2\wit{\nu}   &=& \frac{4}{a} (1+Y^2) (1-a-Y^2-aY^2) N_1 \nonumber \\
-\wit{\delta} &=& 2a(1-Y^2) (1-a+Y^2+aY^2) N_1 \nonumber \\
-\wit{\beta}  &=& 2(Y^2-1) (-1+a+Y^2+aY^2) N_1 \nonumber \\
-\wit{\sig}   &=& \frac{2}{a} (a^2-1)(1-Y^2) (1+Y^2) N_1 \nonumber \\
-\wit{D}      &=& 2(1+Y^2) (-1+a-Y^2-aY^2) N_1
\EEA
A necessary condition for positivity are the relations
\BEQ
\frac{ 2\wit{\nu}}{\wit{\beta}} \geq 0 \;\; , \;\; 
\frac{\wit{D}}{\wit{\delta}} \geq 0
\EEQ
Provided none of the factors in the rates vanishes, we then find 
\BEQ
0 \leq \frac{1}{a} \frac{Y^2 +1}{Y^2 -1} \leq 0 
\EEQ
This means that one of the following conditions must be satisfied:
\BEQ \label{YCases}
i) \;\; Y = \pm \sqrt{\frac{1-a}{1+a}} \;\; , \;\; 
ii) \;\; Y = \pm \sqrt{\frac{a-1}{a+1}} \;\; , \;\;
iii) \;\; Y = \pm i
\EEQ
In the first case, we write $a=i A$ with $A$ real, 
i.e. $Y^2 = (1-iA)/(1+iA)$, and find that
\BEQ \label{S:dek}
\wit{H}_{j,j+1} = \frac{1}{2+A^2} \left( \begin{array}{cccc}
4 & -2A^2 & -2A^2 & 0 \\
-2 & 2+2A^2 & -2 & 0 \\
-2 & -2 & 2+2A^2 & 0 \\
0 & 0 & 0 & 0 \end{array} \right)
\EEQ
does correspond to a stochastic system, which we call model I. We identify
\BEQ
C = 1 \;\; , \;\;  h= -2 \frac{\sqrt{1+A^2}}{2+A^2} \;\; , \;\; 
\eta_1=\eta_2 = - \frac{A^2}{2+A^2} \;\; , \;\;
D_1 = D_2 =1
\EEQ
In the second case, we get
\BEQ
\wit{\nu}=\wit{\beta}=-\wit{\sig}=2
\EEQ
which is not stochastic. Finally, the third case $Y^2=-1$ gives, 
with $a=i A$ and $A$ real
\BEQ \label{S:vot}
\wit{H}_{j,j+1} = \frac{1}{1+A^2} \left( \begin{array}{cccc}
0 & -2A^2 & -2A^2 & 0 \\
0 & 2+2A^2 & 0 & 0 \\
0 & 0 & 2+2A^2& 0 \\
0 & -2 & -2 & 0 \end{array} \right)
\EEQ
which is again stochastic, is called model III, and we identify
\BEQ
C = 1 \;\; , \;\; h = \frac{2A}{1+A^2} \;\; , \;\;
\eta_1 = - \frac{(A-1)^2}{1+A^2} \;\; , \;\;
\eta_2 = - \frac{(A+1)^2}{1+A^2} \;\; , \;\;
D_1 = D_2 =1 \label{3.45}
\EEQ

\noindent {\bf 2. $F_2=0$.} This means that $\wit{\nu}=\wit{\sig}=0$.
Solving for $h$ we find
\BEA
h &=& 4 a Y (-1+a+Y^2+aY^2) N_2 \\
N_2^{-1} &=& 1-2a+a^2-2Y^2+6a^2Y^2+Y^4+2aY^4+a^2Y^4 \nonumber
\EEA
The non-vanishing rates are now found as above and the discussion is
completely parallel.  
We take $Y=\sqrt{ (a-1)/(a+1)}$ and $a=i A^{-1}$ and get the stochastic
quantum Hamiltonian
\BEQ \label{S:koa}
\wit{H}_{j,j+1} = \frac{1}{2+A^2} \left( \begin{array}{cccc}
0 & 0 & 0 & 0 \\
0 & 2+2A^2 & -2 & -2 \\
0 & -2 & 2+2A^2 & -2 \\
0 & -2A^2 & -2A^2 & 4 \end{array} \right)
\EEQ
which is called model II and we identify
\BEQ
C = 1 \;\; , \;\; h = 2 \frac{\sqrt{1+A^2}}{2+A^2} \;\; , \;\; 
\eta_1 = \eta_2 = -\frac{A^2}{2+A^2} \;\; , \;\;
D_1 = D_2 =1 \label{3.48}
\EEQ
This transformation of model II to free fermions had been given before
\cite{Kreb95}. Note that the only difference with model I (\ref{S:dek})
occurs in the sign of $h$, which expresses the fact that the two 
models I and II are related by a particle-hole exchange. 
The other possibilities (analogous to (\ref{YCases})) 
do not give anything new. 

\noindent {\bf 3. $Y=\pm \sqrt{(a-1)/(a+1)}$.} We write again $a=i A^{-1}$ and
find in particular
\BEQ
-\wit{\sig} = 4 \wit{\nu} = 2- h \frac{2+A^2}{\sqrt{1+A^2}}
\EEQ
This can only be made stochastic if $\wit{\nu}=\wit{\sig}=0$. But then
$h=2\sqrt{1+A^2}/(2+A^2)$ and we are back to the case (\ref{S:koa}) 
already found.  

\noindent {\bf 4. $Y=\pm \sqrt{(1-a)/(1+a)}$.} In the same way, we find that
$h=-2\sqrt{1+A^2}/(2+A^2)$ and we recover the system (\ref{S:dek}). 

\noindent {\bf 5. $a^2 =1$.} We find that $\wit{\delta}/\wit{\beta}=-1$. 
Now $\wit{\delta} = h(1-Y^2)/(2Y)$ which only vanishes for $h=0$ or $Y^2=1$. 
In both cases, the transformation $\cal B$ is diagonal and we are back to the
systems studied in \cite{Gryn94}. 

In conclusion, we have found the complete list of parity-symmetric stochastic
systems which can be reduced to a free fermion model with site-independent
interactions through the transformation (\ref{BTrans}). These are given in
(\ref{S:koa}), (\ref{S:dek}), (\ref{S:vot}) and describe particles undergoing
coagulation/decoagulation with the 
coagulation rate equal to the diffusion rate
(in this case, the transformation was already known \cite{Kreb95}), the model
obtained from the latter one when particles and holes are exchanged and
finally the so-called biased voter model, see \cite{Dick95}. This 
correspondence is not immediately obvious and is explained in appendix~C. 

\subsection{Stochastic systems with a drift}

Having found the set of transformations which render the symmetric rates
$\wit{\vph}$ positive, we can now extend the identification with free
fermion systems to models including a drift. 

We illustrate the procedure for the case when $F_1=0$ and then proceed to 
list the results. When $F_1=0$, we have $\wit{\gamma}=0$ and thus
$\wit{\gamma}_L=\wit{\gamma}_R=0$, since both must be positive. From
(\ref{AntiH}), this fixes $y=(D_1-D_2+2g)a(1-Y^2)/(8Y)$. But on the
other hand, we have
\BEA
\wit{\delta}_L + \wit{\delta}_R &=& 2 \wit{\delta} \nonumber \\
\wit{\delta}_L - \wit{\delta}_R + 2y &=& (D_1 - D_2 -2g) 
\frac{a(Y^2 -1)}{4Y} 
\EEA
Thus, solving for $\wit{\delta}_{L,R}$, we find
\BEQ
\wit{\delta}_{L,R} = \wit{\delta} \pm \frac{a(Y^2-1)}{4Y} ( D_1 - D_2)
\EEQ
Similar considerations apply for the other rates. We can now list the
stochastic systems with a drift which can be mapped to a free fermion 
system via (\ref{BTrans}).
\begin{enumerate}
\item Model I with death/creation process (\ref{S:dek}) extends to the
following system with the only non-vanishing rates
\BEQ
\wit{\delta}_{R,L} = A^2 \,\wit{\sig}_{L,R} = A^2 \,\wit{D}_{L,R} =
\frac{2 A^2}{2+A^2} \mp \frac{A^2}{2\sqrt{1+A^2}} (D_1 - D_2)
\EEQ 
and the surface fields are given by
\BEQ
x = \frac{1}{4}(D_1 - D_2 -2g) \frac{1}{\sqrt{A^2+1}} \;\; , \;\;
y = -\frac{1}{4}(D_1 - D_2 +2g) \frac{A^2}{\sqrt{A^2+1}} 
\EEQ
The positivity of the rates restricts $D_1 -D_2$ to a certain interval whose
boundaries depend only on the real parameter $A$. 
\item For the coagulation/decoagulation model II (\ref{S:koa}) we get the
following system with the non-vanishing rates
\BEQ
\wit{\beta}_{L,R} = A^2 \,\wit{\gamma}_{L,R} = A^2 \,\wit{D}_{L,R} = 
\frac{2 A^2}{2+A^2} \pm \frac{A^2}{2\sqrt{1+A^2}} (D_1 - D_2)
\EEQ
and the surface fields are given by
\BEQ
x = -\frac{1}{4}(D_1 - D_2 +2g)  \frac{A^2}{\sqrt{A^2+1}} \;\; , \;\; 
y = \frac{1}{4}(D_1 - D_2 -2g) \frac{1}{\sqrt{A^2+1}} 
\EEQ
For this model, the transformation to free fermions is already 
known \cite{Hinr95}. 
\item Finally, the biased voter model III (\ref{S:vot}) is extended to the
following system with the non-vanishing rates
\BEQ
\wit{\delta}_{L,R} = A^2 \,\wit{\beta}_{L,R} =
\frac{2 A^2}{1+A^2} \pm \frac{A}{2} (D_1 - D_2)
\EEQ
and the surface fields are given by
\BEQ
x = -\frac{1}{4}(D_1 - D_2 +2g) A^{-1} \;\; , \;\;
y = -\frac{1}{4}(D_1 - D_2 +2g) A
\EEQ
\end{enumerate}

In order to calculate the particle number correlators using the free fermion
representation, we have to rewrite the particle number operator 
$\wit{n}_j$ (\ref{NOp})
in the free fermion basis. This is straightforward from the known
transformations,
and thus completely determined from the parity symmetric
case. First, for the system I (\ref{S:dek}), we can rewrite the
parameters in the transformation matrix (\ref{Euler}) in the form
\BEQ \label{3.58}
A = - \tan (2\psi) \;\; , \;\; \phi = i \psi \;\; , \;\; a = i A \;\; , 
\;\; b=1
\EEQ
and we want $n_j = B^{-1} \wit{n}_j B$ with $\wit{n}_j = \left( 
\matz{0}{0}{0}{1}\right)_j$. We find
\BEA
n_j &=& \frac{1}{2} \left( {\bf 1} - \cos 2\psi \, \sig_j^z + 
\sin 2\psi \, \sig_j^y\right)
\nonumber \\
  &=& \frac{1}{2} \left( {\bf 1} - \frac{1}{\sqrt{1+A^2}} \sig_j^z -
\frac{A}{\sqrt{1+A^2}} \sig_j^y \right) \label{3.59}
\EEA
Second, for the system III (\ref{S:vot}), the transformation parameters are
\BEQ
\phi = \frac{i \pi}{4} \;\; , \;\; a = i A \;\; , \;\; b=1
\EEQ
and we find
\BEQ
n_j = \frac{1}{2} \left( {\bf 1} + \sig_j^y \right)
\EEQ
Finally, for the system II (\ref{S:koa}), the transformation parameters are
\BEQ
A = \tan (2\psi) \;\; , \;\; \phi = i \psi \;\; , \;\; a = i A^{-1} \;\; , 
\;\; b=1
\EEQ
and we have
\BEQ
n_j = \frac{1}{2} \left( {\bf 1} - \frac{1}{\sqrt{1+A^2}} \sig_j^z +
\frac{A}{\sqrt{1+A^2}} \sig_j^y \right)
\EEQ

If one wants  to calculate time correlators of the number operators
in the stochastic system then one needs the above equalities,
the state $\langle s| {\cal B}$ and has also to compute the action of 
${\cal B}^{-1}$ on the initial state. 
The calculation of $\langle s| {\cal B}$
reduces to a single site problem since $\langle s|$ is a factorized state, 
$\langle s| = (1,1)^{\otimes L}$ and 
also ${\cal B} = B^{\otimes L}$ factorizes. Later on we shall need the
identity
\BEQ
(1,1) B = (\sqrt{a}b \cosh{\phi} + \sqrt{a^{-1}}b \sinh{\phi}, \sqrt{a}b^{-1} 
\sinh{\phi} +
\sqrt{a^{-1}} b^{-1} \cosh{\phi} )
\EEQ
{}From this one can, in particular, calculate density correlators
 for all four cases
(see section 5).

\subsection{Summary}

To summarize our classification of the 
stochastic systems with site-independent interactions
which may be transformed through a single-site transformation (\ref{HTilde})
into a free fermion system with site-independent couplings, 
we list in table~\ref{tab3} the stochastic systems,  
\begin{table}
\begin{center}
\begin{tabular}{|c|c|cc|cc|} \hline 
model & rates / number operator & $x$ & $y$ & $h$ & $\eta$ \\ \hline 
  & ~~    &     &     &     &        \\[0.5\baselineskip]
 I & $\wit{\delta}_{R,L} = A^2 \wit{\sig}_{L,R} = A^2 \wit{D}_{L,R}$ = & 
 $-{\cal D}_{-}\frac{1}{\sqrt{A^2+1}}$ & $-{\cal D}_+ 
 \frac{A^2}{\sqrt{A^2+1}}$  
 & $-\frac{2\sqrt{A^2+1}}{A^2+2}$ & $-\frac{A^2}{A^2+2}$ \\
 &$A^2 \left[ 2/(A^2+2) \mp (D_1-D_2)/(2\sqrt{A^2+1})\right]$ & & & & \\[\TZ]
 & $2n = 1- \frac{1}{\sqrt{1+A^2}}\sig^z -\frac{A}{\sqrt{1+A^2}}\sig^y$ 
& & & &  \\[0.5\baselineskip]
 & ~~ &  &  &  &  \\ \hline
 & ~~    &     &     &     &        \\[0.5\baselineskip] 
 II & $\wit{\beta}_{L,R} = A^2 \wit{\gamma}_{L,R} = A^2 \wit{D}_{L,R}$ = &
 $-{\cal D}_{+}\frac{A^2}{\sqrt{A^2+1}}$ & 
 $-{\cal D}_{-}\frac{1}{\sqrt{A^2+1}}$
 & $\frac{2\sqrt{A^2+1}}{A^2+2}$ & $-\frac{A^2}{A^2+2}$ \\
 &$A^2 \left[ 2/(A^2+2) \pm (D_1-D_2)/(2\sqrt{A^2+1})\right]$ & & & & \\[\TZ]
 & $2n = 1- \frac{1}{\sqrt{1+A^2}}\sig^z +\frac{A}{\sqrt{1+A^2}}\sig^y$ 
& & & & \\[0.5\baselineskip]
 & ~~ &  &  &  &  \\ \hline
 & ~~    &     &     &     &        \\[0.5\baselineskip]
III & $\wit{\delta}_{L,R} = A^2 \wit{\beta}_{L,R}$ = & 
 $ -{\cal D}_{+} / A$ & $-{\cal D}_{+} A$ & $\frac{2A}{A^2+1}$ & 
 $\frac{A^2-1}{A^2+1}$ \\
 &$A^2 \left[ 2/(A^2+1) \pm (D_1 - D_2)/(2A) \right]$ & & & & \\[\TZ]
 & $2n = 1 - \frac{i}{\sqrt{A^2-1}}\sig^x + \frac{A}{\sqrt{A^2-1}}\sig^y $
 & & & & \\[0.5\baselineskip]
 & ~~ &  &  &  &  \\ \hline
 & ~~    &     &     &     &        \\[0.5\baselineskip] 
 IV & $\wit{\alpha}, \wit{\nu}, \wit{D}_L, \wit{D}_R$ & 
 0 & 0 & $\wit{\alpha}-\wit{\nu}$ & $2\sqrt{\wit{\alpha}\wit{\nu}}$ \\
 & $2\wit{\alpha}+2\wit{\nu}=\wit{D}_L + \wit{D}_R=2$, $2g = D_2 -D_1$ 
 & & & & \\[\TZ]
 & $2n = 1 -\sig^z$ & & & & \\[0.5\baselineskip] \hline
\end{tabular} \end{center} 
\caption[Free fermion models]{Stochastic models which can be 
transformed into free fermions, 
with the normalization $(D_1+D_2)/2=1$.
The rates given are the only non-vanishing ones. $x$ and $y$ are the surface
fields and $h$ and $\eta$ the parameters of the free fermion model. $A$ is a
free real parameter and the short-hand
${\cal D}_{\pm} = \pm(D_1 - D_2 \pm 2g)/4$ is used. We also write the 
expression for the particle number operator $\wit{n}$ transformed 
into the free fermion basis $n=B^{-1}\wit{n}B$ and expressed in terms of
Pauli matrices such that $\eta_1=\eta_2=\eta$ in $H$, see eq.~(\ref{FFHam}).  
\label{tab3}} 
\end{table}
as characterized by the non-vanishing reaction rates as defined in 
table~\ref{tab1}, which can be transformed into the free fermion Hamiltonian 
\BEQ \label{FFHam}
H = - \sum_{j} \left[ 
D_1 \sig_j^- \sig_{j+1}^+ + D_2 \sig_j^+ \sig_{j+1}^- 
+ \eta \left( \sig_j^- \sig_{j+1}^- + \sig_j^+ \sig_{j+1}^+ \right)
+ h_1 \sig_j^z + h_2 \sig_{j+1}^z  -1 \right]
\EEQ
with $h=h_1+h_2$ and $g=h_1-h_2$. Note that we also used the transformation
parameter $b$ in (\ref{Euler}) and chosen to be $b^8=\eta_1/\eta_2$
to enforce $\eta_1=\eta_2=\eta$ in table~\ref{tab3}. 
In table~\ref{tab3}, we also list the values 
$x,y$ for the surface fields which
are needed to guarantee the stochastic interpretation in case of a drift,
when $D_1\neq D_2$, and the particle number operator (\ref{NOp}) when 
transformed into the free fermion basis. 

Model I describes particles undergoing creation and diffusion with the same
rate and death with an arbitrary rate, with and without a parity-breaking
bias. Model II is obtained from this by a particle-vacancy exchange and
describes particles undergoing coagulation and diffusion with the same
rate and decoagulation with an arbitrary rate, with and without a drift. 
The transformation of this model to free fermions had been given before
\cite{Kreb95,Hinr95}. Model III is the biased voter model \cite{Dick95} 
with and without spatial asymmetry. While the voter model is usually defined
in terms of particle reactions involving three neighboring sites, we
remind the reader in appendix~C how to reformulate it in terms of two-sites
reactions. Finally, model IV describes annihilation and birth together with
diffusion with or without a drift and considered in detail 
in \cite{Lush86,Gryn94}. This last model is trivial from our transformation
point of view, because by fixing $h$ and $g$ the original Hamiltonian 
$H$ is already a stochastic matrix itself and the
transformation matrix $B$ will be diagonal.

As expected, the four models share the
property that the parameters $h,\eta$ in (\ref{FFHam}) satisfy $h^2+\eta^2=1$,
which is the well-known stochastic line in the free fermion system 
(\ref{FFHam}). These systems can
therefore be mapped via a duality transformation \cite{Sigg77} onto the
one-dimensional kinetic Ising model with Glauber dynamics. However, since
the transformation to free fermions of model IV is trivial and of models I,II
and III
is not, the correlators to be discussed in the free fermion language to get
the particle number correlations are quite different. 

Another distinction becomes apparent in the presence of a
drift. For models I,II, the divergence terms parametrized
by $x$ and $y$ are automatically generated through the similarity
transformation. 
Although for periodic boundary conditions these terms
cancel, for non-periodic boundary conditions, they will generate surface
fields. In order to find a stochastic system, it will be necessary to add
into (\ref{FFHam}) further surface fields and see how they will transform
under (\ref{BTrans}). As a consequence, boundary conditions will generically
affect the spectrum in an important way. This is different for model III. 
Here, for a fixed value of the parameter $A$, the couplings $x,y$ of the
divergence terms are proportional to each other and the 
parameter $g$ can be chosen in (\ref{FFHam}) such that $x=y=0$ simultaneously. 
Then even for non-periodic
boundary conditions with a drift, the free fermion Hamiltonian
only contains terms which are bilinear in the fermion operators and the
techniques of appendix~A for its diagonalization can be applied immediately.
For model IV the couplings $x$ and $y$ always vanish. 

It is of interest to compare the list of free fermion systems of 
table~\ref{tab3} with the systems obtained through the approach of
Peschel et al. \cite{Pesc95}. They consider 
a diffusive coagulation/decoagulation
system where in addition birth and creation processes may also be present
and identify certain classes of observables for which closed systems
of equations of motion can be established and solved. The exitation spectrum
of the corresponding quantum Hamiltonian is the one of a XY quantum chain
in a {\em site-dependent} transverse field \cite{Pesc95}. Obviously, 
our models I,II are special cases of the systems studied in \cite{Pesc95},
but the voter model is not included there. It remains a challenge
to extend the present transformation technique to integrable Hamiltonians
with site-dependent interactions and investigate which stochastic systems
beyond those studied in \cite{Pesc95} exist (in particular, in what
directions model III can be extended).

Finally, the transformation (\ref{BTrans}) we studied is completely local. 
Although we derived everything here for one-dimensional systems with
translational invariance, the transformations as they stand can also be
applied to models which are either higher dimensional or contain 
non-translational invariant interactions, as they might arise for disordered
or quasiperiodically modulated systems. Since this type of problem has been
much studied for free fermion equilibrium models, see \cite{Iglo93} for
a review, the techniques developed for equilibrium systems 
can be carried over to non-equilibrium problems as well.

\section{Stochastic systems from Hamiltonian with $XXZ$ spectrum}

In this section we give the classification of those stochastic systems
which can be obtained through the  transformation (\ref{BTrans})
from the most general real Hamiltonian having the spectrum of an integrable
generalized Heisenberg $XXZ$ quantum chain.

\subsection{Determination of the transformation matrix}

Our starting point is a matrix $H=\sum_j H_{j,j+1}$ where 
$H_{j,j+1}$ is the $4\times 4$ matrix describing
the nearest-neighbor interactions\footnote{One might also consider a lower
triangular form for $H$, which physically merely amounts to a particle-hole
interchange.}
\BEQ \label{H}
H_{j,j+1} = \left( \begin{array}{cccc} 
h_{11} & h_{12} & h_{13} & h_{14} \\
0 & h_{22} & h_{23} & h_{24} \\
0 & h_{32} & h_{33} & h_{34} \\
0 & 0 & 0 & h_{44} \end{array} \right)
\EEQ
and we look for a matrix $\wit{H}$ through the transformation
\BEQ \label{BTrans1}
\wit{H} = {\cal B} H {\cal B}^{-1} \;\; , 
\;\; {\cal B} = \bigotimes_{j=1}^{L} B_j
\EEQ
with
\BEQ \label{b}
B = \left( \begin{array}{cc} 
b_{11} & b_{12}\\
b_{21} & b_{22} \end{array} \right)
\EEQ
and describing a stochastic system. 

As pointed out in \cite{Alca94} for a matrix of the form 
(\ref{H}) the spectrum
does not depend on the parameters $h_{12}, h_{13}, h_{14}, h_{24}, h_{34}$.
Without these matrix elements, this Hamiltonian can be diagonalized using
standard coordinate Bethe ansatz \cite{Beth31,Yang66}. The eigenvalues
of the system with $L$ sites and periodic boundary conditions are 
\BEA
E_0  &=& h_{11} L \label{ev1} \\
E(p_1,\dots,p_N) =  E_0& -&\sum_{\ell=1}^{N} \left[ (2h_{11}-h_{22}-h_{33})
-(h_{23}+h_{32})\cos p_{\ell}+\mbox{i}(h_{23}-h_{32})\sin p_{\ell}\right]
\nonumber\\
  & & 1 \leq N \leq L/2\label{ev2} \\
E(p_1,\dots,p_N)  =   E_L&-&\sum_{\ell=1}^{N} \left[ (2h_{44}-h_{22}-h_{33})
-(h_{23}+h_{32})\cos p_{\ell}-\mbox{i}(h_{23}-h_{32})\sin p_{\ell}\right]
\nonumber\\
 & & L/2 \leq N \leq L-1\label{ev3} \\
E_L & = & h_{44} L \label{ev4}
\end{eqnarray}
where for fixed $N$ the $p_\ell$ are given by the Bethe ansatz 
equations \cite{Matt}
\BEQ
p_{l}\;=\;\frac{2\pi n_{\ell} + 
\sum_{p_{j}\neq p_{\ell}}\psi_{p_{\ell} p_{j}}}{L}
\label{BAE0}
\EEQ
where the $n_{\ell},n_{j}$ are distinct integers 
and the phase shifts $\psi_{p_{\ell} p_{j}}$ are given by
\BEQ
\label{BAE1}
\cot\left(\frac{\psi_{p_{\ell} p_{j}}}{2}\right) = 
\frac{(h_{11}+h_{44}-h_{22}-h_{33})\sin
\left(\frac{p_{\ell}-p_{j}}{2}\right)}{(h_{23}+h_{32})
\cos\left(\frac{p_{\ell}+p_{j}}{2}\right)-
\mbox{i}(h_{23}-h_{32})\sin\left(
\frac{p_{\ell}+p_{j}}{2}\right)-(h_{11}+h_{44}-h_{22}-h_{33})
\cos\left(\frac{p_{\ell}-p_{j}}{2}\right)}
\EEQ
if $1 \leq N \leq L/2$ and 
\BEQ
\label{BAE2}
\cot\left(\frac{\psi_{p_{\ell} p_{j}}}{2}\right) = 
\frac{(h_{11}+h_{44}-h_{22}-h_{33})
\sin\left(\frac{p_{\ell}-p_{j}}{2}\right)}{(h_{23}+h_{32})
\cos\left(\frac{p_{\ell}+p_{j}}{2}\right)+
\mbox{i}(h_{23}-h_{32})\sin\left(\frac{p_{\ell}+p_{j}}{2}\right)-
(h_{11}+h_{44}-h_{22}-h_{33})\cos\left(
\frac{p_{\ell}-p_{j}}{2}\right)}
\EEQ
if $L/2 \leq N \leq L-1$. 
{}From these equations we can decide how many
steady states the systems has, i.e. how many eigenvalues 0, and whether the
low-lying excitation spectrum has a non-vanishing gap or is gapless.

The first step towards a classification of stochastic 
processes with a spectrum
given by (\ref{ev1}) - (\ref{ev4}) consists 
in determining the matrix $B$. Since $\wit{H}$ is stochastic, 
$\wit{H}_{j,j+1}$ has the left ground state
\BEQ
\bra{s} = (1,1)^{\otimes^2}
\EEQ
which is a {\it product state}.
This implies that also $H$ will have a left ground state as a product state
and related to $\bra{s}$
\BEQ\label{cond2}
\bra{s}B = \bra{\omega} = (\omega_1,\omega_2)^{{\otimes}^2} = 
\left( \omega_1^2, \omega_1 \omega_2,
\omega_1 \omega_2, \omega_2^2 \right)
\EEQ
where $\omega_{1,2}$ are free parameter. It follows that we must have 
\BEQ\label{cond1}
\bra{\omega}H_{j,j+1} = \bra{\omega} (A_j - A_{j+1}) = (0,x,-x,0)
\EEQ
where $A_j$ is an arbitrary matrix acting on site $j$ and $x$ is an 
arbitrary constant (depending on $A$).
Next, we observe from (\ref{cond1}) that we can always arrange to have
$\omega_1 + \omega_2 =1$. It is thus enough to keep only $\omega := \omega_1$
as free parameter. 

Now the implementation of eq.~(\ref{cond1}) allows us to distinguish three
cases.

\begin{description}
\item {\bf Case I: $\omega = 0$.}
In this case $\bra{\omega} = (0,0,0,1)$ and by imposing
(\ref{cond1}) we have:
\BEQ\label{condI}
h_{44} = 0
\EEQ
whereas, from (\ref{cond2}):
\BEQ \label{bI}
B = \left( \begin{array}{cc} 
b_{11} &1- b_{22}\\
-b_{11} & b_{22} \end{array} \right)
\EEQ
Moreover, from the positivity of the spectrum we do have 
the following inequalities:
\BEA \label{ineI}
h_{11} &\ge& 0 \nonumber \\
(h_{23}+h_{32})\cos {k} +h_{22}+h_{33} &\ge& 0  \qquad \forall k\in
[0,2\pi]  
\EEA
where we have used (\ref{ev3}) and (\ref{BAE2}) for $N=1$.

\item {\bf Case II: $\omega = 1$.}
The left eigenvector is $\bra{\omega} = (1,0,0,0)$
and eq. (\ref{cond1}) gives:
\BEA\label{condII}
h_{11} &=& 0 \nonumber \\
h_{12}+h_{13} &=& 0  \\
h_{14} &=& 0 \nonumber
\EEA
By using eq. (\ref{cond2}) the transformation matrix $B$ simplifies to:
\BEQ \label{bII}
B = \left( \begin{array}{cc} 
b_{11} &- b_{22}\\
1-b_{11} & b_{22} \end{array} \right)
\EEQ
Positivity of the spectrum translates, in the same way as above,
in this case into the following
inequalities:
\BEA \label{ineII}
h_{44} &\ge& 0 \nonumber \\
(h_{23}+h_{32})\cos {k} +h_{22}+h_{33} &\ge& 0  \qquad \forall k\in
[0,2\pi]  
\EEA
where we have used (\ref{ev3}) and (\ref{BAE2}) for $N=L-1$.

Further information can be gained by considering how many steady states
exist. If there is a single steady state, we must have
\BEQ \label{bUngl}
0 \leq b_{11} \leq 1 .
\EEQ
The important inequality (\ref{bUngl}) is seen as follows. 
If $h_{11}=0$, the {\it right} ground
state of $H_{j,j+1}$ is $\ket{0} = (1,0,0,0) = (1,0)^{\otimes^2}$. This 
corresponds to the right ground state $\ket{\wit{0}}$ of $\wit{H}_{j,j+1}$
through
\BEQ
\ket{\wit{0}} = (\wit{\rho}, 1-\wit{\rho})^{\otimes^2} = B_1 B_2 \ket{0}
\EEQ
where $\wit{\rho}$ is the mean particle density per site in the steady state
and obviously $0 \leq \wit{\rho} \leq 1$. From the explicit form of the 
matrix $B$, one has directly
\BEQ \label{Dicht}
\wit{\rho} = b_{11}
\EEQ
and (\ref{bUngl}) follows. In addition, eq.~(\ref{Dicht}) allows to calculate
$\wit{\rho}$ once the transformation matrix is found. 

If there are two steady states, one of
the following conditions must be satisfied:
\BEQ
(h_{23}+h_{32})\cos {k} +h_{22}+h_{33} =0  \qquad \hbox{for}
\qquad k=0,\pi
\EEQ
or
\BEQ
h_{44} = 0
\EEQ
In these cases, which we shall not treat explicitly, 
we did not find a similar constraint on the transformation parameter.

\item {\bf Case III: $\omega \neq 0,1$.}
Since $(\omega,1-\omega) = (1,1) {\cal R}$ with
\BEQ
{\cal R} = \left( \begin{array}{cc}
\omega &0\\
0 & 1-\omega \end{array} \right)
\EEQ
condition (\ref{cond1}) can be written as
\BEQ\label{cond3}
\bra{\omega}H_{j,j+1} =
\bra{s}{\cal R}\otimes {\cal R} H_{j,j+1} {\cal R}^{-1}\otimes {\cal R}^{-1} =
(0,x,-x,0)
\EEQ
In this way we have "absorbed" the parameter $\omega$ in the
coefficients of the Hamiltonian and condition (\ref{cond1})
simplifies to:
\BEQ\label{cond4}
(1,1,1,1)H_{j,j+1} = (0,x,-x,0)
\EEQ
{}From eq. (\ref{cond3}) we then obtain:
\BEA\label{condIII}
h_{11} &=& 0 \nonumber \\
h_{12}+h_{13}+h_{22}+h_{33}+h_{23}+h_{32} &=& 0  \\
h_{14}+h_{24} + h_{34} + h_{44} &=& 0 \nonumber
\EEA

For $B$, eq. (\ref{cond2}) becomes $(1,1) B = (1,1)$ and we obtain 
\BEQ \label{bIII}
B = \left( \begin{array}{cc} 
b_{11} & 1-b_{22}\\
1 - b_{11} & b_{22} \end{array} \right)
\EEQ
with $\det{B} = b_{11}+b_{22}-1$. 
Positivity constraints are the same as in case II.
Again, we can distinguish between subcases
in which we have one steady steady state or two steady states
obtaining the same  additional constraints on the parameters that
we obtained for case II.

A final comment is in order. From eqs.~(\ref{cond3}) it is clear that $H$ by 
itself also satisfies the probability conservation condition eq.~(\ref{StoCon})
for a stochastic system. Furthermore, the transformation (\ref{bIII})
has exactly the same form as previously derived \cite{Henk95,Simo95} for the
stochastic similarity transformation between two stochastic systems. Thus,
taking the off-diagonal elements of $H$ to be negative, we shall obtain the
complete list of two-states stochastic systems which can be reduced
through a stochastic similiarity transformation to the integrable model
which just contains the rates $\alpha,\gamma$ and $\delta$, corresponding
to annihilation, coagulation and death processes, respectively. 
\end{description}

By giving the explicit forms for the transformation matrices, we have 
implemented the probability conservation condition (\ref{StoCon}). It remains
to implement the positivity conditions. As we had done before in the
case of the free fermion integrable Hamiltonian $H$, we consider first
the parity-symmetric case, which by itself will completely specify the
transformation parameters $b_{11}$ and $b_{22}$. Subsequently, we can then
generalize to include parity-non symmetric situations, although we defer
from carrying out this explicitly here. It will turn out that
the task of analysing this set of seven coupled inequalities is considerably
simplified by going
over to new parameters, as defined in table~\ref{tab4}. 
\begin{table}
\begin{center}
\begin{tabular}{|c|cc|} \hline
parameter & \multicolumn{2}{c|}{ definition } \\ \hline
$\Delta$ & \multicolumn{2}{c|}{$(h_{23}+h_{32})/P$} \\ 
$\Omega$  & \multicolumn{2}{c|}{$h_{24}+h_{34}$} \\
$Q$ & \multicolumn{2}{c|}{$h_{22}+h_{33}+h_{23}+h_{32}$} \\
$P$ & \multicolumn{2}{c|}{$h_{44} - Q$} \\
$R$ & \multicolumn{2}{c|}{$Q/(2 P)$} \\ \cline{2-3}
 & Case II & Case III \\ \cline{2-3}
$S$ & $\Omega/P$ & $2+ (Q+\Omega)/P$ \\
\hline
\end{tabular}
\caption{Combinations of the coefficients $h_{ij}$ that
occur frequently in the text. With the exception of $S$, which is different
for the cases II and III, all parameters are defined in the same way in all
three cases. \label{tab4}}
\end{center} \end{table}
Also, it will turn out to be convenient in the cases II and III to replace
$b_{11}$ and $b_{22}$ by $q_{1,2}$ as follows

\BEQ \label{bq}
b_{11} = q_1\;\; , \;\; b_{22} = \left\{ 
\matz{ S q_2}{\mbox{\rm ; case II}}{1-b_{11} + S q_2}{\mbox{\rm ; case III}}
\right.
\EEQ

\subsection{Calculation of the stochastic rates}
We now give the rates of the stochastic system as obtained from the 
integrable chain $H$. 
\subsubsection{Case I: integrability through equations of motion}
After performing the similarity transformation (\ref{BTrans1})
on the matrix
\BEQ \label{HI}
H^{I} = \left( \begin{array}{cccc} 
h_{11} & h_{12} & h_{13} & h_{14} \\
0 & h_{22} & h_{23} & h_{24} \\
0 & h_{32} & h_{33} & h_{34} \\
0 & 0 & 0 & 0 \end{array} \right)
\EEQ
with $B$ given by (\ref{bI}) we obtain the transformed matrix
$\wit{H^{I}}$. The probability conservation condition
(\ref{StoCon}), expressed, by eqs. (\ref{ProCon}) are here
automatically satisfied and we have only to worry about the positivity
conditions of the elements $\wit{h}^{I}_{ij}$ for all $i\neq j$.

In this case, however, it turns out that the transformed elements
$\wit{h}^{I}_{ij}$ do satisfy the following pair of equations:
\BEA\label{GunI}
D^{I}_1&:=& \wit{h}^{I}_{34} -
\wit{h}^{I}_{21}-\wit{h}^{I}_{41}-
\wit{h}^{I}_{12}-\wit{h}^{I}_{32}+\wit{h}^{I}_{23}
+\wit{h}^{I}_{43}+\wit{h}^{I}_{14} = 0
\nonumber \\ 
D^{I}_2 &:=& \wit{h}^{I}_{24} -  
\wit{h}^{I}_{31}-\wit{h}^{I}_{41}-
\wit{h}^{I}_{13}-\wit{h}^{I}_{23}+\wit{h}^{I}_{32}
+\wit{h}^{I}_{42}+\wit{h}^{I}_{14}= 0
\EEA
When these conditions are both satisfied, 
one has a closed system of $k$ equations
for the equal-time $k$-point correlation functions \cite{Schu95}.
Since it can be checked immediately from the rates of the stochastic system
whether or not the conditions eqs.~(\ref{GunI}) are satisfied, the description
either through an integrable quantum chain or through the equations of motion
offer the same amount of information. While it might be easier to extract
relaxation times from the quantum Hamiltonian, the full time-dependence of
observables like the density is probably more easily extracted 
from the equations of motion. 

For practical applications it might be useful to repeat eqs.~(\ref{GunI})
where the $\wit{h}_{ij}$ are replaced by the physical rates
\BEA
\wit{\beta}_R+\wit{\gamma}_L+2\wit{\alpha} +\wit{D}_R &=&
\wit{\delta}_R+\wit{\sigma}_R+2\wit{\nu}+\wit{D}_L \nonumber \\
\wit{\beta}_L+\wit{\gamma}_R+2\wit{\alpha} +\wit{D}_L &=&
\wit{\delta}_L+\wit{\sigma}_L+2\wit{\nu}+\wit{D}_R 
\EEA
The two conditions coincide in the parity-symmetric case when they read
\BEQ \label{CIPs}
\wit{\delta}+\wit{\sigma}+2\wit{\nu} = \wit{\beta}+\wit{\gamma}+2\wit{\alpha}
\EEQ
and we note that the value of the diffusion constant $\wit{D}$ does not
enter into the integrability condition.

\subsubsection{Case II}
The matrix to be transformed is now 
\BEQ \label{HII}
H^{II} = \left( \begin{array}{cccc} 
0 & h_{12} & -h_{12} & 0 \\
0 & h_{22} & h_{23} & h_{24} \\
0 & h_{32} & h_{33} & h_{34} \\
0 & 0 & 0 & h_{44} \end{array} \right)
\EEQ
and the transformation matrix $B$ is given by (\ref{bII}). Probability
conservation is already implemented. In view of the result of the previous
subsection, it is of interest to check for integrablity through the
equations of motion.  
The expressions from eqs.~(\ref{GunI}) become in
this case:
\BEA
D^{II}_1 &=& - \left( h_{12} + h_{24}\right)/ ( S q_2 )
\nonumber \\ 
D^{II}_2 &=& \left( h_{12}-h_{34}\right)/ ( S q_2 )
\label{GunII}
\EEA
and we see that the integrability from case I is only recovered if 
simultaneously $\Omega=0$ and $h_{12}=h_{34}$ (provided $q_2$ remains
finite). 

Using the notation of table~\ref{tab4}, we can now write down the 
parity-symmetric rates
\BEA
\wit{\alpha} &=& -\frac{P}{2 q_2} q_1^2 \left( q_2 - q_1 \right) \nonumber \\
\wit{\beta} &=& \frac{P}{q_2} \left( q_1-1\right)
\left[ q_1 \left(q_1 -1\right) -q_2 \left(q_1 + R\right) \right] \nonumber \\
\wit{\gamma} &=& -\frac{P}{q_2} q_1 \left[ q_1 \left( q_1 -\frac{1}{2} \right)
-q_2 \left( q_1 +R \right) \right] \nonumber \\
\wit{\delta} &=& \frac{P}{q_2} q_1 \left[ q_1\left( q_1 -1\right) -q_2\left(
q_1 -1 -R \right) \right] \nonumber \\
\wit{\nu} &=& -\frac{P}{2 q_2} \left( q_1 -1\right)^2 \left( q_2+1-q_1\right)
\label{SymmHII} \\
\wit{\sigma} &=& -\frac{P}{q_2} \left( q_1 -1\right) \left[ \left( q_1-
\frac{1}{2}\right)\left( q_1 -1\right) -q_2\left( q_1 -1-R\right)\right] 
\nonumber \\
\wit{D} &=& -\frac{P}{q_2} \left[ q_1\left(q_1 -\frac{1}{2}\right)\left(q_1 -1
\right) -q_2 q_1 \left( q_1 -1\right) +\frac{1}{2} q_2 \Delta \right]
\nonumber 
\EEA
We remark that the parameters are defined such that $P$ always emerges
as a prefactor. This implies of course that $P\neq 0$ and we shall consider
later what happens when $P=0$. In order to get a stochastic system, the
seven rates of eq.~(\ref{SymmHII}) must all be positive. We postpone the
analysis of these inequalities to the next section and discuss first the
rates for case III.

\subsubsection{Case III}
The matrix to be transformed is
\BEQ \label{HIII}
H^{III} = \left( \begin{array}{cccc} 
0 & h_{12} & -h_{12}-h_{22}-h_{33}-h_{23}-h_{32} &
-h_{24}-h_{34}-h_{44} \\
 0 & h_{22} & h_{23} & h_{24} \\
0 & h_{32} & h_{33} & h_{34} \\
0 & 0 & 0 & h_{44} \end{array} \right)
\EEQ
and the transformation matrix $B$ is given by (\ref{bIII}).
Again, eqs.~(\ref{ProCon}) are automatically satisfied
with $ \wit{C_1}=\wit{C_4}=0$ and  $\wit{C_2}=-\wit{C_3}=
h_{22}+h_{32}+h_{12}$. We begin by looking at the expressions
$D^{III}_{1,2}$, see (\ref{GunI}), which for integrability 
through the equation of motions for the particle number correlators
whould have to vanish and which become in this case
\BEA
D^{III}_1 &=& 
\left( h_{23}-h_{32}-h_{44}-h_{12}-h_{24}\right)/( S q_2)
\nonumber \\ 
D^{III}_2 &=&  \left(h_{22}+2h_{32}+h_{33}+h_{12}-h_{44}-h_{34}\right)/(S q_2)
\label{GunIII}
\EEA

Now, we observe that with the parametrization of table~\ref{tab4} and 
eq.~(\ref{bq}), the expression for rates in the {\it parity-symmetric} case
are exactly the same as in case II and already given in eqs.~(\ref{SymmHII}). 
Therefore for the analysis of the positivity, it is enough to consider
one of the cases only. As a bonus, we shall afterwards have the freedom
to work with the integrable matrices $H$ of either case II or case III,
depending of the applications of interest. 

We stress, however, that the coincidence of the rates for the cases II and III
only holds when $P\neq 0$. When $P=0$, both cases must be discussed 
separately.

\subsection{Analysis of the positivity conditions}

We now adress the problem of how to characterize those stochastic systems
which can be transformed via (\ref{BTrans}) to the integrable Hamiltonian
$H^{II}$ or $H^{III}$. We restrict attention to the case when the elements
of $H$ as well as the transformation parameters $q_{1,2}$ are 
{\it real}.\footnote{We had already seen in the discussion of case III above 
that this condition might be the one needed for applications anyway. In 
addition our results from section 3, where all the elements of the free
fermion Hamiltonian turned out to be real at the end makes it plausible that
this should be the most important case.} We thus have to satisfy
simultaneously the seven inequalities
\BEQ \label{RU}
\wit{\alpha} \geq 0 \;\; , \;\;
\wit{\beta} \geq 0 \;\; , \;\;
\wit{\gamma} \geq 0 \;\; , \;\; 
\wit{\delta} \geq 0 \;\; , \;\; 
\wit{\nu} \geq 0 \;\; , \;\; 
\wit{\sigma} \geq 0 \;\; , \;\;
\wit{D} \geq 0
\EEQ
and the rates are given explicitly in eqs.~(\ref{SymmHII}). 
Since the parameter $\Delta$ appears only in $\wit{D}$, we can always satisfy
the corresponding inequality by suitably choosing $\Delta$ and it remains
to consider the other six inequalities. 

While at first sight this problem looks rather difficult, it can be solved,
following the ideas presented in section 3 after having made a few 
observations. From eqs.~(\ref{SymmHII}), we first note that $P/q_2$ only
appears as a scale factor and thus only its {\it sign} is going to be
relevant for the positivity discussion. Second, after rescaling 
the rates depend only on the
transformation parameters $q_1$ and $q_2$, which characterize the
similarity transformation $B$, and the parameter $R$, which characterizes
the Hamiltonian $H$. We think of the rates as functions of these three
parameters, e.g. $\wit{\alpha} = \wit{\alpha}(q_1,q_2,R)$. We now fix $R$
and look at the values of the function $\wit{\alpha}$ in $q_1 - q_2$ space. 
There will be regions $G_+^{(\alpha)}$ for which $\wit{\alpha}$ will 
be positive and regions for which $\wit{\alpha}$ will be negative. 
The boundaries of the regions $G_+^{(\alpha)}$ are characterized by the
curves $\wit{\alpha}(q_1, q_2;R) =0$. Mapping out the boundary curves for
a fixed value of $R$ will rapidly furnish a geometrical idea of the shape
of the region $G_+^{(\alpha)}$. This process is repeated for all six rates
and the finally the desired region in $q_1 - q_2$ space where all rates
are positive is the intersection
\BEQ
G_+ = G_+^{(\alpha)} \cap G_+^{(\beta)} \cap G_+^{(\gamma)}
\cap G_+^{(\delta)} \cap G_+^{(\nu)} \cap G_+^{(\sigma)}
\EEQ
The graphical inspection of these regions will then suggest how to satisfy
in a simple way the inequalities (\ref{RU}). Technically, the proof is
much shortened by concentrating on the case with a single steady state
which makes the inequalities (\ref{bUngl}) available. Then the following
distinction is sensible: 
$i)\, q_1 =0$, $ii)\, q_1 =1$ and $iii)\, 0 < q_1 < 1$.
(If eq.~(\ref{bUngl}) does not apply, it can be shown that no further
stochastic systems exist.) 

\subsubsection{$q_1 =0$}
In this case the parity-symmetric rates become
\BEA
\wit{\alpha} &=& \wit{\gamma} = \wit{\delta} = 0 \;\; , \;\; 
\wit{\beta} = P R \;\; , \;\; 
\wit{\nu} = -\frac{P}{2 q_2} \left( q_2 +1 \right) \nonumber \\
\wit{\sigma} &=& \frac{P}{2 q_2} \left( 2 q_2 \left( 1+R\right) +1\right)
\;\; , \;\; 
\wit{D} = -\frac{1}{2} P \Delta
\EEA
It is easy to see that for these rates, the positivity conditions
can be satisfied. However, the integrability of this model is
physically trivial, since only particle creation reactions are allowed. 

\subsubsection{$q_1 =1$}
In this case the parity-symmetric rates become
\BEA
\wit{\beta} &=& \wit{\nu} = \wit{\sigma} = 0 \;\; , \;\; 
\wit{\alpha} = \frac{P}{2 q_2} \left( 1 - q_2\right) \;\; , \;\; 
\wit{\gamma} = \frac{P}{2 q_2} \left( 2 q_2 \left(1+R\right) -1 \right)
\nonumber \\
\wit{\delta} &=& P R \;\; , \;\; 
\wit{D} = -\frac{1}{2} P \Delta
\EEA
The discussion of positivity is completely analogous to the case above. 
Again, since only particle annihilation reactions are allowed, the 
integrability of the system is trivial and we just recover the model reviewed
in section 2. 

\subsubsection{$0 < q_1 < 1$}
We now turn to the discussion of the remaining cases. Again, the different
possibilities for the signs of $P$ and $q_2$ must be treated separately. \\
{\bf 1. $P=-1$ and $q_2 > 0$.} From the positivity of the rates, provided that
neither $q_1 =0$ nor $q_1 =1$, we get the following inequalities
\BEA
\wit{\alpha} \geq 0 &\Longrightarrow& q_1 -q_2 \leq 0 \nonumber \\
\wit{\beta} \geq 0 &\Longrightarrow& q_1(q_1-1)-q_2(q_1+R) \geq 0 \nonumber \\
\wit{\gamma} \geq 0 &\Longrightarrow& q_1(q_1-\frac{1}{2})-q_2(q_1+R) \geq 0
\nonumber \\
\wit{\delta} \geq 0 &\Longrightarrow& q_1(q_1-1)-q_2(q_1-1-R) \leq 0 
\nonumber \\
\wit{\nu} \geq 0 &\Longrightarrow& q_1 -q_2 -1 \leq 0  \\
\wit{\sigma} \geq 0 &\Longrightarrow& (q_1-\frac{1}{2})(q_1-1)-q_2(q_1-1-R)
\leq 0 \nonumber \\
\wit{D} \geq 0 &\Longrightarrow& q_1(q_1-\frac{1}{2})(q_1-1)-q_2\left(
q_1(q_1-1) -\Delta/2\right) \geq 0 \nonumber
\EEA
{}From these, we observe that 
\BEQ \label{RBed}
R \leq -\frac{1}{2}
\EEQ
To see this, we note that from $\wit{\beta}\geq 0$ and $\wit{\delta}\geq 0$
we have the inequalities $q_1(q_1-1) \geq q_2(q_1+R)$ and $q_1(q_1-1) \leq
q_2(q_1-1-R)$, respectively. Combining these, we find that
$q_2(q_1+R) \leq q_2(q_1-1-R)$ and since $q_2>0$, the assertion follows. Next,
we have the following 

\noindent {\bf Lemma 1:} {\it If $P=-1$ and $q_2 >0$ and $0<q_1<1$, the
conditions
\BEQ
\wit{\beta} \geq 0 \;\; , \;\; \wit{\nu} \geq 0 \;\; , \;\; 
\wit{\sigma} \geq 0 \;\; , \;\; \wit{D} \geq 0
\EEQ
are sufficient for positivity.} \\
\noindent {\bf Proof:} Since $0<q_1<1$, we have 
$q_1(q_1-1) \leq (q_1-1/2)(q_1-1)$. Now, $\wit{\sigma}\geq 0$ and 
$\wit{\delta}\geq 0$ are equivalent to the inequalities
$(q_1-1/2)(q_1-1)\leq q_2(q_1-1-R)$ and $(q_1-1)q_1 \leq q_2(q_1-1-R)$,
respectively. It is now obvious that $\wit{\sigma}\geq 0$ implies
$\wit{\delta}\geq 0$. On the other hand, we also have the inequality
$q_1(q_1-1/2) \geq q_1(q_1-1)$. Now, $\wit{\beta}\geq 0$ and 
$\wit{\gamma}\geq 0$ are equivalent to the inequalities
$q_1(q_1-1)\geq q_2(q_1+R)$ and $q_1(q_1-1/2) \geq q_2(q_1+R)$, respectively.
It follows that $\wit{\beta}\geq 0$ implies that $\wit{\gamma}\geq 0$. 
Finally, $\wit{\alpha}\geq 0$ and $\wit{\nu}\geq 0$ are equilavent to
the inequalities $q_1 \leq q_2$ and $q_1 \leq q_2+1$, respectively, which
completes the proof. \hfill q.e.d. 

It remains to characterize $G_+$ from the remaining rates. The result is
\BEQ
G_{+,1} = \left\{ \left( q_1, q_2, R\right) \left| 
1+R \leq q_1 \leq -R \; ; \;
q_2 \geq \max\left( \frac{q_1(q_1-1)}{q_1+R}, 
\frac{(q_1-1/2)(q_1-1)}{q_1-1-R}, q_1-1\right) \; ; \; 
R \leq -\frac{1}{2} \right. \right\}
\EEQ
{\bf 2. $P=-1$ and $q_2 < 0$.} Now we have the following set of inequalities
\BEA
\wit{\alpha} \geq 0 &\Longrightarrow& q_1 -q_2 \geq 0 \nonumber \\
\wit{\beta} \geq 0 &\Longrightarrow& q_1(q_1-1)-q_2(q_1+R) \leq 0 \nonumber \\
\wit{\gamma} \geq 0 &\Longrightarrow& q_1(q_1-\frac{1}{2})-q_2(q_1+R) \leq 0
\nonumber \\
\wit{\delta} \geq 0 &\Longrightarrow& q_1(q_1-1)-q_2(q_1-1-R) \geq 0 
\nonumber \\
\wit{\nu} \geq 0 &\Longrightarrow& q_1 -q_2 -1 \geq 0  \\
\wit{\sigma} \geq 0 &\Longrightarrow& (q_1-\frac{1}{2})(q_1-1)-q_2(q_1-1-R)
\geq 0 \nonumber \\
\wit{D} \geq 0 &\Longrightarrow& q_1(q_1-\frac{1}{2})(q_1-1)-q_2\left(
q_1(q_1-1) -\Delta/2\right) \leq 0 \nonumber
\EEA
The treatment is completely analogous to the first case. Again, from
$\wit{\beta}$ and $\wit{\delta}$ we find that $R\leq -\frac{1}{2}$. Then we
prove the \\ 
\noindent {\bf Lemma 2:} {\it If $P=-1$ and $q_2 <0$ and $0<q_1<1$, the
conditions
\BEQ
\wit{\alpha} \geq 0 \;\; , \;\; \wit{\gamma} \geq 0 \;\; , \;\; 
\wit{\delta} \geq 0 \;\; , \;\; \wit{D} \geq 0
\EEQ
are sufficient for positivity.} \\
Finally, the region $G_+$ where all rates are positive is found to be
\BEQ
G_{+,2} = \left\{ \left( q_1, q_2, R\right) \left| 
1+R \leq q_1 \leq -R \; ; \;
q_2 \leq \min\left( \frac{q_1(q_1-1/2)}{q_1+R}, 
\frac{q_1(q_1-1)}{q_1-1-R}, q_1\right) \; ; \; 
R \leq -\frac{1}{2} \right. \right\}
\EEQ
{\bf 3. $P=+1$ and $q_2 >0$.} In fact this case does not give anything new.
By assumption we have $0 < q_1 < 1$. From $\wit{\nu}\geq 0$ it follows that
$0< q_2 \leq q_1 -1 < 0$, which is impossible, because $q_2=0$ would
correspond to a singular transformation matrix $B$. \\
{\bf 4. $P=+1$ and $q_2 <0$.} Also here no new system arises. From
$\wit{\alpha}\geq 0$ it follows that $0 < q_1 \leq q_2 <0$ which is 
impossible. The only way to find stochastic systems for $P$ positive is
to consider the cases $q_1=0,1$ which have been dealt with before. 

In conclusion, the only non-trivial parity-symmetric 
stochastic system which can be reduced through (\ref{BTrans}) 
to a real integrable quantum chain with the spectrum of the XXZ Hamiltonian
is the class of models with rates given in eq.~(\ref{SymmHII}) and the
domain of positivity
\BEQ \label{Gplus}
G_+ = G_{+,1} \cup G_{+,2}
\EEQ
We remark that in the $q_1 - q_2$ plane, the regions $G_{+,1}$ and $G_{+,2}$
are mapped onto each other under the transformation $(q_1, q_2) \rar
(1-q_1, -q_2)$, or geometrically speaking, by point symmetry through the
point $q_1 =\frac{1}{2}, q_2=0$. 

The positivity condition in (\ref{Gplus}) is supplemented, for $P=-1$, 
but independently of the sign of $q_2$, by the inequality
\BEQ \label{Dplus}
\Delta \geq 2 q_1 (q_1 -1) \frac{q_2 -q_1 +1/2}{q_2}
\EEQ
Eqs.~(\ref{Gplus},\ref{Dplus}) specify completely, for the non-trivial case
$P=-1$, the restrictions which have to be put on the values of the free
parameters $q_1, q_2, R, \Delta$. 

\subsubsection{The case $P=0$}

We complete the analysis by stating the results in the case when $P=0$. For
the proofs, we refer the reader to appendix~D. \\
{\bf Case II:} Here positivity requires that $\Omega=Q=0$ and consequently,
all reaction rates vanish. The only processes which survive are diffusion to
the right and to the left. \\
{\bf Case III:} If there is just a single steady state, we find the stochastic
system
\BEQ 
\label{HIIIa}
\wit{H}_{j,j+1} = \frac{1}{2}\left( \begin{array}{cccc} 
g_{1} & -b_{11}h_{44} & -b_{11}h_{44}& 0 \\
(b_{11}-1)h_{44}& g_{2} &
(h_{23}+h_{32}) &-b_{11}h_{44} \\ 
(b_{11}-1)h_{44} & (h_{23}+h_{32}) & g_{3} 
&-b_{11}h_{44} \\ 
0 & (b_{11}-1)h_{44} &(b_{11}-1)h_{44}& g_{4}
\end{array} \right) 
\EEQ
where the $g_{\ell}$ are determined such that the sum of each of the column
vector of $\wit{H}_{j,j+1}$ vanishes. From a physical point of view, however,
the integrability of the model is obvious, since it describes particles
moving diffusively and, since $\wit{\gamma}=\wit{\delta}=r_-$ and
$\wit{\beta}=\wit{\sigma}=r_+$, decay radioactively with rate $r_-$ and
are created independently of their neighbors with rate $r_+$. We remark
that in this case, it is easily checked that the integrability conditions
through the equations of motion, viz. $D_{1}^{III}=D_{2}^{III}=0$, are
satisfied.  

\subsection{Some applications} 

We are now in a position to give a more physical characterization of the
new integrable stochastic systems obtained. We shall discuss here only
the class of models of which the rates are explicitly given in (\ref{SymmHII})
with the domain $G_+$ (\ref{Gplus}), together with (\ref{Dplus}),
gives the range of variables where
the stochastic rates are positive, and shall leave out completely all the
other systems, since their behaviour is easily obtained without reference
to the XXZ chain. 

Let us begin with a discussion
of the meaning of the condition $R\leq -\frac{1}{2}$. Recalling the definition
of $R$ from table~\ref{tab4} and the fact that $P$ must be negative, we get
the inequality $Q \geq -P = Q-h_{44}$, from which it follows that both
$h_{44}\geq 0$ and $Q>h_{44}$ are necessarily satisfied. These are also
necessary conditions for $H$ to be a stochastic matrix by itself. In order
to understand eq.~(\ref{RBed}) further, 
we consider the example of $H^{III}$ in
the parity-symmetric case and we also assume that $H^{III}$ is itself
stochastic. Then the matrix elements of $H$ can be written in terms of
chemical reaction rates, using the notation of table~\ref{tab4} as
\BEQ
h_{22}=h_{33}=D+\delta \;\; , \;\; h_{23}=h_{32}=D \;\; , \;\; 
h_{44} =2(\alpha+\gamma)
\EEQ
We then find that $Q=2\delta$ and $P=2(\alpha+\gamma+\delta)$ and finally
\BEQ
R = \frac{Q}{2P} = -\frac{1}{2} \frac{\delta}{\delta-\alpha-\gamma} 
\leq -\frac{1}{2}
\EEQ
because $Q>h_{44}$ implies $\delta>\alpha+\gamma$. So we see that in this
particular example eq.~(\ref{RBed}) simply means that also $H$ is 
stochastic. 

Next, we ask for a characterization of integrability through similarity
transformation to the XXZ chain solely in terms of the chemical reaction
rates. In eqs.~(\ref{SymmHII}), the six rates (without the diffusion constant
$\wit{D}$) are expressed in terms of the four parameters $q_1, q_2, P$ and $R$.
So we should expect to find two relations between the rates. Indeed, 
tedious calculations lead to the following relations
\begin{eqnarray}
&& \left\{-4\,\wit{\nu}^{3}+\left[ 8\,\wit{\delta}+4\wit{\beta}-12
\wit{\sigma}\right]\wit{\nu}^{2}
+\left[\left (8\wit{\sigma}-2\wit{\beta}\right)\wit{\delta}
-\left(\wit{\beta}-2\wit{\sigma}\right) \left(\wit{\beta}-6
\wit{\sigma}\right)\right]\wit{\nu}-\wit{\sigma}
\left(\wit{\beta}-2\wit{\sigma}\right)^{2}\right\}
\wit{\alpha} \nonumber \\
&+& 4\wit{\alpha}^{2}\wit{\nu}^{2}-\wit{\delta}^{2}\wit{\nu}^{2}+
\left(2\wit{\delta}^{3}+\left(\wit{\beta}-2\wit{\sigma}\right )
\wit{\delta}^{2}\right)\wit{\nu}
+\wit{\sigma}\left(-
\wit{\sigma}+\wit{\beta}\right)\wit{\delta}^{2} = 0 \label{XXZCeins}
\end{eqnarray}
and
\begin{eqnarray}
& & \left\{
-4\wit{\nu}^{3}+\left[
4\wit{\gamma}+4\wit{\delta}-12\wit{\sigma}\right]\wit{\nu}^{2}
+2\left(\wit{\gamma}-2\wit{\sigma}\right)\wit{\sigma}^{2}+
\left[\wit{\delta}^{2}+4\wit{\delta}\wit{\sigma}
+6\left(\wit{\gamma}-2\wit{\sigma}\right)\wit{\sigma}
\right]\wit{\nu}\right\}\wit{\alpha} \nonumber \\
&+& \wit{\gamma}^{2}\wit{\sigma}^{2}-\wit{\delta}^{2}\wit{\sigma}^{2}+
\left(4\wit{\nu}^{2}+4\wit{\nu}\wit{\sigma}+\wit{\sigma}^{2}
\right)\wit{\alpha}^{2}+\left(\wit{\gamma}^{2}-\wit{\delta}^{2}\right )
\wit{\nu}^{2}
+\left(2\wit{\gamma}^{2}\wit{\sigma}+\wit{\delta}^{3}+
\left(\wit{\gamma}-2\wit{\sigma}\right)\wit{\delta}^{2}\right)\wit{\nu} = 0 
\nonumber \\
& & ~ \label{XXZCzwei}
\end{eqnarray}
While in case I, the integrability condition could be expressed as a linear
relation (\ref{CIPs}) 
between the rates, we have here two quartic relations. We note that,
as in case I, the value of the diffusion constant $\wit{D}$ does not enter
into the integrability conditions. For a given reaction-diffusion system,
eqs.~(\ref{XXZCeins},\ref{XXZCzwei}) can be used to check whether or not
the model is integrable through the transformation into an XXZ chain. 

Since the type of integrability we consider here mainly involves the
spectrum of the Hamiltonian $H$, it is particularly easy to extract the
relaxation time towards the steady state and the mean particle density
in the steady state. The relaxation time $\tau$ is the inverse of the
mass gap $Q$ and thus given by
\BEQ
\frac{1}{\tau} = Q = 2 P R
\EEQ
In particular, since both $P$ and $R$ are negative, it follows that the
relaxation time $\tau$ is always finite. A tedious calculation gives $\tau$
in terms of the reaction rates

\begin{eqnarray}
\tau^{-1} &=& -2{\frac {
\left (-2\wit{\sigma}-6
\wit{\nu}+\wit{\beta}\right )
\wit{\alpha}-2\wit{\nu}^{2}+\left(\wit{\beta}-4\wit{\sigma}\right)
\wit{\nu}-2\wit{\delta}\wit{\sigma}+\wit{\beta}
\wit{\sigma}-2\wit{\sigma}^{2}
-\wit{\delta}^{2}}{
\left(-2\wit{\sigma}-4\wit{\nu}+\wit{\beta}
\right)\wit{\alpha}-\wit{\delta}\wit{\nu}-\wit{\delta}
\left(\wit{\delta}+\wit{\sigma}\right)}}
\nonumber \\
&\times& \frac{
\left(\left(-8\wit{\nu}^{2}+\left(-12\wit{\sigma}+2\wit{\beta}+2\wit{\delta}
\right)\wit{\nu}+2\wit{\sigma}
\left(-2\wit{\sigma}+\wit{\beta}\right)\right)\wit{\alpha}
-2\wit{\delta}\left(3\wit{\delta}+\wit{\beta}\right)\wit{\nu}-2
\wit{\delta}\wit{\sigma}\left(\wit{\delta}+\wit{\beta}\right)\right)}
{ \left(2\wit{\alpha}\wit{\nu}+2\wit{\nu}^{2}
+\left(-\wit{\beta}+4\wit{\sigma}-\wit{\delta}\right)
\wit{\nu}-\wit{\sigma}\left(-2\wit{\sigma}-\wit{\delta}+\wit{\beta}\right)
\right)}
\nonumber \\ & & ~
\end{eqnarray}
We remark that in this form there is no longer an explicit dependence on
neither the coagulation rate $\wit{\gamma}$ nor the diffusion constant
$\wit{D}$. 

Finally, the mean particle density in the steady state is found from
$\wit{\rho}_{\infty} = q_1$. In terms of the rates, it reads
\BEQ
\wit{\rho}_{\infty} = {\frac {\left(-2\wit{\sigma}-4\wit{\nu}+
\wit{\beta}\right)
\wit{\alpha}-\wit{\delta}\wit{\nu}-
\wit{\delta}\left(\wit{\delta}+\wit{\sigma}\right)}{\left(-2\wit{\sigma}
-6\wit{\nu}+\wit{\beta}\right)\wit{\alpha}-2\wit{\nu}^{2}+
\left(\wit{\beta}-4\wit{\sigma}\right)
\wit{\nu}-\wit{\delta}^{2}-2\wit{\delta}\wit{\sigma}+\wit{\beta}
\wit{\sigma}-2\wit{\sigma}^{2}}}
\EEQ
which is also not explicitly dependent on $\wit{\gamma}$ or $\wit{D}$. 

\subsection{Summary}

In this section, we have found the parity-symmetric stochastic processes
which are integrable through a similarity transformation of the type
(\ref{BTrans}) onto a real XXZ quantum chain. Up to rescaling, we find a 
non-trivial class of integrable systems, which depends on three parameters. 
These systems approach their steady steady state exponentially fast. Their
rates are explicitly given in eqs.~(\ref{SymmHII}) and the constraints on the
free parameters $q_{1,2}, P$ and $R$ are given in (\ref{Gplus}). 
Since, working within the setting of case III, the integrable model
is for the parity-symmetric case identical to the known system
eq.~(\ref{HMat}), we can read off from table~\ref{tab2} the time
dependence of one- and two-point correlators in the $t\rar\infty$ limit.

\section{Calculation of the density for uncorrelated
 initial states with translational invariance}

We will now illustrate the use of the similarity
transformations developed in the previous sections.
We will calculate the mean density $\tilde{\rho}(t)$ for the various
cases presented in table \ref{tab3} for an uncorrelated initial 
state with density $\tilde{\rho}_0$. Since the mean density $\wit{\rho}$
is translation invariant, the result will {\em not} depend on the
diffusion bias $D_L - D_R$. Since the bias is proportional to $D_1 -D_2$,
it is enough to set $D_1 = D_2 =1$, which fixes the time scale as well,
see eq.~(\ref{3.12}). This calculation is greatly simplified using the
free fermion formulation given by the Hamiltonian eq.~(\ref{Hsigma}). 

The initial state is of the product form
\beq
\ket{\tilde{\rho}_0}\;=\;\bigotimes_{j=1}^{L}\left(
\begin{array}{c}1-\tilde{\rho}_0\\ \tilde{\rho}_0
\end{array}
\right)_{j}\enspace.
\label{5.1}
\eeq
We wish to calculate the matrix element 
\BEQ
\bra{s}\,\tilde{n}_{x}\exp(-\tilde{H}t)\,\ket{\tilde{\rho}_0}\,=\,
\,\bra{s}\,B\,n_{x}\exp(-Ht)\,B^{-1}\,\ket{\tilde{\rho}_0}
\EEQ 
averaged over all sites. In section 3, we had 
already calculated $n_{x}$ for the four systems of table \ref{tab3}
and related the parameters of the 
stochastic Hamiltonians to the parameters of the
free fermion Hamiltonian (\ref{Hsigma}).
Since the transformation (\ref{BTrans}) is local and $\ket{\wit{\rho}_0}$ 
is a product state, the
state $B^{-1}\,\ket{\tilde{\rho}_0}$ is also a product state.
As pointed out in the previous section, the normalization of a
state is arbitrary and therefore we can always represent 
$\ket{\rho_0}\,=\,B^{-1}\,\ket{\tilde{\rho}_0}$ in the form
(\ref{5.1}) with $\tilde{\rho}_0$ substituted by $\rho_0$.
Also, $\bra{s}\,B\,=\,\bra{\omega}$,
where $\bra{\omega}\,=\,\bigotimes_{j=1}^{L}\,( \omega \, ,\, 1-\omega)$.
The ratios $\frac{\rho_0}{1-\rho_0}$ and
$\frac{1-\omega}{\omega}$ can be found for each separate case
of section 3
by applying $B^{-1}$ to $\ket{\tilde{\rho}_0}$ and $B$ to $\bra{s}$.

The operator $n_{x}$ was
written for models I-III in terms of the Pauli spin operators $\spy{x}$ and
$\spz{x}$ in section 3 (for model IV, trivially $n_{x}=\tilde{n}_{x}$).
Due to the form of $\bra{\omega}$ we have that 
\BEQ
\bra{\omega}\,\spy{x}\;=\;\mbox{i}\,
\frac{1-\omega}{\omega}\,\bra{\omega}\,(1-\tilde{n}_{x})\;-\;
\mbox{i}\,\frac{\omega}{1-\omega}\,\bra{\omega}\,\tilde{n}_{x}
\label{o5.3} 
\EEQ	
(if $\omega\neq 0,1$). Furthermore, 
$\spz{x}\,=\,1\,-\,2\tilde{n}_{x}$. Therefore the calculation of
the density is reduced to the calculation of the matrix element
$\sum_{x}\,\bra{\omega}\,\tilde{n}_{x}\,\exp(-Ht)\,\ket{\rho_0}$.
In order to do this, we write the Hamiltonian (\ref{Hsigma})
in terms of spinless
fermion operators by means of a Jordan-Wigner transformation, see
appendix~A.
Then we express the fermion creation and annihilation operators at
a given site $x$ in terms of their discrete Fourier transforms to
momentum space. As is well known from the properties of
the Jordan-Wigner transformation, we have antiperiodic (periodic)
boundary conditions in the sector with an even (odd) number of particles,
e.g. \cite{Lieb61,Lush86,Gryn94}.
In terms of the spin language, each new fermion corresponds to a
spin flip, the ferromagnetic state with all spins up being the fermion
vacuum. For simplicity, we shall work in the even sector, since this
simplification does not affect the results in the thermodynamic limit.
The Hamiltonian (\ref{Hsigma}) becomes
\begin{equation}
H\;=\;(1-h)L\,+\,\sum_{p>0}\,(\,2h\,-\,2\cos
\left(\frac{2\pi p}{L}\right)\,)\,
(\crea{b}{p}\anni{b}{p}\,+\,\crea{b}{-p}\anni{b}{-p})
\,-\, 2\sin\left(\frac{2\pi p}{L}\right)\,
\,(\eta_{2}\crea{b}{-p}\crea{b}{p}\,+\,\eta_{1}\anni{b}{p}\anni{b}{-p})
\label{5.2}
\end{equation}
where $p$ runs over the half-integers $1/2,\ldots,L/2-1/2$
because of the antiperiodic boundary conditions. The number operator
is written in terms of fermion operators as 
$\sum_{x}\,\tilde{n}_{x}=\sum_{p}\crea{b}{p}\anni{b}{p}$.

Finally it can be shown \cite{Lush86,Schu95a}
that the projection of the states $\bra{\omega}$ and
$\ket{\rho_0}$ over the even sector can be written as 
\beq
\bra{\omega}^{even}\;=\;\bra{0}\,\prod_{p>0}\,(\,1\;+\;f_{\omega}^{2}\cot\,
\left(\,\frac{\pi p}{L}\,\right)\,\anni{b}{p}\,\anni{b}{-p}\,)
\label{5.3}
\eeq
with $f_{\omega}\,=\,\frac{1-\omega}{\omega}$ and
\beq
\ket{\rho_0}^{even}\:=\:{\cal N}^{-1}\prod_{p>0}\,
(1\:+\:\mu^{2}\,\cot\,\left(\,\frac{\pi p}{L}
\,\right)\,\crea{b}{-p}\,\crea{b}{p}\,)\,
\ket{0}
\label{5.4}
\eeq
where \(\mu\:=\:\rho_0/(1-\rho_0)\) and the normalization ${\cal N}=
\langle\,\omega\,\ket{\rho_0}^{even}$.

So we have reduced our problem to the calculation of 
$\bra{\omega}^{even}\,\crea{b}{p}\anni{b}{p}\exp(-Ht)\,\ket{\rho_0}^{even}$.
Notice that both states $\bra{\omega}^{even}$ and 
$\ket{\rho_0}^{even}$ belong
to a translation invariant subspace of the original Hilbert space with
zero momentum. In this subspace, the operators $\anni{b}{p}\anni{b}{-p}$,
$\crea{b}{-p}\crea{b}{p}$ and $1-\crea{b}{p}\anni{b}{p}
-\crea{b}{-p}\anni{b}{-p}$ are isomorphic to a 
Pauli spin algebra \cite{Schu95a},
 corresponding respectively to 
$\spup{p}$, $\spdo{p}$ and $\spz{p}$. One can therefore
develop the term $\exp(-Ht)$ using the known rules for the
exponential of an expression involving Pauli spin matrices.
First notice that
\beq
\sum_{p}\bra{\omega}^{even}\,\crea{b}{p}\,\anni{b}{p}\,
\exp(-Ht)\,\ket{\rho_0}^{even}\;=\;
2\sum_{p>0}\,f_{\omega}^{2}\cot\,\left(\,\frac{\pi p}{L}\right)\,
\bra{\omega}^{even}\,\anni{b}{p}\,
\anni{b}{-p}\,\exp(-Ht)\,\ket{\rho_0}^{even}
\label{5.5}
\eeq
since 
\BEQ
\bra{\omega}^{even}\,\left(\,\crea{b}{p}\;+\;f_{\omega}^{2}\cot\,\left(\,
\frac{\pi p}{L}\,\right)\,\anni{b}{-p}\,\right)\,=\,0
\EEQ
which follows directly from eq.~(\ref{5.3}). 

Since $H=\sum_{p>0}
H_{p}$, $\bra{\omega}^{even}\,H=0$
and $\bra{\omega}^{even}$ is a product state also in momentum space, we
have $\bra{\omega}^{even}\,H_{p}\,=\,0$ separately. Therefore we
can write
\begin{eqnarray}
\sum_{x}\,\bra{\omega}^{even}\,\tilde{n}_{x}\,\exp(-Ht)\,\ket{\rho_0}^{even}
&=&\,2\sum_{p>0}\,f_{\omega}^{2}\cot\,\left(\,\frac{\pi p}{L}\right)\,
\bra{\omega}^{even}\,\anni{b}{p}\,
\anni{b}{-p}\,\exp(-Ht)\,\ket{\rho_0}^{even}\nonumber\\
&=&
2\sum_{p>0}\,f_{\omega}^{2}\cot\,\left(\,\frac{\pi p}{L}\right)\,
\bra{\omega}^{even}\,\spup{p}\,\exp(-H_{p}t)\,\ket{\rho_0}^{even}
\label{5.6}
\end{eqnarray}
with $H_{p}\;=\;2(1-h\cos\left(\,\frac{\pi p}{L}\right))\;-\;
(\,\alpha_{p}\spx{p}\,+\,\beta_{p}\spy{p}\,+\,\gamma_{p}\spz{p}\,)$.
The constants 
$\alpha_{p}$, $\beta_{p}$ and $\gamma_{p}$ are given by
$\alpha_{p}=(\eta_{1}+\eta_{2})\sin\left(\,\frac{\pi p}{L}\right)$,
$\beta_{p}=\mbox{i}\,(\eta_{1}-\eta_{2})\sin\left(\,\frac{\pi p}{L}\right)$
and $\gamma_{p}=(2h-2\cos\left(\,\frac{\pi p}{L}\right))$. If we define
a vector {\boldmath $a$} by the components 
$(\alpha_{p},\beta_{p},\gamma_{p})$
we have
\beq
\exp(\mbox{{\boldmath $a$}}\cdot\mbox{{\boldmath $\sigma_{p}$}}\,t)\;=\;
\cosh(\mid\mbox{{\boldmath $a$}}\mid \,t)\,+\,
\frac{\mbox{{\boldmath $a$}}\cdot
\mbox{{\boldmath $\sigma_{p}$}}}
{\mid\mbox{{\boldmath $a$}}\mid}\,
\sinh(\mid\mbox{{\boldmath $a$}}\mid \,t)
\label{5.7}
\eeq
where $\mbox{{\boldmath $\sigma_{p}$}}\;=
\;(\spx{p},\spy{p},\spz{p})$ and 
$\mid\mbox{{\boldmath $a$}}\mid\;=\;\sqrt{\alpha_{p}^{2}+\beta_{p}^{2}
+\gamma_{p}^{2}}$. Substituting the definitions above and using the identity
$h^{2}+\eta_{1}\eta_{2}\;=\;1$ one finds that 
$\mid\mbox{{\boldmath $a$}}\mid\;=
\;2-2h\cos\left(\,\frac{2\pi p}{L}\right)$. 
Rexpressing $\spx{p}$ and $\spy{p}$
in terms of $\spup{p}$, $\spdo{p}$, one gets
\begin{eqnarray}
\exp(\mbox{{\boldmath $a$}}\cdot\mbox{{\boldmath $\sigma_{p}$}}\,t)&=&
\cosh\left(\,(\,2-2h\cos\left(\,\frac{2\pi p}{L}\right)\,)\,t\,\right)\;+\;
\sinh\left(\,(\,2-2h\cos
\left(\,\frac{2\pi p}{L}\right)\,)\,t\,\right)\times\nonumber\\
& &\mbox{}\times \left(\,\frac{h-\cos\left(\,\frac{2\pi p}{L}\right)}{
1-h\cos\left(\,\frac{2\pi p}{L}\right)}\,\spz{p}\;+\;
\frac{\eta_{2}\sin\left(\,\frac{2\pi p}{L}\right)}
{1-h\cos\left(\,\frac{2\pi p}{L}\right)}\,
\spdo{p}\,+\,\frac{\eta_{1}\sin
\left(\,\frac{2\pi p}{L}\right)}
{1-h\cos\left(\,\frac{2\pi p}{L}\right)}\,\spup{p}\,\right)\enspace.
\label{5.8}
\end{eqnarray}
Developing $\exp(-H_{p}t)$ in (\ref{5.6}) with the help of (\ref{5.8}),
using the algebra of the $\sigma$ matrices and the 
property of the $\ket{\rho_0}^{even}$ state
\beq
\spdo{p}\ket{\rho_0}^{even}\;=\;\left(\,\mu^{-2}\tan
\left(\,\frac{\pi p}{L}\right)
\,-\,\mu^{-4}\tan^{2}\left(\,\frac{\pi p}{L}\right)\,\spup{p}\,\right)
\ket{\rho_0}^{even}
\label{5.9}
\eeq
one gets after a little algebra the expression
\begin{eqnarray}
\sum_{x}\,\bra{\omega}^{even}\,\tilde{n}_{x}\,\exp(-Ht)\,\ket{\rho_0}^{even}
\!\!\!\!&=&\!\!\!\!
2\sum_{p>0}f_{\omega}^{2}\cot\,\left(\,\frac{\pi p}{L}\right)\,
\bra{\omega}^{even}\,\spup{p}\ket{\rho_0}^{even}\label{5.10}\\
& &\mbox{}\times\left[\,1+\,(\mu^{-2}\eta_{2}-(1+h))
\frac{1-e^{-4(1-h\cos\left(\,\frac{2\pi p}{L}\right))t}}
{1-h\cos\left(\,\frac{2\pi p}{L}\right)}
\sin^{2}\left(\,\frac{\pi p}{L}\right)\right]. \nonumber
\end{eqnarray}

Now it can be easily shown that $\bra{\omega}^{even}
\,\spup{p}\ket{\rho_0}^{even}
\,=\,\frac{\mu^{2}\,\cot\,(\,\frac{\pi p}{L}\,)}{
1+\mu^{2}f_{\omega}^{2}\cot^{2}\,(\,\frac{\pi p}{L}\,)}\,$. Substituting
this expression in (\ref{5.10}) we get
\begin{eqnarray}
\sum_{x}\,\bra{\omega}^{even}\,\tilde{n}_{x}\,\exp(-Ht)\,\ket{\rho_0}^{even}
\!\!\!\!&=&\!\!\!\!
2\sum_{p>0}\frac{\mu^{2}\,f_{\omega}^{2}\,
\cot^{2}\,(\,\frac{\pi p}{L}\,)}{
1+\mu^{2}f_{\omega}^{2}\cot^{2}\,(\,\frac{\pi p}{L}\,)}\label{5.11}\\
& &\mbox{}\times\left[\,1+\,(\mu^{-2}\eta_{2}-(1+h))
\frac{1-e^{-4(1-h\cos\left(\,\frac{2\pi p}{L}\right))t}}
{1-h\cos\left(\,\frac{2\pi p}{L}\right)}
\sin^{2}\left(\,\frac{\pi p}{L}\right)\right].\nonumber
\end{eqnarray}

As stated above the values of $f_{\omega}$ and of $\mu$ can be found from
the application of the transformation matrix to the original states. However,
one can extract $f_{\omega}^{2}$ from the condition 
$\bra{\omega}^{even}\,H_{p}=0$. We obtain 
$f_{\omega}^{2}\,=\,\frac{1-h}{\eta_{2}}$, this result being valid for
all the four cases of section 3. Separating the time-independent part
of (\ref{5.11}) and using this condition, one obtains after dividing by 
$L$ and taking the thermodynamic limit
\begin{eqnarray}
\rho(t)&=&\frac{1}{2\pi}\,\int_{-\pi}^{\pi}\,dp\,\frac{1}{1+\frac{1+h}{1-h}
\tan^{2}\left(\frac{p}{2}\right)}\nonumber\\
& &\mbox{}-\,\frac{1}{2\pi}\,\int_{-\pi}^{\pi}
\,dp\,\frac{f_{\omega}^{2}(\eta_{2}\,-\,(1+h)\mu^2)}
{\mu^{2}f_{\omega}^{2}\,+\,\tan^{2}\left(\frac{p}{2}\right)}\,
\frac{e^{-4(1-h\cos p)t}}{1-h\cos p}\,
\sin^{2}\left(\frac{p}{2}\right).
\label{5.12}
\end{eqnarray}
Performing the first integral, one obtains
\begin{equation}
\rho(t)\,=\,\frac{1}{1+\sqrt{\frac{1+h}{1-h}}
}\,-\,\frac{1}{2\pi}\,\int_{-\pi}^{\pi}
\,dp\,\frac{f_{\omega}^{2}(\eta_{2}\,-\,(1+h)\mu^2)}
{\mu^{2}f_{\omega}^{2}\,+\,\tan^{2}\left(\frac{p}{2}\right)}\,
\frac{e^{-4(1-h\cos p)t}}{1-h\cos p}\,
\sin^{2}\left(\frac{p}{2}\right)
\label{5.13}
\end{equation}
which is the central result of this section. 

When $h=\pm 1$, the generic 
long-time behaviour of the density is $\rho(t) \simeq
\rho_{\infty} + {\cal A} t^{-1/2}$, where $\cal A$ is some constant. On the
other hand, when $|h|<1$, the generic long time behaviour is
$\rho(t) \simeq \rho_{\infty} + {\cal A'} t^{-3/2} \exp(-t/\tau)$,
where the relaxation time is
\BEQ
\tau^{-1} = 4 ( 1-h)
\EEQ
and ${\cal A}'$ is another constant. 
To proceed further we need the values of $h$, $\eta_{2}$, 
$\mu^{2}$ and $f_{\omega}$ for each of the four cases of table \ref{tab3}. 
 We have
\begin{enumerate}
\item For model I,
we use eqns.~(\ref{3.58},\ref{3.59}) together with eq.~(\ref{o5.3}) to express
the action of the Pauli spin matrices in the state $\bra{\omega}$ in terms
of $\tilde{n}_{x}$. We obtain for the particle density
\beq
\tilde{\rho}(t)\;=\;\frac{2}{\sqrt{1+A^{2}}}\,\rho(t)\,-
\,\frac{1}{\sqrt{1+A^{2}}}
\label{5.14}
\eeq
where $\rho(t)$ is given by (\ref{5.13}) with $h$ and 
$\eta_{2}$ as a function of $A$ were defined in section 3 and 
\beq
f_{\omega}^2\;=\;-\,\left[\frac{\sqrt{1+A^{2}}+1}{A}\right]^2 \:\:,\;\;
\mu^{2}\;=\;-\,\frac{A^{2}}{(1+\sqrt{1+A^{2}})^{2}}\,\left[\frac{1+
\tilde{\rho_0}\sqrt{1+A^{2}}}{1-\tilde{\rho_0}\sqrt{1+A^{2}}}\right]^{2}
\label{5.15}
\end{equation}
where $\tilde{\rho}_0$ is the initial particle density. 
Using (\ref{5.13}) and (\ref{5.14}),
we get the steady state particle density 
\beq
\tilde{\rho}_{\infty}\;=\;\frac{1}{1+A^{2}}\enspace.
\label{5.16}
\eeq

Model I describes particles undergoing diffusion and interacting with each
other through death and creation processes, as defined in table~\ref{tab1}. 
Two extreme limits can immediately be understood intuitively. 
When the death processes are absent ($h\rar -1$ or $A\rar 0$), 
the steady state density is $\rho_{\infty}=1$ 
and the approach to the steady state is algebraic. 
On the other hand, if {\em only} the death process is 
present ($h\rar 0$ or $A\rar\infty$), $\rho_{\infty}=0$ and the approach
to the steady state is exponential. While in the first case, particles must
diffuse to create local configurations which are favorable for more reactions
taking place and which involves time scales of order $t^{1/2}$, in the second
case diffusion is absent and particles are just taken out of the system
with a certain rate. Thus a finite cluster of particles will disappear
within a finite time which is proportional to the cluster size and the
death rate. 

\item For model II, where we have symmetric diffusion with
coagulation and decoagulation, we repeat what we have done for the case above.
This model had already been studied 
in detail before \cite{Kreb95,Hinr95}. We obtain for the density 
\beq
\tilde{\rho}(t)\;=\;\frac{2}{\sqrt{1+A^{2}}}\,
\rho(t)\,-\,\frac{1}{\sqrt{1+A^{2}}}\,+\,1
\label{5.17}
\eeq
with $\rho(t)$ given as above but with $h$ and $\eta_{2}$ given by
(\ref{3.48}) and
\beq
f_{\omega}^{2}\;=\;-\,\left[\frac{A}{\sqrt{1+A^{2}}+1}\right]^2 \:\:,\;\;
\mu^{2}\;=\;-\left[ \frac{\sqrt{1+A^2}+1}{A}\right]^2 
\left[ \frac{ 1-(1-\wit{\rho}_0)\sqrt{1+A^2}}{1+(1-\wit{\rho}_0)\sqrt{1+A^2}}
\right]^2 .
\label{5.18}
\end{equation}
The steady state density is
\beq
\tilde{\rho}_{\infty}\;=\;1\;-\;\frac{1}{1+A^{2}}
\label{5.19}
\eeq
We point out that under a particle-hole transformation $\wit{\rho} \rar
1-\wit{\rho}$ 
(and consequently $\mu \rar 1/\mu$ and $f_{\omega}^2 \rar 1/f_{\omega}^2$)
we just recover the results found for model I.

\item For model III, where we have symmetric diffusion 
with decoagulation and death,  we repeat the steps of the 
previous cases. We obtain for the density
\beq
\tilde{\rho}(t)\;=\;-\frac{2A}{A^{2}-1}\,\rho(t)\,+\,\frac{1}{1+A}
\label{5.20}
\eeq
with $h$ and $\eta_{2}$ given by (\ref{3.45}) and 
\beq
f_{\omega}^2\;=\;-\,\left[\frac{A-1}{A+1}\right]^2 \:\:,\;\;
\mu^{2}\;=\;-\,\left(\frac{A\tilde{\rho}_0-(1-\tilde{\rho}_0)}
{A\tilde{\rho}_0+(1-\tilde{\rho}_0)}\right)^{2},
\label{5.21}
\end{equation}
In order to calculate the equilibrium density we have to distinguish
three different cases, i.e. $\mid A \mid\,<\,1$, 
$\mid A\mid\,=\,1$ and $\mid A \mid\,>\,1$. 
In the first case the steady state density is $\rho_{\infty}=1$, 
this result being
easy to understand since the rate of decoagulation ($\wit{\beta}=2$) 
is larger than the death rate ($\wit{\delta}=2A^{2}$). 
On the other hand, when $\mid A\mid\,>\,1$, $\rho_{\infty}=0$.
When $\mid A\mid\,=\,1$ the two processes equilibrate 
each other and  $\rho_{\infty}=1/2$. From (\ref{3.45}) we see that when 
$A\,=\,\pm 1$, $h=\pm 1$ and the decay towards the steady state 
is algebraic. For all other values of
$A$ the decay is exponential. 

This is in agreement with numerical studies in the biased voter model,
see \cite{Dick95} and references therein. The exponential decay of the
biased voter model can be intuitively understood using the same argument
as presented for model I. For the symmetric voter model, the algebraic
decay is expected since this model is dual \cite{Sigg77} 
to the Glauber-Ising model \cite{Glau63} at zero temperature. 

\item Finally, model IV just corresponds to model of 
symmetric diffusion with pair
creation and annihilation. 
We simply have $\tilde{\rho}(t)=\rho(t)$ with 
$\rho(t)$ given by (\ref{5.13}), with
$h=\tilde{\alpha}\,-\,\tilde{\nu}$ and $\eta_{1}\,=\,2\tilde{\alpha}$,
$\eta_{2}\,=\,2\tilde{\nu}$, $2\tilde{\alpha}$ and 
$2\tilde{\nu}=2-2\tilde{\alpha}$ being
respectively the rates of for pair annihilation and pair creation,
see table~\ref{tab1}. The constant 
$f_{\omega}\,=\,1$ and $\mu=\tilde{\rho}_0/(1-\tilde{\rho}_0)$.

If we take $h\,=\,\pm 1$, then we
have pure pair annihilation and pure pair creation, respectively, 
and we recover previously known results \cite{uralt,Lush86,Kreb95}. 
The decay to the steady state of the particle density 
is algebraic and of the form $t^{-1/2}$. 

For $0< h <1$, we obtain an unexpected result. For $\mu=0$, the
time-dependent density had been calculated before \cite{Gryn94} and
we reproduce their result. The long-time behaviour of the density is
generically $\rho(t)-\rho_{\infty} \sim t^{-1/2} \exp(-t/\tau)$. On the
other hand, if $0<\mu\leq 1$, we get from (\ref{5.13}) that
$\rho(t)-\rho_{\infty}\sim t^{-3/2} \exp(-t/\tau)$. In addition,
we have because of the particle-hole symmetry that the result is invariant
under the transformation $h\rar -h, \mu\rar 1/\mu$. So the long-time
behaviour of the system does depend in this case on the initial 
condition.\footnote{This comment also applies to models I, II and III if
$\mu=0$ and $0<h<1$ or $\mu=\infty$ and $-1<h<0$.} 

\end{enumerate}

In summary, the main result of this section is eq.~(\ref{5.13}), together
with eqs.~(\ref{5.14}, \ref{5.17}, \ref{5.20}) which yield the particle 
densities in models
I, II and III. While we have limited ourselves to extract but the leading
behaviour for large times, further terms could be readily obtained. 

\section{Conclusions}

In this paper, we have addressed the problem of explicitly calculating
the long-time behaviour of time-dependent averages of several types
of reaction-diffusion systems in situations (here in low dimensions)
where the presence of strong fluctuation effects prevents the use of
simple kinetic equations. We have seen that, in particular in one
dimension, the relationship of the master equation with the quantum
Hamiltonians of known integrable quantum chains can be fruitfully
employed. Our aim has been to show how to obtain, through a certain
similarity transformation of the form eq.~(\ref{BTrans}), the equivalent
stochastic systems as characterized through their quantum Hamiltonian.
Technically, the difficulty lies in satisfying simultaneously the set
of coupled inequalities coming from the positivity of the rates which enter
into the master equation, but from the examples presented here it should
be clear how to proceed for more general models. 

We have obtained the following:
\begin{enumerate}
\item Starting from the most general free fermion Hamiltonian with
site-independent nearest-neighbor interactions, eq.~(\ref{FFHam}), we have
found all equivalent (non-trivial) stochastic systems, as listed in
table~\ref{tab3}. The temporal behaviour of the averages of {\em all}
systems can be found from a {\em single} calculation as shown in
section 5 and we have seen that a single parameter $h$ describes the
physics. Specifically, we have for the density $\rho(t)$ at large times
\BEQ
\rho(t) - \rho_{\infty} \sim 
\left\{ \begin{array}{lcr} t^{-1/2} & ; & h=1 \\
t^{-3/2} \exp(-t/\tau) & ; & 0<h<1 \end{array} \right.
\EEQ
(provided the initial density $\rho_0 \neq 0$) where $\rho_{\infty}$
is the steady-state density and $\tau=\frac{1}{4} (1-h)^{-1}$ is the
relaxation time. Because of the particle-hole symmetry, the same form
holds for the density of holes when $-1\leq h <0$. 
It is known that there exist larger classes of reaction-diffusion systems
which have the same Hamiltonian spectrum as a free fermion chain with
site-dependent interactions \cite{uralt,Pesc95}. The explicit transformation
generalising (\ref{BTrans}) between these systems remains to be found.
Another interesting extension would be the consideration of these
stochastic systems on quasiperiodically modulated lattices. 

\item As a second example, we have started from the most general
Hamiltonian which has the same spectrum as the XXZ chain (with
site-independent interactions) and is thus treatable thorugh the
Bethe ansatz. Up to a rescaling in the rates, we have seen in section 4
that there is a three-parameter family of stochastic systems equivalent
to the integrable XXZ chain. The conditions for this type of integrability
can be stated exclusively in terms of the reaction rates. Furthermore, we
have seen that if we require parity symmetry (no left-right bias in the
reaction rates) this system is equivalent (working with the Hamiltonian
$H^{III}$ in section 4) to the known reaction-diffusion system
eq.~(\ref{HMat}) with only irreversible reactions. The long-time
behaviour of the model (\ref{HMat}) depends on the two parameters
$\delta$ and $\Delta=1+\delta-\alpha-\gamma$ and is given in table~\ref{tab2}. 
While for $\delta=0$, densities and correlators decay algebraically in time,
we find that for $\delta\neq 0$, the relaxation times for multiparticle
correlators depend on wether
or not there exist low-lying bound states in the multiparticle sectors of the
XXZ chain. 

This observation should remind us to be careful when extrapolating results
an insights from many ongoing calculations, 
which often implicitly contain a free fermion
condition (which is usually formulated as specifying an infinite reaction
rate on the encounter of two particles), to more general situations, where
non-trivial bound states of the quantum chain may reflect themselves in the
values of the relaxation times. 

\end{enumerate}

\section*{Acknowledgements}
We are grateful to  G.M. Sch\"utz for valuable comments
during the early stages of this work. 
It is also a pleasure to thank R.B. Stinchcombe and R. Wilman 
for useful discussions. 
MH and EO were supported by a grant of the EC program 
`Human Capital and Mobility'. JES is supported by the grant 
PRAXIS XXI/BD/3733/94 - Portugal.

\newpage
\appendix                    
                             
\appsection{A}{Non-hermitian free fermion Hamiltonians}

Here we discuss how the standard techniques \cite{Lieb61} of reducing the 
diagonalization of $2^L \times 2^L$ free
fermion Hamiltonian matrices to the diagonalization of a much smaller
matrix can be extended to non-hermitian cases as well. 
For periodic boundary conditions, the complete diagonalization can be achieved
by a sequence of Jordan-Wigner, Fourier and Bogoliubov transformations,
as detailed in \cite{Lush86,Gryn94,Hinr95}. 
Here we want a general formulation which 
treats all boundary conditions and space-dependent couplings on the
same level. 

The problem
of diagonalizing a non-hermitian quantum Hamiltonian such as (\ref{Hsigma})
is through a Jordan-Wigner transformation
\BEQ
\sig_m^+ = c_m \exp\left( i\pi\sum_{j=1}^{m-1} c_j^{\dag} c_j \right) 
\;\; , \;\;
\sig_m^{-} = \exp\left( i\pi\sum_{j=1}^{m-1} c_j^{\dag} c_j \right) c_m^{\dag}
\EEQ 
reduced to diagonalizing the following quadratic form
\BEQ
H = -\sum_{n,m=1}^L \left[ c_n^{\dag} A_{nm} c_m 
+ \frac{1}{2} c_{n}^{\dag} B_{nm} c_m^{\dag} 
+\frac{1}{2} c_{n} D_{nm} c_{m} \right]
\EEQ
where $A,B,D$ are $L\times L$ matrices. For the system (\ref{Hsigma}) one
has for periodic boundary conditions
\BEQ
A = \left( \begin{array}{cccccc} 
h   & D_2 &        &        & & -D_1 (-1)^{\cal N} \\
D_1 & h   & D_2    &        & &                    \\
    & D_1 & h      & D_2    & &                    \\
    &     & \ddots & \ddots & \ddots   &           \\
    &     &        &  D_1   &  h & D_2             \\
-D_2 (-1)^{\cal N} & & &    & D_1 & h 
\end{array} \right)
\EEQ
\BEQ
B = \eta_2 \left( \begin{array}{cccccc}
0  & 1  &        &        & & (-1)^{\cal N} \\
-1 & 0  & 1      &        & &               \\
   & -1 & 0      & 1      & &               \\
   &    & \ddots & \ddots & \ddots &        \\
   &    &        & -1     & 0      & 1      \\
-(-1)^{\cal N} & & &      & -1     & 0
\end{array} \right) 
\;\; , \;\;
D = \eta_1 \left( \begin{array}{cccccc}
 0 & -1 &        &        & & -(-1)^{\cal N}\\
 1 & 0  & -1     &        & &               \\
   &  1 & 0      & -1     & &               \\
   &    & \ddots & \ddots & \ddots &        \\
   &    &        &  1     & 0      & -1     \\
(-1)^{\cal N}  & & &      & 1      & 0
\end{array} \right)                     
\EEQ
where ${\cal N}=\sum_{j=1}^{L} c_m^{\dag} c_m$ is the fermion number
operator and the $c_m$ satisfy the usual anticommutation relations. 
For more general nearest-neighbor interactions and/or boundary conditions, 
the precise form of
the matrices $A,B,D$ will be affected but not the diagonalization technique
to be described. 

Following \cite{Lieb61}, we seek new fermionic variables $\xi_{q}$ 
which render $H$ diagonal
\BEQ \label{A:Hxi}
H =\sum_q \Lambda_q \left( \xi_q^+ \xi_q - \frac{1}{2} \right)
\EEQ
and which are defined through the following canonical transformation
\BEQ
\xi_q^+ = \sum_m \left(\alpha v_{qm}c_m^{\dag}+\alpha^{-1}u_{qm}c_m\right)
\;\; , \;\;
\xi_q = \sum_m \left(\alpha u_{qm}^*c_m^{\dag}+\alpha^{-1}v_{qm}^*c_m\right)
\EEQ
where $u,v$ are $L\times L$ matrices and $\alpha$ is a free parameter. 
For $\alpha=1$ we recover the canonical transformation for hermitian $H$
\cite{Lieb61}. We stress that $\xi_q^{+} \neq \xi_q^{\dag}$, the hermitian
conjugate of $\xi_q$, when 
$\alpha\neq 1$. But all what we need are the anticommutation relations
\BEQ
\anti{\xi_{q}^+}{\xi_p} = \delta_{q,p} \;\; , \;\;
\anti{\xi_q^+}{\xi_p^+} = \anti{\xi_q}{\xi_p} = 0
\EEQ
which imply that
\BEQ
\sum_m \left( u_{qm} u_{pm}^* + v_{qm} v_{pm}^* \right) = \delta_{q,p}
\;\; , \;\;
\sum_m \left( u_{qm} v_{pm} + v_{qm} u_{pm} \right) = 0
\EEQ
We can thus use $\xi_q^+$ and $\xi_q$ as particle creation and annihilation 
operators to generate the entire spectrum of $H$ from the one-particle
energies $\Lambda_q$. 
Now, from (\ref{A:Hxi}) we must have $\comm{\xi_q}{H}=\Lambda_q \xi_q$
and this leads to
\BEQ
\Lambda_q \alpha u_{qm}^* = 
\sum_n \left( \alpha u_{qn} A_{mn} - \alpha^{-1} v_{qn} B_{nm} \right)
\;\; , \;\;
\Lambda_q \alpha^{-1} v_{qm}^* = -
\sum_n \left( \alpha u_{qn} D_{nm} + \alpha^{-1} v_{qn} A_{nm} \right)
\EEQ
{}From the structure of the Hamiltonian (\ref{Hsigma}), it is apparent that
$B = - (\eta_2/\eta_1) D$. We now choose $\alpha = (\eta_2/\eta_1)^{1/4}$.
Introducing a vector notation $(\lvk{u_q})_n = u_{qn}$, we then have
\BEQ
\Lambda_q \left( \lvk{u_q}^* , \lvk{v_q}^* \right) = 
\left( \lvk{u_q}^* , \lvk{v_q}^* \right)
\left( \matz{A^T}{-\alpha^2 D}{\alpha^2 D}{-A} \right)
\EEQ
which reduces the problem to the diagonalization of a $2L \times 2L$ matrix. 

A further simplification can be achieved when $A$ is symmetric, $A=A^T$
(which in the context of the model (\ref{Hsigma}) means $D_1=D_2$). 
We then find, analogously to \cite{Lieb61}
\BEQ
\Lambda_q^2 \left( \lvk{u_q}^* + \lvk{v_q}^* \right) = 
\left( \lvk{u_q}^* + \lvk{v_q}^* \right) {\cal M} \;\; , \;\;
{\cal M} = \left( A +\alpha^2 D\right)\left( A-\alpha^2 D\right)
\EEQ
and have reduced the problem to the diagonalization of an $L\times L$ matrix.
In particular, since $\cal M$ is hermitian, it follows in this case that all
$\Lambda_q$ are real and can be taken to be positive. The hermitian case
treated in \cite{Lieb61} is recovered for $\alpha=1$. 

\appsection{B}{Some remarks on the transformation (\ref{HTilde})}

We add some technical remarks on the transformation (\ref{HTilde}) used to
find stochastic systems from the free fermion Hamiltonian (\ref{Hsigma}). 

\subsection{The case $D_1+D_2=0$}

When $D_1+D_2=0$, the analysis of section~3 has to be repeated. First,
probability conservation (\ref{StoCon}) is implemented by fixing $\eta_{1,2}$
\BEQ
\eta_{1,2} = \pm h \, (\eta(a,Y))^{\pm 1} \;\; , \;\;
\eta(a,Y) = \left( \frac{1-a+Y+aY}{-1+a+Y+aY} \right)^2
\EEQ
The parity symmetric matrix elements are
\BEA
-2\wit{\alpha} &=& -a^2 h (1-a+Y^2+aY^2) \cdot G_1\cdot M \nonumber \\
-2\wit{\nu}    &=& a^{-1} h (-1+a+Y^2+aY^2)\cdot G_2\cdot M \nonumber \\
-\wit{\gamma}  &=& \frac{1}{2} h a (a^2-1) (Y^2-1)\cdot G_1\cdot M 
\nonumber \\
-\wit{\delta}  &=& h a (Y^2-1) (1-a+Y^2+aY^2)\cdot G_3\cdot M \nonumber \\
-\wit{\beta}   &=& -h (Y^2-1) (-1+a+Y^2+aY^2)\cdot G_3\cdot M \nonumber \\
-\wit{\sig}    &=& \frac{1}{2a} h (a^2-1) (Y^2-1)\cdot G_2\cdot M 
\nonumber \\
-\wit{D}       &=& -\frac{1}{2} h (a^2-1) (Y^2-1)^2\cdot G_2\cdot M \label{B2}
\EEA
where 
\BEA
G_1 &=& 1-2a+a^2+6Y^2-2a^2Y^2+Y^4+2aY^4+a^2Y^4 \nonumber \\
G_2 &=& 1-2a+a^2-2Y^2+6a^2Y^2+Y^4+2aY^4+a^2Y^4 \nonumber \\
G_3 &=& 1-2a+a^2+Y^2+2aY^2+a^2Y^2 \nonumber \\
M^{-1} &=& Y (1-a+Y+aY)^2 (-1+a+Y+aY)^2
\EEA
and we see that $h$ only enters into the matrix elements as a constant factor
and thus plays no role whatsoever. The parity nonsymmetric matrix elements are
\BEA
\wit{\delta}' &=& a (D_1 -g) \frac{Y^2-1}{4Y} \nonumber \\
\wit{\sig}'   &=& a^{-1} (D_1+g) \frac{Y^2-1}{4Y} \nonumber \\ 
\wit{\gamma}' &=& a (D_1+g) \frac{Y^2-1}{4Y} \nonumber \\ 
-\wit{\beta}' &=& a^{-1} (D_1-g) \frac{Y^2-1}{4Y} \nonumber \\
\wit{D}'      &=& - D_1 \frac{Y^2+1}{2Y} 
\EEA 

Second, we apply the reality criterion. Since both 
$\wit{\delta}'/\wit{\beta}'$ and $\wit{\gamma}'/\wit{\sig}'$ must be real, 
it follows that $a^2$ is real. In fact, we find that the reality conditions
(\ref{aYFaelle}) also hold here. Third, one localizes the curves in
$(a,Y)$ space where the parity symmetric rates change sign. Mapping out the
corresponding areas, we find as before that the positivity conditions
(\ref{StoCon}) for the rates cannot simultaneously be satisfied. Finally,
one might try to use the factorized structure of the matrix elements in
(\ref{B2}) to put selectively some rates to zero. Carrying out the calculation,
we find that no stochastic systems occur, since there will be always 
pairs of parity-symmetric rates which are not positive multiples of each other.

In conclusion, the case $D_1+D_2=0$ does not lead to any stochastic system.

\subsection{Alternative forms of the transformation matrix} 

Here we briefly discuss a few alternative transformation matrices.
Begin with the transformation matrix
\BEQ
B = \left( \matz{x}{y}{0}{x^{-1}} \right)
\EEQ
where $x$ and $y$ are free parameters. It is enough here to consider the
unbiased case $D_1 = D_2 =1$. Probability conservation is then implemented
through $g=0$ and
\BEQ
\eta_1 = (h+1) (1+xy)^2 / x^4  \;\; , \;\; \eta_2 = (1-h) x^4 /(1+xy)^2
\EEQ
For the rates, we find in particular
\BEQ
\wit{\sig} = -\wit{\beta} = (h-1) xy / (1+xy)^2
\EEQ
and, from the positivity conditions, we must have either $xy=0$ or $h=1$.
In the first case, the transformation $B$ is singular for $x=0$ and trivial
for $y=0$. In the second case, $\eta_2=0$ and we find
\BEQ \label{HAC}
\wit{H} + 1 = \left( \begin{array}{cccc}
0 & 0 & 0 & -2(1+2xy) \\
0 & 1 & -1 & 2 xy \\
0 & -1 & 1 & 2 xy \\
0 & 0 & 0 & 2 \end{array} \right)
\EEQ
which is indeed stochastic for $-\frac{1}{2} \leq xy \leq 0$. We recover the
annihilation-coagulation model \cite{Henk95,Simo95} discussed in section~2. 
As discussed in section~3, we can include surface fields to allow for biased
reaction-diffusion rates.
Alternatively, one might try
\BEQ
B = \left( \matz{x}{-y^{-1}}{y}{0} \right)
\EEQ
The only stochastic form is (\ref{HAC}) with $y\rar 1/y$. Considering  the
inverses of the above transformation matrices corresponds to the 
particle-vacancy exchange $A \leftrightarrow \es$. 

Finally, we consider (\ref{Hsigma}) with $D_1=D_2=D$ and $\eta_1=\eta_2=\eta$.
The transformation matrix is
\BEQ
B = \left( \matz{b_{11}}{b_{12}}{b_{21}}{b_{22}} \right)
\EEQ 
Let $\wit{C}_m =\sum_{k=1}^{4} (\wit{H}_{j,j+1})_{k,m}$. A necessary condition
for probability conservation is $\wit{C}_2 = \wit{C}_3$. This implies
\BEQ \label{B.11}
\frac{ (b_{11}+b_{21}) ( b_{21}+b_{22} ) 2g }{\det B} = 0
\EEQ
The case $g=0$ was discussed in section~3. If $b_{11}+b_{21}=0$, the two
other conditions from probability conservation require that
\BEQ
h - \eta \left(\frac{b_{22}}{b_{11}}\right)^2 = 
h - \eta \left(\frac{b_{12}}{b_{11}}\right)^2 =
h + \eta \frac{b_{12}b_{22}}{b_{11}^2}
\EEQ
This implies that either $b_{12}^2 = b_{22}^2$ or that $\eta=0$. In the
first case, $b_{12}=-b_{22}$ renders $B$ singular and from $b_{12}=b_{22}$,
it follows $\eta=0$ anyway. So we take the second case. From $\wit{\alpha}$
and $\wit{\gamma}$, it then follows that $D=0$ as well. The resulting
Hamiltonian is
\BEQ
\wit{H} = \left( \begin{array}{cccc} 
-2h V & 2h_2 U & 2h_1 U & 0 \\
2h_2 V & -2h_1 V - 2h_2 U & 0 & 2h_1 U \\
2h_1 V & 0 & -2h_1 U - 2h_2 V & 2h_2 U \\
0 & 2h_1 V & 2h_2 V & -2h U \end{array} \right)
\EEQ
where 
\BEQ
U = \frac{b_{12}}{b_{12}+b_{22}} \;\; , \;\; V = \frac{b_{22}}{b_{12}+b_{22}}
\EEQ
This is stochastic provided $h_1, h_2 \leq 0$ and $b_{12}\cdot b_{22}\geq 0$. 
Physically, the system is trivial, since it consists of a set
of non-interacting particles which are spontaneously created and decay
spontaneously. Similarly, if from (\ref{B.11}) we have $b_{21}+b_{22}=0$, 
we get $\eta=D=0$ and 
\BEQ
\wit{H} = - \left( \begin{array}{cccc} 
-2h V & 2h_2 U & 2h_1 U & 0 \\
2h_2 V & -2h_1 V - 2h_2 U & 0 & 2h_1 U \\
2h_1 V & 0 & -2h_1 U - 2h_2 V & 2h_2 U \\
0 & 2h_1 V & 2h_2 V & -2h U \end{array} \right)
\EEQ 
with
\BEQ
U = \frac{b_{11}}{b_{11}+b_{21}} \;\; , \;\;
V = \frac{b_{21}}{b_{11}+b_{21}}
\EEQ
which is stochastic if $h_1, h_2 \geq0$ and $b_{11}$ and $b_{21}$ have 
the same sign. The physics is again trivial.   
                             
\appsection{C}{Quantum chain formulation for the biased voter model}
                             
Conventionally, the biased voter model \cite{Dick95} 
is defined as a Markov process
where the sites on the lattice are either in state 0 or state 1. If site
$j$ is in state 0, it changes to state 1 at a rate $\lmb r_1$ 
where $r_1$ is the
number of nearest neighbors of the site $j$ which are in state 1. Similarly,
sites in state 1 change to state 0 at rate $r_0$. Thus the elementary
processes involve three sites rather than two as it is in the
quantum Hamiltonians we consider here. Now consider the stochastic quantum
Hamiltonian $H = \sum_{j} H_{j,j+1}$ with two-body matrix
\BEQ
H_{j,j+1} = \left( \begin{array}{cccc}
0 & -1 & -1 & 0 \\
0 & \lmb+1 & 0 & 0 \\
0 & 0 & \lmb+1 & 0 \\
0 & -\lmb & -\lmb & 0 \end{array} \right)
\EEQ
{}From this we construct the three-sites contribution
\BEQ
H_{j-1,j,j+1} = {\bf 1}\otimes H_{j,j+1} + H_{j-1,j}\otimes {\bf 1}
\EEQ 
which completely describes the transitions of the central site $j$. The
transitions of the central site occur with the rates
\BEQ
\begin{array}{ccccc}
\es A \es \rar \es \es \es & 2 &~~;~~& A\es A \rar A A A & 2\lmb \\
\es A A   \rar \es \es A   & 1  &~~;~~& \es\es A\rar\es A A & \lmb \\
A A \es \rar A   \es \es   & 1  &~~;~~& A\es\es \rar A A\es & \lmb \\
A A A \rar A \es A         & 0     &~~;~~& \es\es\es\rar\es A\es & 0
\end{array}
\EEQ
Now, if one identifies $A$ with the state 1 and $\es$ with the state 0, one
reproduces for the central site $j$ the rules of the biased voter model
as specified above. Carrying out the sum over all lattice sites yields the
same rules for all sites $j$. 

\appsection{D}{The case $P=0$ for the XXZ chain}

We give here the details of the calculations leading to the results stated
in section 4. The cases II and III must be considered separately. 

\subsection{Case II}

The following notation is used. We write $b_1 = b_{11}$ and $b_2 = b_{22}$
and use the parameters $Q$ and $\Omega$ from table~\ref{tab4}. Then the
parity-symmetric rates are
\BEA
\wit{\alpha} &=& \frac{\Omega}{2} \frac{b_1^3}{b_2} \nonumber \\
\wit{\beta} &=& \frac{b_1 -1}{2 b_2} \left[ 2\Omega b_1(b_1-1)-b_2 Q\right]
\nonumber \\
\wit{\gamma} &=& -\frac{b_1}{2 b_2} \left[ 2\Omega b_1 (b_1 -\frac{1}{2})
-b_2 Q \right] \nonumber \\
\wit{\delta} &=& \frac{b_1}{2 b_2} \left[ 2\Omega b_1 (b_1 -1) + b_2 Q\right]
\nonumber \\
\wit{\nu} &=& \frac{\Omega}{2} \frac{(b_1-1)^3}{b_2} \label{CIIPNull}\\
\wit{\sigma} &=& -\frac{b_1-1}{2 b_2}\left[ 2\Omega (b_1-\frac{1}{2})(b_1-1)
+b_2 Q\right] \nonumber \\
\wit{D} &=& -\frac{1}{2 b_2} \left[ 2\Omega (b_1-\frac{1}{2})(b_1-1) +b_2
(h_{23}+h_{32}) \right] \nonumber 
\EEA
Now, we can always arrange to have $\Omega$ positive. To see this, recall that
the transformation matrix $B$ viewed as function of $b_2$ satisfies
\BEQ
B(b_2) = B(-b_2) \left( \begin{array}{cc} 1 & 0 \\ 0 & -1 \end{array}\right)
\EEQ
Therefore a change of sign in $b_2$ changes the sign of the matrix elements
$h_{12}, h_{24}, h_{34}$. Since $h_{12}$ does not occur in 
eqs.~(\ref{CIIPNull}), we can always absorb the sign of $\Omega$ into $b_2$. 
Next, from the positivity of $\wit{\alpha}$ and $\wit{\nu}$ one has the
inequalities $\Omega b_1 b_2 \geq 0$ and $\Omega (b_1-1) b_2\geq 0$. 
The following two cases are to be distinguished: \\
{\bf 1. $b_1 \geq 1$ and $b_2 \geq 0$.} From $\wit{\beta}\geq 0$ we get
$b_1(b_1-1) \geq Q/(2\Omega) b_2$ or $b_1 \geq 1+(Q/(2\Omega)(b_2/b_1) =:
1+ {\cal U}$. On the other hand, from $\wit{\gamma}\geq 0$ we have
$b_1(b_1-\frac{1}{2}) \leq Q/(2\Omega) b_2$ or
$1 \leq b_1 \leq \frac{1}{2} +(Q/2\Omega)(b_2/b_1) = \frac{1}{2}+{\cal U}$. 
This implies that ${\cal U} \geq \frac{1}{2} > 0$. Now, taking the two
inequalities together, we have
$1+{\cal U} \leq b_1 \leq \frac{1}{2} +{\cal U}$ which is impossible. The only
escape is to set $\Omega=Q=0$. \\
{\bf 2. $b_1 \leq 0$ and $b_2\leq 0$.} From the positivity of 
$\wit{\gamma}$ and $\wit{\sigma}$ we get $2\Omega (b_1-x)(b_1-1/2) \mp b_2 Q
\leq 0$, and $x=0,1$, respectively. Adding these two relations gives 
$(b_1-1)^2\leq 1/2$,
which would imply $0\geq b_1 =1/2$, which is impossible. Again the only
escape is to set $\Omega=Q=0$. 

But if $\Omega=Q=0$, all rates with the exception of $\wit{D}$ vanish, which
means that the only possibility is pure diffusion. 

\subsection{Case III}

Using the same notation as above for case II, we find the 
parity-symmetric rates 
\BEA
-2\wit{\alpha} &=&
b_1^2\left(-b_1(\Omega+Q)\right)/(b_1+b_2-1)\nonumber\\
-2\wit{\nu}    &=& 
(1-b_1)^2\left(-b_1(\Omega+Q)+\Omega+Q \right)
\nonumber\\     
-2\wit{\gamma}  &=& b_1 \left (
2(\Omega+Q)b_1^2-b_1(\Omega+2Q)-b_2Q +Q\right)/(b_1+b_2-1)
\nonumber\\    
-2\wit{\delta} &=&-b_1 \left (
2(\Omega+Q)b_1^2-b_1(2\Omega+Q)+b_2 Q
-Q\right)/(b_1+b_2-1)\nonumber\\   
 -2\wit{\beta}   &=&(1-b_1) \left (
2(\Omega+Q)b_1^2-b_1(2\Omega+3Q)-b_2Q +Q\right)/(b_1+b_2-1)
  \\  
-2\wit{\sig}    &=& (b_1-1) \left (
2(\Omega+Q)b_1^2-b_1(3\Omega+2Q)+b_2 Q
+\Omega\right)/(b_1+b_2-1) \nonumber \\  
-2\wit{D}       &=&
\left[ 2b_1^3(\Omega+Q)-b_1^2(3\Omega+3Q)
+b_1(\Omega+Q+h_{23}+h_{32}) \right.\nonumber\\
&+&\left. (b_2-1)(h_{23}+h_{32}) \right]/(b_1+b_2-1) \nonumber
\label{SymmHIII} 
\EEA
For a single steady state we have $0 < b_1 < 1$ and we get the inequalities
$(Q+\Omega)b_1 \leq 0$ from $\wit{\alpha}\geq 0$ and 
$(Q+\Omega)(b_1-1) \leq 0$ from $\wit{\nu}\geq 0$. 
This implies that $h_{44}=Q=-\Omega$. Inserting into the
rates, one sees that the remaining positiviy conditions become
\BEQ
(b_1+ b_2-1) > 0 \;\; , \;\; (h_{23}+h_{32})\le 0  \label{PosHIIIa} 
\EEQ
and one recovers the stochastic Hamiltonian given in section 4. 

\appsection{E}{The long time behaviour of correlation functions}

In this appendix we give a brief 
discussion of the results presented in
Table 2, namely we show how the 
long time behaviour of the one and two-point correlation
functions may be obtained by a 
truncation of the equations of motion
for these quantities which becomes exact at
long times \cite{Henk95}. It is easy to see that 
the long time behaviour of the one
and two-point equal time 
correlation functions is determined by the lowest lying 
state in the one and two-particle sectors 
respectively, which is non-orthogonal
to the initial state, provided that
there is not a state in a sector with 
a larger number of particles
that has a lower energy. We have seen in 
section 2 that the spectrum of
Hamiltonian (\ref{HMat}) is identical to the 
spectrum of $H_{XXZ}(h,\Delta,\delta)$.
So the study of the spectrum of the one 
and two-particle sector of
$H_{XXZ}$ allows one to obtain directly the time 
decay of these two correlation
functions.
In particular it can be shown, and we will 
do this below, that there is a line of
bounds states in the two particle sector
for any $\Delta\neq0$.  The presence of those 
states has to be taken into account 
in the discussion of the behaviour of
the two point correlation functions.
The  energy of lowest state in 
the one-particle sector is, as given
by (\ref{ev2}), equal to $E=2\delta$ 
($p_1=0$). So the decay of the
one-particle correlation function 
will be determined by this quantity.
The proof of the existence of a 
line of bound states in the two-particle sector is also
straightforward.
By definition a bound state 
between two particles, 1 and 2 say,
exists when their momenta 
acquire an imaginary part. 
We see from equation (\ref{BAE0}) that such an imaginary
part must be included in the phase shift $\psi_{p_1p_2}$
and since we are interested in the thermodynamic
limit we therefore put $\psi_{p_1p_2}=
\mbox{i}\phi L$. Hence $p_{1}=\frac{2\pi}{L}n_1+i\phi$ and
$p_{2}=\frac{2\pi}{L}n_2-i\phi$.
Substituting these two equations in (\ref{BAE1}) with $\psi_{p_1p_2}$
of the above  form one obtains
\be
\frac{(1+\delta-\alpha-\gamma)
\sin[\,\frac{\pi}{L}(n_1-n_2)+i\phi\,]}{
\cos[\,\frac{\pi}{L}(n_1+n_2)\,]-(1+
\delta-\alpha-\gamma)\cos[\,\frac{\pi}{L}(n_1-n_2)+i\phi\,]}
= \cot\left(\frac{iL\phi}{2}\right) \rar \mp i
\label{E1}
\ee
for $L$ large. The minus (plus) sign
applies when $\phi$ is positive (negative),
and  we have substituted $h_{11}$,
$h_{22}$, $h_{33}$, $h_{44}$, $h_{23}$
and $h_{32}$ by their values as given 
in (\ref{HMat}). We shall from now on, and as
in section 2, use the abbreviation 
$\Delta=1+\delta-\alpha-\gamma$. We can, after
some trivial manipulations write (\ref{E1}) as
$\cos[\,\frac{\pi}{L}(n_1+n_2)\,]=\Delta\,
e^{-\mid\phi\mid}e^{\pm i\frac{\pi}{L}(n_1-n_2)}$.
{}From this equation it follows that
$n_1-n_2=k\pi$ where $k$ is an integer. It also follows that
if $\mid\Delta\mid\,<1$ then the maximum value $
\cos[\,\frac{\pi}{L}(n_1+n_2)\,]$ 
can achieve is $\mid\Delta\mid\,<1$.
So there is no bound state with zero 
total momentum within this
range of $\Delta$. The energy of the 
bound state can be calculated from
(\ref{ev2}) after the appropriate 
substitutions and the use of the
cosine addition formula  and is equal to
\be
E_{b}=4(1+\delta)-2\Delta-\frac{2}{\Delta}
\cos^{2}\left[\,\frac{\pi}{L}(n_1+n_2)\,\right]\enspace.
\label{E2}
\ee
Now one can ask when this energy 
will be lower than the lowest lying
state in the continuum namely the
state with energy $E=4\delta$
($p_1=p_2=0$). This will happen
if $4-2(\Delta+\frac{1}{\Delta}
\cos^{2}\left[\,\frac{\pi}{L}(n_1+n_2)\,\right])<0 $ subjected
to the restriction $\cos[\,\frac{\pi}{L}(n_1+n_2)\,]
=(-1)^k\Delta\,e^{-\mid\phi\mid}$. From these
equations one obtains $\Delta>1$ and 
$E=4\delta -2(\Delta-2+\Delta^{-1})$
for the energy of the lowest lying state in the line of bound states
($n_1=-n_2$).  So for this range of the parameter the
long time behaviour of two point 
correlation functions will be determined by
the energy of this state. For the 
other values of $\Delta$ the decay
is determined by the state in the 
continuum with lowest energy
($E=4\delta$). These are the results 
presented in Table 2.

Let us now show how these results arise in
the context of the equations of motion 
for the correlation functions, following \cite{Henk95}.
First let us consider the one-point 
correlation function. The equation
of motion for the expectation value is
 obtained by using the Heisenberg 
equation of motion $\frac{d}{dt}
\bra{s}\nac{j}\ket{\Psi}=\bra{s}[H,\nac{j}]\ket{\Psi}$  
where $H$ is given by (\ref{HMat})
and we are taking the average
in the initial state $\ket{\Psi}$ 
in the way indicated by (\ref{AVE}).
Using the properties of $\bra{s}$ one obtains
\begin{eqnarray}
\frac{d}{dt}\ave{\nac{j}}&=&\ave{\nac{j+1}}+
\ave{\nac{j-1}}-2(1+\delta)\ave{\nac{j}}\nonumber\\
& &\mbox{}+(\,\delta-\gamma-2\alpha)
\,(\,\ave{\nac{j}\nac{j+1}}+
\ave{\nac{j-1}\nac{j}}\,)
\label{E3}
\end{eqnarray}
where we are using the abbreviation 
$\ave{\nac{j}}=\bra{s}\nac{j}\ket{\Psi}$.
Now one knows that the energy of the 
two particle lowest lying state (either
the state with $p_1=p_2=0$ for $\Delta\leq 1$ or the bound state
for $\Delta>1$) is higher than the energy of
the lowest lying state of the one 
particle sector (which has an energy equal
to $E=2\delta$ ($p_1=0$)). So the last
term of the r.h.s of (\ref{E3}) has a faster 
decay than the remaining ones 
since the decay of this term is determined
by a higher value of the energy.
Hence we can drop it at large times \cite{Henk95}. 
Assuming a translationally invariant 
initial state with $\ave{n_{j}(0)}=\rho_0$,
one finds for the average particle 
density at $t$ large 
$\ave{\nac{j}}=\rho_{0}e^{-2\delta t}$, 
which is the result presented in
Table 2.
The equation for the two point 
correlation function is obtained in a similar
way. Taking the expectation value 
of the Heisenberg equation of motion
for $\nac{j}\nac{l}$ ($l>j$) one obtains
\begin{eqnarray}
\frac{d}{dt}\ave{\nac{j}\nac{l}}&=&\ave{\nac{j+1}
\nac{l}}+\ave{\nac{j-1}\nac{l}}+
\ave{\nac{j}\nac{l-1}}\nonumber\\
& &\mbox{}+\ave{\nac{j}\nac{l+1}}
-4(1+\delta)\ave{\nac{j}\nac{l}}\nonumber\\
& &\mbox{}+(\,\delta-\gamma-2\alpha)
\,(\,\ave{\nac{j}\nac{j+1}\nac{l}}+
\ave{\nac{j-1}\nac{j}\nac{l}}\nonumber\\
& &\mbox{}+\ave{\nac{j}\nac{l}\nac{l+1}}+
\ave{\nac{j}\nac{l-1}\nac{l}}\,)
\label{E4}
\end{eqnarray}
for $l\neq j+1$ and
\begin{eqnarray}
\frac{d}{dt}\ave{\nac{j}\nac{j+1}}&=&\ave{\nac{j-1}\nac{j}}
+\ave{\nac{j}\nac{j+2}}-2(1+\delta+\gamma+\alpha)
\ave{\nac{j}\nac{j+1}}\nonumber\\
& &\mbox{}+(\,\delta-
\gamma-2\alpha)\,(\,\ave{\nac{j}\nac{j+1}\nac{j+2}}+
\ave{\nac{j-1}\nac{j}\nac{j+1}})
\label{E5}
\end{eqnarray}
for $l=j+1$. Now one proceeds as before. 
It is easy to see that there
is no bound state involving three
particles in the three particle sector.
This is so because then all 
the momenta of the three particles would have 
an imaginary part and the energy of the state
would be complex.  This is impossible given
that $H_{XXZ}$, which determines the
 spectrum, is hermitian. So the energy 
of the lowest lying state in the three
particle sector is higher than the
energy of the lowest lying state in
the two particle sector. Therefore the
three particle correlators decay faster
then the two-particle correlators
and can at large times be dropped
from equations (\ref{E4}) and (\ref{E5}).
If we assume, as above a translational 
invariant initial state then the
truncated equations for the two-point
correlation function become, where $C(r,t)=\ave{\nac{i}\nac{i+r}}$
\beq
\frac{d}{dt}C(r,t)\;=\;-4(1+\delta)
\,C(r,t)\,+\,2(\,C(r+1,t)\,+\,C(r-1,t)\,)
\label{E6}
\eeq
for $r\neq1$ and
\beq
\frac{d}{dt}C(1,t)\;=\;-2(1+\delta+\gamma+\alpha)\,C(1,t)\,
+\,2\,C(2,t)
\label{E7}
\eeq
for $r=1$.
Now equation (\ref{E7}) can be given the form (\ref{E6}) if we
define $C(0,t)=\Delta\,C(1,t)$. So now we have to solve equation
(\ref{E6}) with this boundary condition. The solution is
\beq
C(r,t)\;=\;e^{-4(1+\delta)t}\,\sum_{y=1}^{\infty}\,
[\,a_{y}I_{r-y}(4t)\,+\,b_{y}I_{r+y-1}(4t)\,]
\label{E8}
\eeq
where $I_n(4t)$ is the modified Bessel function of order $n$,
the $a_y$'s are constants defined by the initial 
distribution\footnote{If $\delta=2\alpha+\gamma$, 
eqs. (\ref{E6},\ref{E7}) are exact at all times. Then the $a_y$ are 
given by the initial conditions of the full physical problem. Otherwise,
this relation is considerably more complicated.}
and $b_{y}=\Delta\,a_{y}-(1-\Delta^2)\,
\sum_{k=1}^{y-1}\,\Delta^{y-1-k}\,a_k$.
We substitute this in (\ref{E8})
and interchange the sums between $y$ and $k$.
We get after a few manipulations
\begin{eqnarray}
C(r,t)&=&e^{-4(1+\delta)t}\,\sum_{y=1}^{\infty}\,
[\,a_{y}I_{r-y}(4t)\,+\,\Delta\,a_{y}I_{r+y-1}(4t)\,]\nonumber\\
& &\mbox{}-\,(1-\Delta^{2})\,e^{-4(1+\delta)t}\,
\sum_{k=1}^{\infty}\,a_k\Delta^{-k}\,\sum_{y=-\infty}^{
+\infty}\,I_{y+r-1}(4t)\,\Delta^{y-1}\nonumber\\
& &\mbox{}+
\,(1-\Delta^{2})\,
e^{-4(1+\delta)t}\sum_{k=1}^{\infty}\,a_k\,\sum_{y=-\infty}^{
0}\,I_{y+r+k-1}(4t)\,\Delta^{y-1}
\label{E9}
\end{eqnarray}
where we have completed the sum in $y$ in the second term and
accordingly subtracted the third term. Since 
$e^{-4t}I_n(4t)\leq1$ then the sum over $y$ in the third term is 
convergent if $\mid \Delta\mid\,>1$. Hence, provided that 
the sum over $a_k$ is convergent this term is bounded
by $(1-\Delta^2)\,e^{-4\delta t}\frac{\mid
\sum_{k=1}^{\infty}\,a_k\mid}{\mid\Delta-1\mid}$
when $\mid \Delta\mid\,>1$. One can then explicitly perform
the sum over $y$ in the second term, using the generating
function for modified Bessel functions. 
Hence if $\mid \Delta\mid\,>1$
one gets 
\begin{eqnarray}
C(r,t)&=&e^{-4(1+\delta)t}\,\sum_{y=1}^{\infty}\,
[\,a_{y}I_{r-y}(4t)\,+\,\Delta\,a_{y}I_{r+y-1}(4t)\,]\nonumber\\
& &\mbox{}-\,(1-\Delta^{2})\,
\Delta^{-r}e^{-[\,4\delta-2(\Delta-2+\Delta^{-1})\,]t}\,
\sum_{k=1}^{\infty}\,a_k\Delta^{-k}\nonumber\\
& &\mbox{}+
\,(1-\Delta^{2})\,
e^{-4(1+\delta)t}\sum_{k=1}^{\infty}\,a_k\,\sum_{y=-\infty}^{
0}\,I_{y+r+k-1}(4t)\,\Delta^{y-1}\enspace.
\label{E10}
\end{eqnarray}
One sees that at long times the first 
and third terms of this
equation decay like $e^{-4\delta t}$. 
Hence if $\Delta>1$ the decay
of the two-point correlation functions 
is dominated by the second term
and the long time behaviour of the two 
point correlation function
would be proportional to 
$e^{-[\,4\delta-2(\Delta-2+\Delta^{-1})\,]t}$.
Also we see that the second term does 
not appear if $\mid\Delta\mid\,<1$,
which is in agreement with the result 
obtained by the Bethe Ansatz that
there is no bound state with total zero 
momenta in this range of the
parameter $\Delta$, otherwise such state 
would appear in the decay of
correlation functions from a 
translational invariant state. Hence we
again obtain the results of Table 2.

\appsection{F}{Role of the boundary conditions on the Hamiltonian spectrum} 

We discuss the relationship between the quantum Hamiltonian spectra
for free and periodic boundary conditions. The main point already becomes
apparent when just considering asymmetric diffusion without any further
reactions \cite{Gwa92,Henk94,Kim95}. While this problem is still very
technical for the case of a finite particle density, the basic mechanism
becomes apparent in a simple way already for a finite number of particles.
To make this paper more self-contained, we recall the argument here, following
\cite{Henk94}.  

For free boundary conditions on a chain of $L$ sites, 
the quantum Hamiltonian reads
\BEQ \label{eq:ADHR}
H' = - \frac{1}{4\Delta}
\sum_{j=1}^{L-1} \left[ \sig_{j}^{x} \sig_{j+1}^{x}
+ \sig_{j}^{y} \sig_{j+1}^{y} + \Delta \sig_{j}^{z} \sig_{j+1}^{z}
 - \frac{q - q^{-1}}{2} \left(
\sig_{j}^{z} - \sig_{j+1}^{z} \right)  - \Delta \right]
\EEQ
where
\BEQ \label{eq:DelQ}
\Delta = \frac{q+q^{-1}}{2} \;\; , \;\; q = \sqrt{\frac{1-\eps}{1+\eps}}
\EEQ
This is the well-known XXZ quantum chain which is symmetric under
the quantum group $U_{q}SU(2)$ \cite{Kiri88,Pasq90}. 
It follows from the Bethe ansatz that for $\eps\neq 0$ the energies for
$L$ large become ${E_L}' \sim 1- \Delta^{-1}$ in 
each sector\footnote{Since $H'$ commutes with the quantum group generators
$S_{\pm}$ and $S_{z}$, for each state $\ket{M}$ in the $M$-particle sector with
energy $E'$ there a $(N+M)$-particle state $\ket{N+M} \sim S_{-}^N \ket{M}$ with
the same energy, see \cite{Sand94}.} with $N$ particles (see also below). 
We stress that the quantum Hamiltonian of
the diffusion process on a periodic
lattice {\em cannot} be obtained by simply taking
periodic boundary conditions in
eq.~(\ref{eq:ADHR}). Rather, from
the master equation $H$ can be written down directly \cite{Gwa92}
\BEQ \label{eq:GS}
H = -\frac{1}{4} \sum_{j=1}^{L} \left[
\vec{\sig}_j \cdot \vec{\sig}_{j+1} -1
+i\eps \left( \sig_{j}^{x}\sig_{j+1}^{y}-\sig_{j}^{y}\sig_{j+1}^{x}
\right)\right]
\EEQ
where $\sig^{x,y,z}$ are Pauli matrices and $\eps$ is related to the 
diffusion bias via equation (\ref{eq:DelQ}). 
It is well known that the low-lying
spectrum of $H$ is gapless (see also below). 
 
To understand the relation between $H$ and $H'$, we introduce
$\sig^{\pm}=\frac{1}{2}(\sig^{x}\pm i \sig^{y})$, and
consider the non-singular operator
\BEQ \label{eq:U}
U = \exp \left( \pi g \sum_{j=1}^{L} j \sig_{j}^{z} \right) \;\; , \;\;
U \sig_{j}^{\pm} U^{-1} = e^{\pm 2\pi g j} \sig_{j}^{\pm}
\EEQ
Choosing $q=e^{2\pi g}$, we obtain
\BEQ \label{eq:H}
H'' = U H U^{-1} = - \frac{1}{2(q+q^{-1})}
\sum_{j=1}^{L} \left[ \sig_{j}^{x} \sig_{j+1}^{x}
+ \sig_{j}^{y} \sig_{j+1}^{y} + \Delta \sig_{j}^{z} \sig_{j+1}^{z}
- \Delta \right]
\EEQ
which is indeed (almost) the Hamiltonian $H'$.
The distinction comes from the boundary conditions.
The surface field $(q-q^{-1})(\sigma_1^z-\sigma_L^z)/8\Delta$
is absent in $H''$ and one has
\BEQ \label{eq:BC}
\sig_{L+1}^{\pm} = q^{\mp L} \sig_{1}^{\pm} \;\; , \;\;
\sig_{L+1}^{z} = \sig_{1}^{z}
\EEQ
which make $H''$ non-hermitian, as $H$ already is.
It is these unusal boundary conditions which give rise to the
different properties of $H$ and $H'$, as we now show.
  
The calculation of the spectrum of $H''$ proceeds  
via the Bethe ansatz \cite{Alca88}. First, we consider the
sector with $N=1$ particle. Then the energies are in this sector
\BEQ\label{eq:energy}
E = 1- \Delta^{-1} \cos \theta \;\; , \;\;
\theta = 2\pi \left(i g + \frac{n}{L}\right)
\EEQ
where $n$ is an integer from the set
$\{ 0, \pm 1,\ldots, \pm(L/2 -1),L/2\}$. For $L$ large, the
energies become
\BEQ\label{eq:gap}
E = 1 - \frac{\cos\left( 2\pi \left(i g +\frac{n}{L}\right)\right)}
{\cosh (2\pi g)} 
\simeq  2\pi^2 \left(\frac{n}{L}\right)^{2}
+ 2\pi i \tanh(2\pi g) \frac{n}{L} + \ldots
\EEQ
We see that the real part $\Re E \sim L^{-2}$ and the imaginary part
$\Im E \sim L^{-1}$. 
 
Next, we take the sector $N=2$. From the Bethe ansatz \cite{Alca88}
we have
\BEA
&& E = (2 \Delta - cos \theta -\cos \theta')\Delta^{-1} \\
&& \theta L -2\pi i g L= 2\pi I - \Theta(\theta,\theta') \;\; , \;\;
\theta' L -2\pi i g L= 2\pi I' - \Theta(\theta',\theta) \nonumber
\EEA
where $I,I'$ are distinct half-integers from the set
$\{\pm\frac{1}{2},\pm\frac{3}{2},\ldots,\pm\frac{L-1}{2}\}$ and
\BEQ
\Theta(\theta,\theta') = 2 \arctan
\left( \frac{\Delta \sin\left( \frac{1}{2}
(\theta-\theta')\right)}{\cos\left(\frac{1}{2}(\theta+\theta')\right)
-\Delta \cos\left(\frac{1}{2}(\theta-\theta')\right)} \right)
\EEQ
Define $\widetilde{\theta}=\theta-2\pi i g$,
$\widetilde{\theta'}=\theta'-2\pi i g$. Then, for small values of
the arguments
\BEQ
\Theta(\widetilde{\theta},\widetilde{\theta'}) \simeq
2 \arctan\left( i \coth(2\pi g) \cdot
\frac{\widetilde{\theta}-\widetilde{\theta'}}{
\widetilde{\theta}+\widetilde{\theta'}} \right)
\EEQ
which is of order unity. Therefore
\BEQ
\widetilde{\theta} = \frac{2\pi}{L} a \;\; , \;\;
\widetilde{\theta'} = \frac{2\pi}{L} a'
\EEQ
where $a,a'$ are of order ${\cal O}(1)$.
It follows that the same cancellation
as observed in the sector $N=1$ also takes place
here and also that the observed
scaling of the energies does not change. Finally, for $N$ arbitrary
\BEQ\label{eq:spectrum}
E = \Delta^{-1} \left( N \Delta - \sum_{n=1}^{N} \cos \theta_n \right)
\EEQ
with
\BEQ
\theta_m L -2\pi i g L=
2\pi I_m - \sum_{n=1}^{N} \Theta(\theta_m,\theta_n)
\EEQ
and the same argument can be repeated. It follows that for any value
of $\eps$, the spectrum shows in the $L\rar\infty$ limit no energy gap. 

While this argument shows that for a finite number of particles, the real
part of the energies scales as $L^{-2}$, a much more detailed analysis
of the Bethe ansatz equations shows \cite{Gwa92,Kim95} that for a finite 
density, one rather has a scaling $\Re E \sim L^{-3/2}$. 
This arguments also goes through when particle reactions 
are added \cite{Henk94}.
In the case of more than one species of particles, the relation between $H$ and
$H'$ becomes even more involved, as explained in \cite{Dahm95}. 

A simple interpretation of this result is as follows. 
For periodic boundary conditions,
the particles are free to chase each other. In fact, 
through a Galilei transformation
on can go into the frame where the diffusion becomes 
left-right symmetric, see \cite{Henk94b}. It is
not surprising that it should take a long time for 
the system to reach a steady state (in fact, $\Im E$ is proportional to the
steady state current density)
and that the relaxation time should grow beyond all 
bounds with increasing system
size $L$. On the other hand, for free boundary 
conditions the particles cannot leave the
system. Rather, they pile up on one of the boundaries. 
The time to reach a steady
state is of the order of the time a single particle needs 
to cross the system from one
boundary to the other and should be finite.

\newpage
\zeile{2}
\noindent {\bf Figure captions}
\begin{description}
\item[Fig. 1] Test of the positivity of the reaction rates in the 
case when the transformation parameters $a$ and $Y$ are both real and
$h^2 > 1$. In each subfigure, the shaded regions correspond to the domains
where one of the following reaction rates is negative with
(a) $\wit{\alpha}$ (b) $\wit{\gamma}$ (c) $\wit{\nu}$ and (d) $\wit{\sigma}$.
\item[Fig. 2] Test of the positivity of the reaction rates in the case when
$a$ and $Y$ are both real and $h=\frac{1}{2}\sqrt{3}$. Shaded regions 
correspond to one of the reaction rates being negative with
(a) $\wit{\alpha}$ (b) $\wit{\gamma}$ (c) $\wit{\nu}$ and (d) $\wit{\sigma}$.
\item[Fig. 3] Test of the positivity of the reaction rates when the 
transformation parameters $A$ and $q$ are real and $h^2 > 1$. Shaded regions 
correspond to one of the reaction rates being negative with
(a) $\wit{\alpha}$ (b) $\wit{\gamma}$ (c) $\wit{\nu}$ and (d) $\wit{\sigma}$.
\end{description}
 
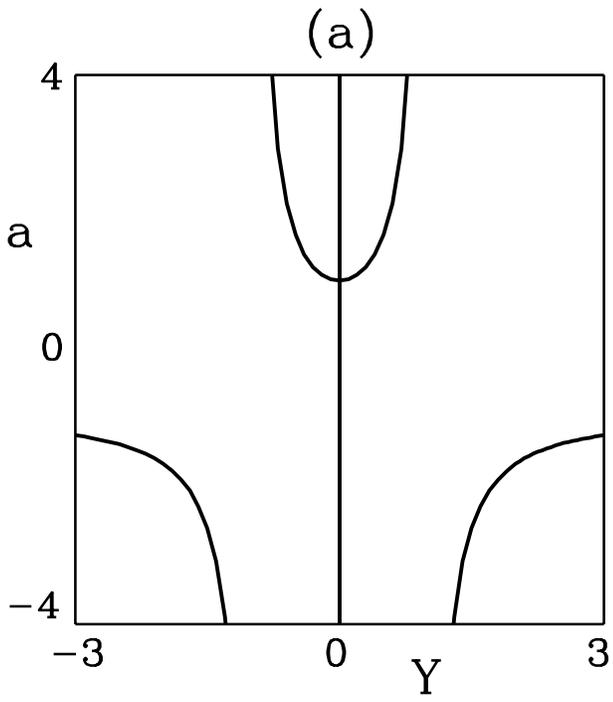
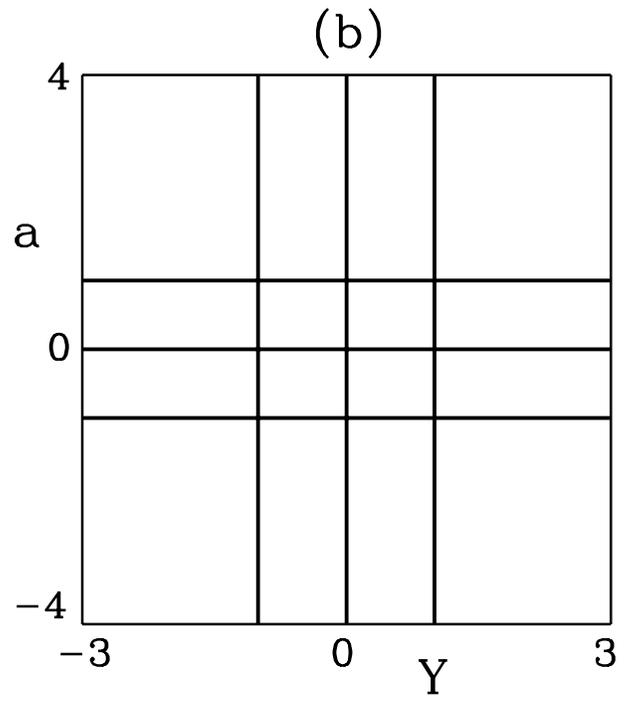
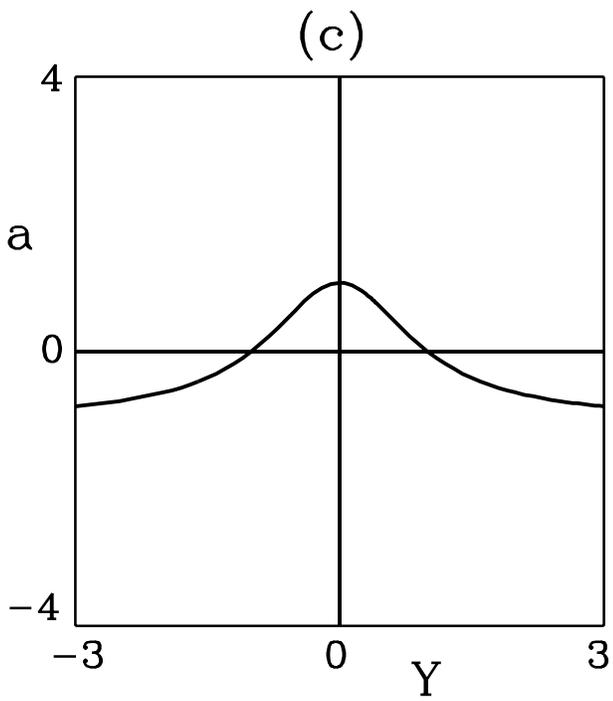
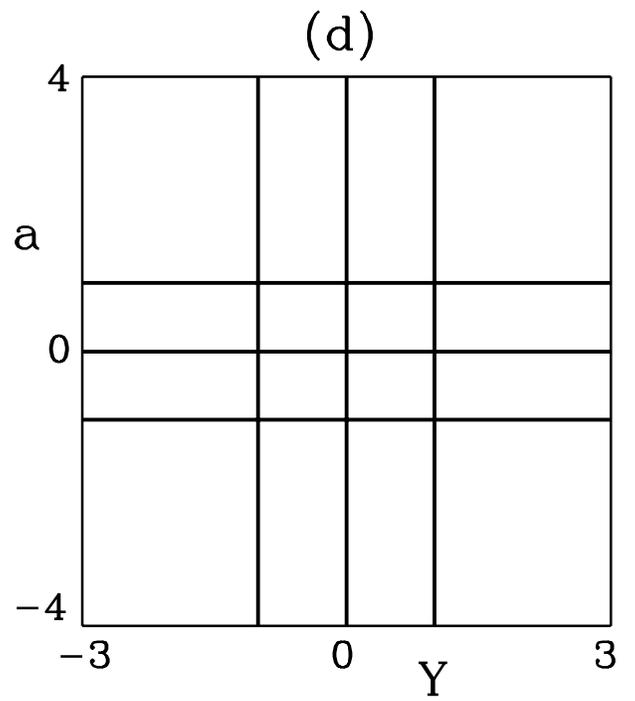
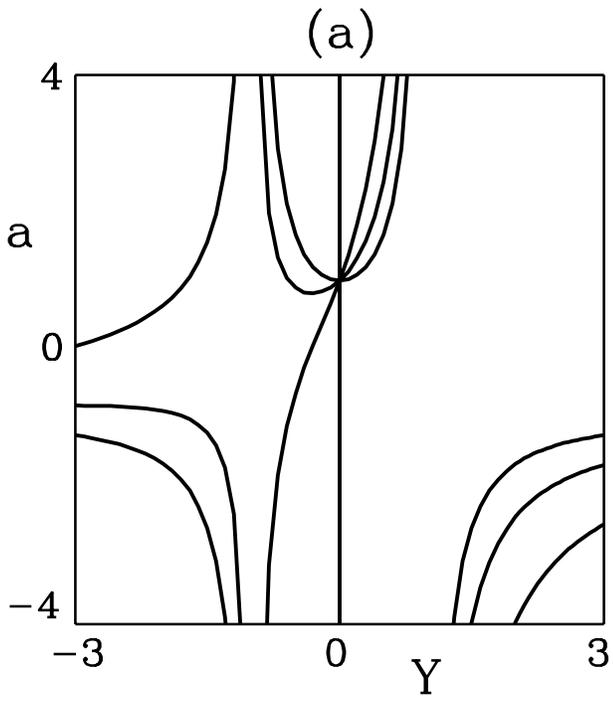
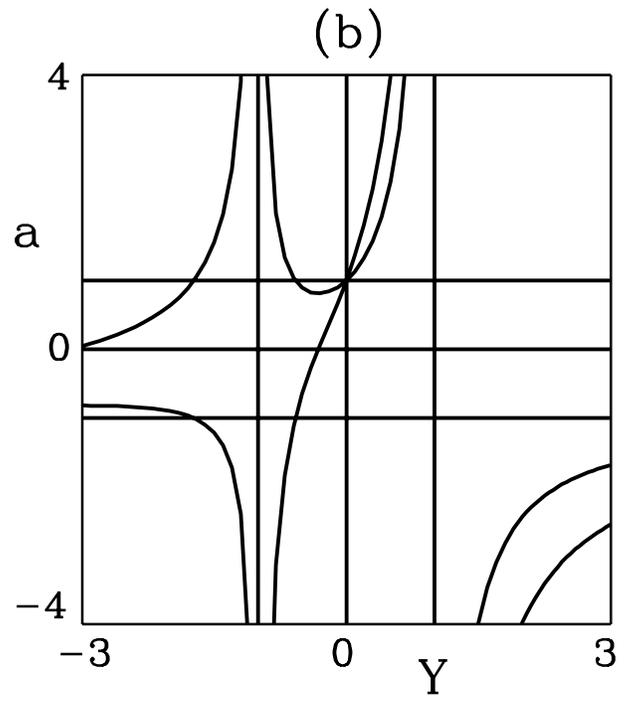
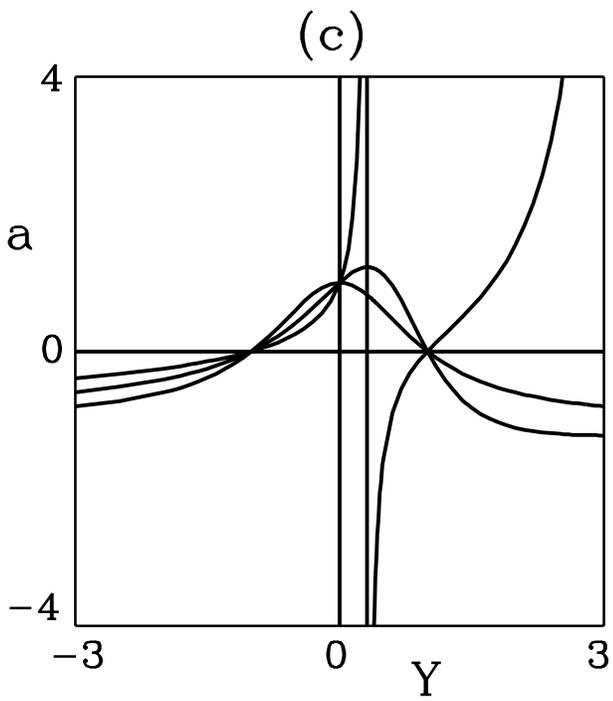
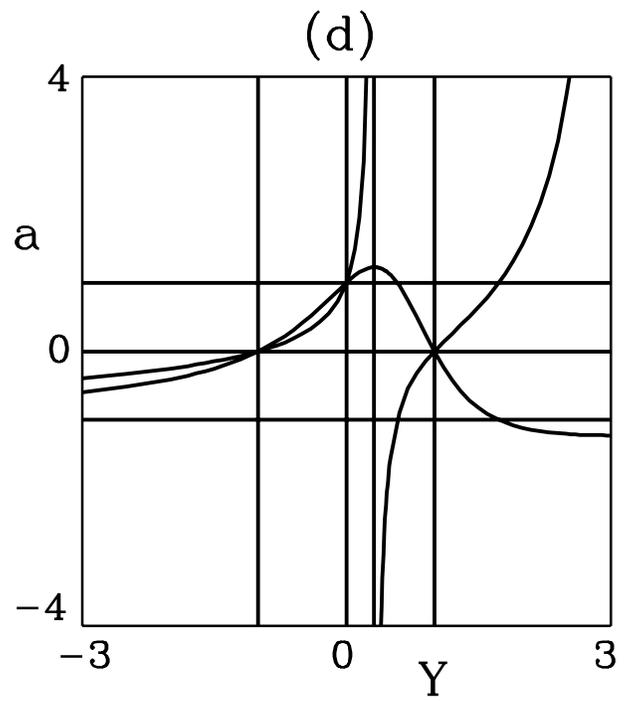
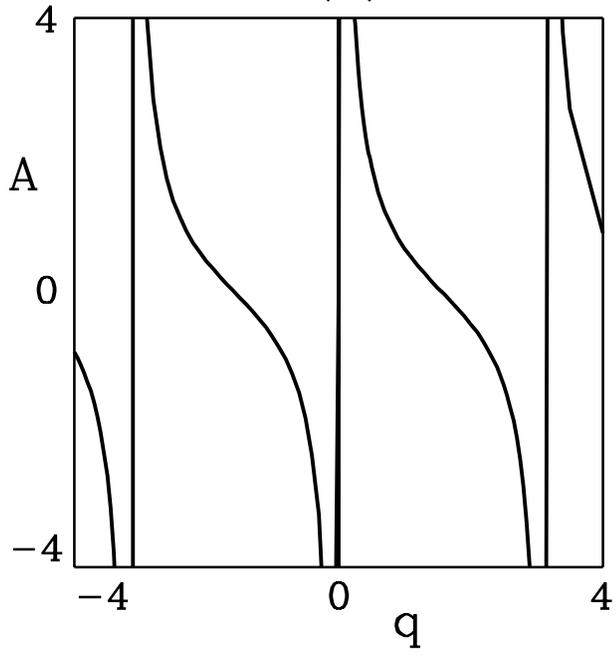
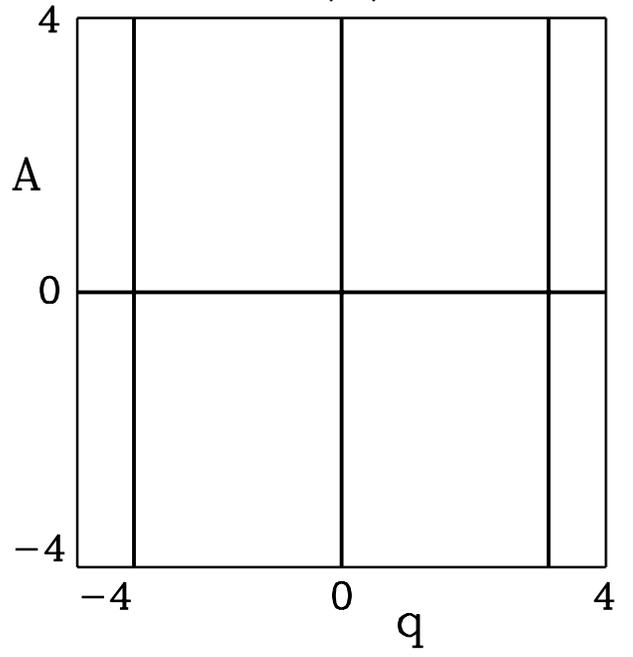
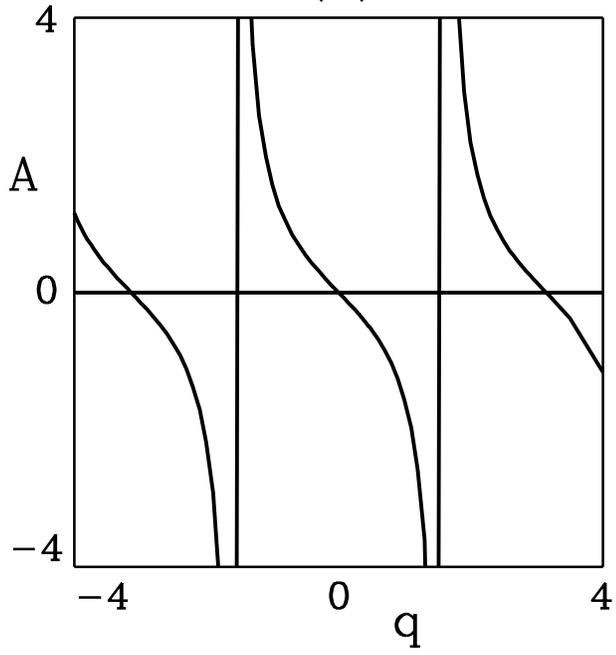
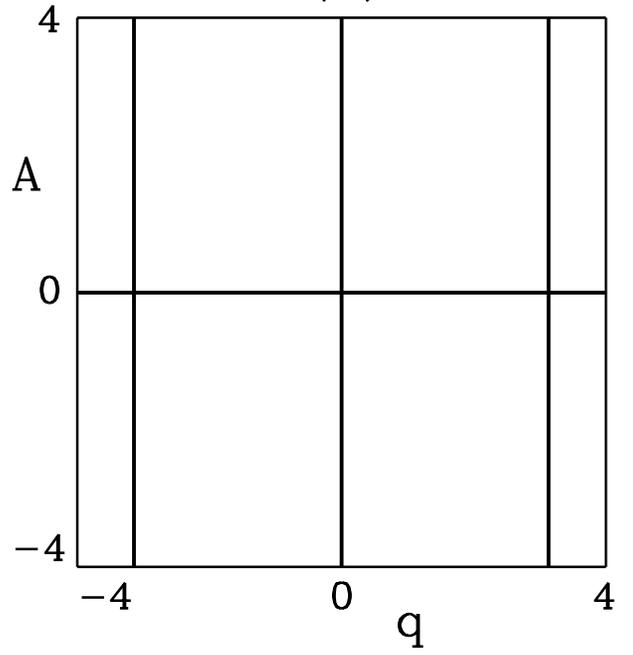
\end{document}